\documentclass[ reprint,
superscriptaddress,
 amsmath,amssymb,
 aps,
 prfluids,
,onecolumn, 9pt]{revtex4-2}

\usepackage{graphicx}
\usepackage{dcolumn}
\usepackage{bm}
\usepackage{gensymb}
\usepackage{stmaryrd}
\usepackage{xcolor}
\usepackage{multirow}
\usepackage{array}
\usepackage{siunitx}
\usepackage{orcidlink}

\newcommand{\bnabla}{\bm{\nabla}}
\newcommand{\bcdot}{\bm{\cdot}}

\newcommand{\bcdoubledot}{\bm{:}}

\DeclareMathAlphabet{\mathsfbi}{OT1}{\sfdefault}{bx}{sl}
\DeclareMathVersion{sfletters}
\SetSymbolFont{letters}{sfletters}{OML}{ntxsfmi}{b}{it}

\makeatletter
\newcommand{\bmsbilow}[1]{%
  \text{\mathversion{sfletters}$\m@th#1$}%
}

\DeclareRobustCommand{\tensor}[1]{%
  \begingroup
  \ifcat\noexpand #1\relax
    \edef\greek@test{\detokenize{#1}}%
    \edef\greek@test{\expandafter\@cdr\greek@test\@nil}%
    \edef\greek@test{\expandafter\@car\greek@test\@nil}%
    \edef\x{\the\lccode\expandafter`\greek@test}%
    \edef\y{\number\expandafter`\greek@test}%
    \ifnum\x=\y\relax
      \bmsbilow{#1}%
    \else
      \mathsfbi{#1}%
    \fi
  \else
    \mathsfbi{#1}%
  \fi
  \endgroup
}
\makeatother

\newcolumntype{M}[1]{>{\centering\arraybackslash}m{#1}}

\begin{document}

\preprint{APS/123-QED}

\title{Nonlinear Three-Dimensional Electrohydrodynamic Interactions of Viscous Dielectric Drops}

\author{Michael A. McDougall\,\orcidlink{0009-0005-2967-9437}}
\email{michael.mcdougall@strath.ac.uk}

\affiliation{%
 Department of Mathematics and Statistics, University of Strathclyde, Livingstone Tower,\\ 26 Richmond Street, Glasgow G1 1XH, United Kingdom
}%

\author{Stephen K. Wilson\,\orcidlink{0000-0001-7841-9643}}
 \email{sw3197@bath.ac.uk}

 \affiliation{%
Department of Mathematical Sciences,
University of Bath,\\
Claverton Down,
Bath BA2 7AY,
United Kingdom}%

 \affiliation{%
 Department of Mathematics and Statistics, University of Strathclyde, Livingstone Tower,\\ 26 Richmond Street, Glasgow G1 1XH, United Kingdom
}%
 
 \author{Debasish Das\,\orcidlink{0000-0003-2365-4720}}
 \email{debasish.das@strath.ac.uk}
 
\affiliation{%
 Department of Mathematics and Statistics, University of Strathclyde, Livingstone Tower,\\ 26 Richmond Street, Glasgow G1 1XH, United Kingdom
}%

\date{\today}

\begin{abstract}
When a drop of a leaky dielectric fluid is suspended in another fluid and subjected to a uniform DC electric field, it becomes polarized, leading to tangential electric stresses that drive fluid motion both inside and outside the drop. In the presence of a second drop, the dynamics of the first drop are altered due to electrohydrodynamic interactions with the second, causing the drops to translate due to dielectrophoretic forces and hydrodynamic interactions. We present a semi-analytical nonlinear three-dimensional small deformation theory for a pair of identical, widely-separated leaky dielectric drops suspended in a weakly conducting fluid. This theory is valid under conditions of large drop separation, high drop viscosity, and high surface tension, ensuring that the drops remain nearly spherical. For the first time, we develop a model within the Taylor--Melcher leaky dielectric framework that incorporates both transient charge relaxation and convection. This allows the model to capture the transition to Quincke rotation, a symmetry-breaking phenomenon in which drops begin to spontaneously rotate in sufficiently strong fields. We derive and numerically integrate coupled nonlinear ordinary differential equations for the dipole moments, shapes, and positions of the drops. Our results show good quantitative agreement with previous numerical and experimental work in the limit of zero charge relaxation and convection, and the model reduces to existing models for isolated drops and interacting solid spheres in the relevant limits. We also discuss the hysteresis in the onset of Quincke rotation of isolated drops observed in experiments. Various trajectories for pairs of drops undergoing Quincke rotation are presented, along with results for fixed drops, including their shape, the nature of electrohydrodynamic interactions (attractive or repulsive), and the conditions leading to Quincke rotation. In particular, it is shown that the onset of Quincke rotation for a pair of drops is qualitatively different from that for an isolated drop due to electrohydrodynamic interactions and from that for a pair of solid spheres due to straining flows present only in drops.

\end{abstract}

\maketitle

\section{\label{sec:level1}Introduction}

Application of an electric field across the interface between two immiscible fluids with dissimilar conductivities and electric permittivities leads to a discontinuity in the electric field, generating an electric stress on the interface. This stress can drive both interfacial deformation and fluid motion \citep{melcher1969electrohydrodynamics}. These phenomena are not only of fundamental scientific interest, but also have practical applications in the development of electrosprays \citep{collins2008electrohydrodynamic}, inkjet printers \citep{basaran2013nonstandard}, microfluidic devices such as electrohydrodynamic pumps \citep{seyed2005electrohydrodynamic}, and in improving techniques for drop coalescence \citep{eow2002electrostatic} and emulsion separation \citep{ptasinski1992electric}. To better understand the physics underlying these applications, theoretical studies have explored various paradigmatic systems involving viscous drops.

The dynamics of an isolated drop suspended in another fluid and exposed to an electric field, particularly in the case where both the drop and surrounding fluid are uncharged leaky dielectric fluids, have been extensively studied. Early theoretical models treated both the drop and the surrounding fluid as perfect dielectrics, in which case the electric stress is always normal to the drop surface and is balanced by surface tension \citep{okonski_thacher,allan1962particle}. These theories predicted that at steady state, the drop would deform into an axisymmetric prolate spheroid, with both fluids remaining stationary \citep{o1953distortion,garton1964bubbles}. These predictions generally matched experimental observations. However, some experiments revealed that initially spherical drops could deform into oblate shapes \citep{allan1962particle}, a phenomenon not explained by the early theories.

In a seminal work, \citet{taylor1966studies} argued that treating the fluids as perfect dielectrics is unrealistic, as even a small amount of conductivity can significantly alter the response of the drop to the applied field. If the fluids can support even a very small but nonzero current, the electric field drives free charges toward the interface from both domains, and the corresponding current discontinuity leads to a surface charge distribution. This distribution leads to tangential electric stresses on the drop surface, which must be balanced by hydrodynamic stresses, resulting in fluid motion both inside and outside the drop that can lead to oblate drop shapes. The assumptions used in Taylor's work were later formalized into the Taylor--Melcher leaky dielectric model \citep{melcher1969electrohydrodynamics,saville1997electrohydrodynamics}, which is now widely used for studying the electrohydrodynamics of weakly conducting (i.e., leaky dielectric) fluids. Readers are referred to \citet{papageorgiou2019} and \citet{vlahovska2019} for recent reviews on the topic.  

In his work, Taylor derived a discriminating function given by
\begin{equation}
    \mathcal{F}=(S-1)^{2}+S(1-SQ)\frac{16+19\lambda}{5(1+\lambda)},
\end{equation}
which provides a leading order (in the drop deformation) prediction of the steady drop shape; specifically, $\mathcal{F}>0$, $\mathcal{F}=0$, and $\mathcal{F}<0$ correspond to prolate, spherical, and oblate drops, respectively. Here, $S=\sigma^{+}/\sigma^{-}$, $Q=\epsilon^{-}/\epsilon^{+}$, and $\lambda=\mu^{-}/\mu^{+}$ represent the ratios of electrical conductivity, permittivity (dielectric constant), and dynamic viscosity, respectively, with the minus subscript referring to the properties of the drop and the plus subscript to those of the surrounding fluid. While predictions of the drop shape based on the discriminating function were qualitatively correct, they failed to accurately quantify the amount of deformation observed in experiments \citep{torza1971electrohydrodynamic}. Subsequent work, such as Ajayi's \citep{ajayi1978note} extension of Taylor's work to include higher-order terms in the drop deformation, did not significantly improve the agreement with experimental results. A shortcoming of both Ajayi's \citep{ajayi1978note} and Taylor's \citep{taylor1966studies} analyses is the neglect of transient charge relaxation and charge convection by interfacial fluid flow, an approximation that is justifiable in weak fields but fails in stronger fields. Isolated drops with strong electrohydrodynamic straining flows have been the subject of considerable interest, in part due to the observation of charge shocks \citep{lanauze2015,das2017sims,peng2024} and streaming \cite{brosseau2017,wagoner2021} at the equator of the drop.

In the late 19th century, studies by \citet{weiler} and \citet{quincke1896ueber} reported the spontaneous rotation of solid spheres and cylinders in sufficiently strong electric fields, a symmetry-breaking instability now known as Quincke rotation. \citet{turcu1987} and \citet{jones1984quincke} later described the mechanism behind this instability in spheres. In a uniform electric field $\bm{E}$, the charge distribution on the sphere can be approximated by a dipole moment $\bm{P}$. The dipole moment aligns with the field if the charge relaxation time $\tau=\epsilon /\sigma$ of the sphere material is smaller than that of the surrounding fluid. Conversely, if the relaxation time of the surrounding fluid is smaller, the dipole aligns antiparallel to the field. In the former case, corresponding to systems where $SQ<1$, any small rotational displacement of the dipole is stabilized by an electric torque of the form $\bm{P\times E}$. However, in the latter case, corresponding to $SQ>1$, the configuration becomes unstable above a critical electric field strength $E_{c,s}$, given by:

\begin{equation}\label{critfield}
E_{c,s}=\sqrt{\frac{2\mu^{+}(2+Q)(1+2S)}{3\epsilon^{+}\tau_{MW}(SQ-1)}},
\end{equation}
where 

\begin{equation}\tau_{MW}=\frac{\epsilon^{-}+2\epsilon^{+}}{\sigma^{-}+2\sigma^{+}}\end{equation}
is the Maxwell--Wagner relaxation time \citep{das2013electrohydrodynamic}, representing the time scale for polarization of the surface of the sphere upon application of the electric field.  In this scenario, a perturbation to the dipole gives rise to an electric torque $\bm{P\times E},$ which begins to rotate the drop and gives rise to a counteracting viscous torque. In equilibrium, the sphere attains a steady angular velocity with the dipole moment tilted away from the direction of the applied field. The direction of the rotational axis, which depends on the initial perturbation, is always perpendicular to the applied field. At steady state, the charge convection around the sphere due to its rotation is balanced by the charge accumulation due to the current discontinuity between the solid and fluid phases. Significant attention has been devoted to studying the dynamics of both solid spheres \citep{das2013electrohydrodynamic,Dolinsky} and viscous drops \citep{he2013,das2021three,salipante2013electrohydrodynamic,salipante2010electrohydrodynamics,firouznia2023spectral} undergoing Quincke rotation. The theoretical analysis of drops in electric fields is considerably more challenging than than of solid spheres due to the deforming interface and the straining component of the induced flow field, which introduce additional nonlinearities in the interfacial charge convection. 

Modelling the electrohydrodynamic interactions between two or more drops is important for understanding the microstructure of emulsions subject to electric fields \citep{varshney2012self,varshney2016multiscale,tadavani2016effect}. The first attempt to model electrohydrodynamic interactions between drops was made by Sozou \citep{sozou1975electrohydrodynamics}, who used a bispherical coordinate system to calculate the induced flow fields of a pair of stationary, identical, spherical drops aligned with the applied field. Later, Baygents, Rivette, and Stone \citep{baygents1998electrohydrodynamic} investigated the same problem using a boundary integral method, which allowed them to include the effects of drop deformation. Their simulations revealed that dielectrophoretic (DEP) forces, which are attractive in this configuration, cause the drops to move together, while the deformation of the drops can be oblate or prolate depending on the value of $\mathcal{F}$. However, if $SQ<1$ and the initial separation of the drops is sufficiently large, electrohydrodynamic flows (which scale as $\mathcal{O}(R^{-2})$ in the limit of widely separated drops, $R\to\infty,$ where $R$ is the distance between the drops) can overcome the attraction of the DEP forces (which scale as $\mathcal{O}(R^{-4})$) and cause the drops to repel. In this situation, $\mathcal{F}$ is positive, and the drops deform into prolate spheroids. More recently, \citet{zabarankin} explored the case of dissimilar drops using a spherical harmonic re-expansion technique, and was able to obtain solutions of arbitrary asymptotic accuracy in $R^{-1}$. These calculations showed that dissimilar drops can become oblate and repel one another if they are both much more viscous than the surrounding fluid (i.e., in the limit $\lambda_{1},\lambda_{2}\rightarrow\infty,$ where $\lambda_{1}$ and $\lambda_{2}$ are the viscosity ratios of the two drops) and one drop (but not both) is more conductive than the surrounding fluid (i.e., if $(S_{1}-1)(S_{2}-1)<0$, where $S_{1}$ and $S_{2}$ are the conductivity ratios of the two drops). Note that these works on drop interactions were restricted to axisymmetric configurations with the drops aligned with the applied field. \citet{sorgentone_kach_khair_walker_vlahovska_2021} were the first to investigate the interactions between identical drops aligned arbitrarily relative to the applied field in three dimensions. Their boundary element simulations revealed complex drop trajectories under the combined electrohydrodynamic and DEP and interactions depending on the electrical properties of the fluids and the initial positions of the drops. They also developed an analytical model, valid in the limits of large drop separation and small drop deformation, which qualitatively captured the drop interactions predicted by their numerical simulations for sufficiently widely separated drops. A quasi-steady version of their analysis assuming spherical drops was further validated by comparison with the experimental results of \citet{kach2022prediction}, who found that the theory described the qualitative features of their observed drop trajectories well, despite the fairly small drop separations occurring in their experiments. In the same study, \citet{kach2022prediction} extended the quasi-steady theory of \citet{sorgentone_kach_khair_walker_vlahovska_2021} to drops with different sizes and/or material properties and to systems of three or more drops. Subsequently, \citet{Sorgentone_Vlahovska_2022} studied dissimilar drops using boundary element simulations and identified conditions under which a pair of dissimilar drops can undergo self-sustained propulsion.
~\citet{kach2023nonequilibrium} have recently extended their previous work \citep{kach2022prediction} to simulate emulsions of 1000 drops in two and three dimensions and observed a wide range of collective behaviours, including sheet and chain formation and clustering. However, all of these studies on drop interactions neglect charge convection, precluding the prediction of Quincke rotation.
Recently, \citet{dong24} conducted numerical simulations to explore the possibility of self-sustained propulsion of pairs of identical counter-rotating drops due to Quincke rotation, as found previously for solid spheres by \citet{das2013electrohydrodynamic}. They found that the motion of the pair of drops varied nonlinearly with the viscosity ratio of the drops, changing direction as it increased. However, these simulations were limited to two dimensional drops. 

In this work, we present, for the first time, a three-dimensional theory that captures electrohydrodynamic interactions between a pair of drops of a viscous leaky dielectric fluid, retaining the effects of transient charge relaxation and charge convection so as to allow for Quincke rotation. This work is organized as follows. In \S\ref{problemformulation}, we describe the problem under consideration, including the governing equations and boundary conditions. In \S\ref{problemsolution}, we describe our approach to solving the problem. In \S \ref{results}, we present the results of our theory, first for freely suspended drops and then for drops fixed in space. Finally, in \S\ref{conclusions}, we summarize the key findings and suggest possible directions for future research.

\section{Problem formulation}\label{problemformulation}
\begin{figure}
    \centering
    \includegraphics[width=0.8\textwidth]{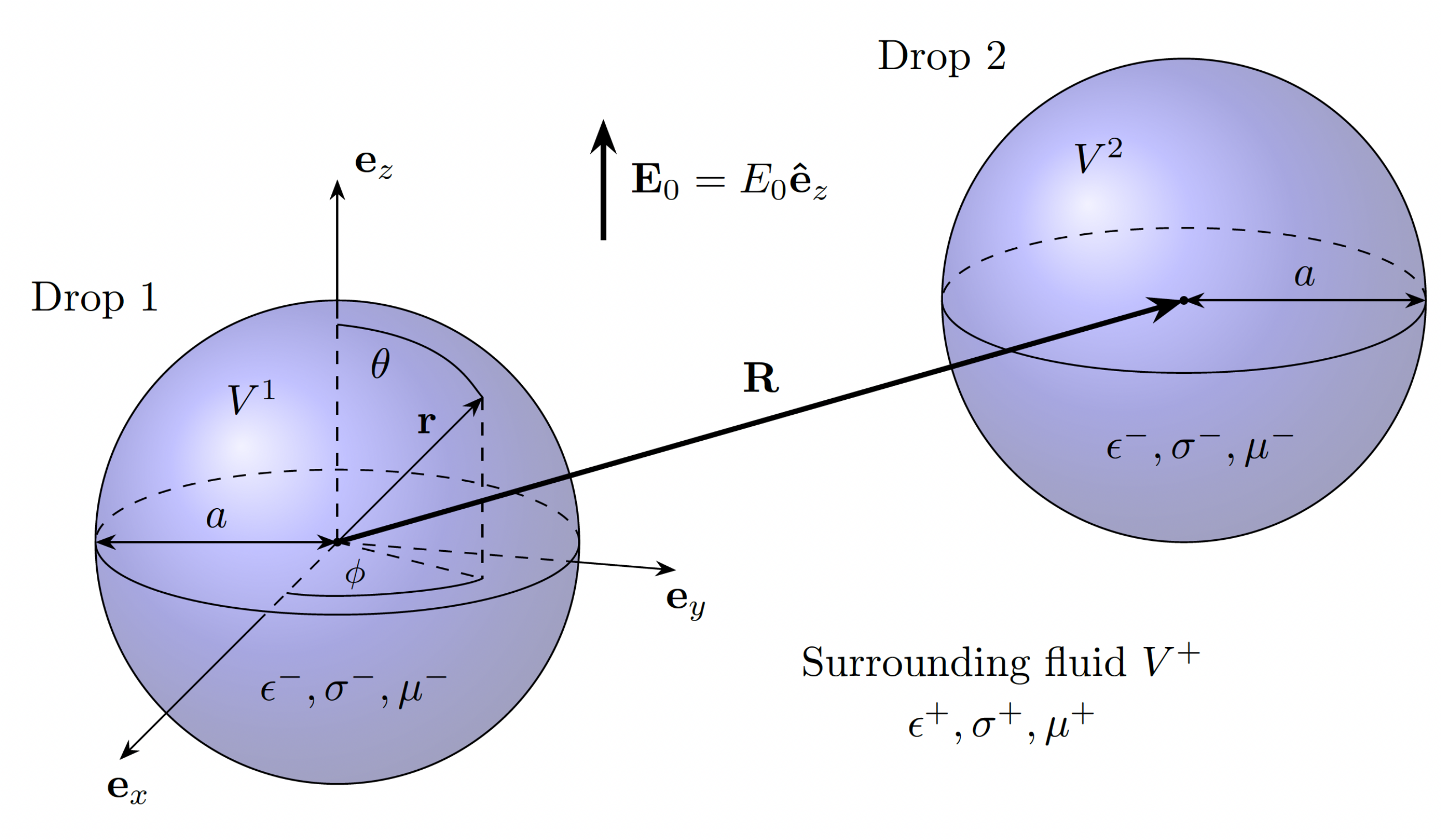}
    \caption{Two identical drops of a leaky dielectric fluid suspended in another leaky dielectric fluid and exposed to a uniform electric field $\bm{E}_{0}=E_{0}\bm{\hat{e}}_{z}.$}
    \label{problemdiagram}
\end{figure} 
Consider two initially spherical leaky dielectric drops with initial radius $a$ and identical material properties suspended in another leaky dielectric fluid and exposed to a uniform DC electric field $\bm{E}_{0},$ as shown in Figure~\ref{problemdiagram}. We use a spherical coordinate system (radial distance $r$, polar angle $\theta$, azimuthal angle $\phi$) with origin at the centre of one of the drops, denoted by drop $1$. The position of the centre of the second drop, denoted by drop $2$, is given by the separation vector  $\bm{R} = (r=R,~\theta=\theta_R,~ \phi=\phi_R)$. Electrical permittivity (dielectric constant), electrical conductivity and viscosity are denoted by $\epsilon, \sigma$ and $\mu,$ respectively, and $+$ and $-$ superscripts denote the values of these parameters in the surrounding fluid and the drops, respectively. We denote the unbounded domain of the surrounding fluid by $V^{+},$ the drop interfaces by $S_{1}$ and $S_{2},$ and the domains of the two drops by $V_{1}$ and $V_{2},$ respectively. We will also define $V^{-}$ to denote the domain of either drop 1 or 2, since the same governing equations apply in each drop. We assume that the drops have no net charge and are neutrally buoyant. Upon application of an external electric field, both drops polarise and charge distributions develop on their interfaces due to the discontinuity of electrical properties there. Assuming no charges exist in either of the drops or the surrounding fluid, the electric potential $\phi^{\pm}(\bm{r},t)$ satisfies Laplace's equation in all three domains,
 \begin{equation}\label{laplaceeqn}\nabla^{2}\phi^{\pm}(\bm{r},t)=0\quad\mbox{for }\bm{r}\in V^{\pm}.\end{equation}
 The electric field is denoted by $\bm{E}^{\pm}(\bm{r},t)=-\bnabla\phi^{\pm},$ and the charge distribution $q_{i}(\bm{r},t)$ on the interface of drop $i$ is given by Gauss's Law:
\begin{equation}\label{gauss}
    q_{i}(\bm{r},t)=\llbracket\epsilon\bm{E}\bcdot\bm{n}_{i}(\bm{r},t)\rrbracket_{i} \quad\mbox{for }\bm{r}\in S_{i},\quad i=1,2.
\end{equation}
Here, the notation $\llbracket f(\bm{r},t)\rrbracket_{i}=f^{+}(\bm{r},t)-f^{-}(\bm{r},t)$ for $\bm{r}\in S_{i}$ is used to denote the discontinuity of a field variable $f$ over the interface of drop $i$, and $\bm{n}_{i}$ is the outward-pointing unit vector normal to the interface of drop $i$. The electric potential and the tangential electric field are continuous across both drop interfaces:
\begin{equation}
    \llbracket\phi\rrbracket_{i}=0, \quad \llbracket\bm{E}\bcdot(\tensor{I}-\bm{n}_{i}\bm{n}_{i})\rrbracket_{i}=\bm{0}  \quad \mbox{for }\bm{r}\in S_{i},\quad i=1,2.
\end{equation}
The Taylor--Melcher leaky dielectric model posits that the evolution of the interfacial charge distribution on the interface of drop $i$ is due to two mechanisms: the discontinuity of current crossing the drop interface as a result of the differing electrical conductivities and charge convected by the fluid velocity at the drop interface. Hence, $q_{i}$ obeys the transport equation 
\begin{equation}\label{chargecons}
    \frac{\partial q_{i}}{\partial t}+\llbracket\sigma\bm{E}\bcdot\bm{n}_{i}\rrbracket_{i}+\bnabla_{s,i}\bcdot(q_{i}\bm{v}^{\pm}_i)=0 \quad \mbox{for }\bm{r}\in S_{i},\quad i=1,2.
\end{equation}
Here, $\bm{v}^{\pm}_i(\bm{r},t)$ is the 
fluid velocity and $\bnabla_{s,i}=(\tensor{I}-\bm{n}_{i}\bm{n}_{i})\bcdot\bnabla$ is the surface gradient operator for the interface of drop $i.$ Note that the fluid velocity $\bm{v}^{\pm}$ itself depends on the charge density, making~\eqref{chargecons} nonlinear and therefore challenging to solve analytically. The fluid flow is governed by the Stokes equations in each domain:
\begin{equation}\label{stokes}
    \mu^{\pm}\nabla^{2}\bm{v}^{\pm}=\bnabla p^{\pm}, \quad \bnabla\bcdot\bm{v}^{\pm}=0 \quad \mbox{for }\bm{r}\in V^{\pm},
\end{equation}
where $p^{\pm}(\bm{r},t)$ is the pressure. The fluid velocity is continuous across the interfaces of the drops: 
\begin{equation}\label{nopenetration}
\llbracket\bm{v}\rrbracket_{i}=\bm{0}\quad\mbox{for }\bm{r}\in S_{i},\quad i=1,2.
\end{equation}
The presence of the electric field generates electric stresses on the interfaces of the drops that are balanced by hydrodynamic stresses and surface tension, i.e.,
\begin{equation}\label{stressbalance}
\llbracket\bm{f}^{E}\rrbracket_{i}+\llbracket\bm{f}^{H}\rrbracket_{i}=2\gamma\kappa_{i}\bm{n}_{i}\quad \mbox{for }\bm{r}\in S_{i},\quad i=1,2,
\end{equation}
where $\llbracket\bm{f}^{E}\rrbracket_{i}(\bm{r},t)$ and $\llbracket\bm{f}^{H}\rrbracket_{i}(\bm{r},t)$ are the discontinuities in the electric and hydrodynamic tractions across the interface of drop $i$, respectively, $\gamma$ is the (constant) surface tension coefficient, and $\kappa_{i}(\bm{x},t)=\frac{1}{2}(\bnabla_{s}\bcdot\bm{n})_i$ is the mean surface curvature of drop $i$.

$\\$
The length and time scales for the problem are taken to be the initial drop radius $a$ and the Maxwell--Wagner relaxation time $\tau_{MW}$, respectively.  After scaling the electric, hydrodynamic and capillary stresses with $\epsilon^{+}E_{0}^{2},\mu^{+}/\tau_{MW}$ and $\gamma/a,$ respectively, two dimensionless parameters appear in the stress balance~\eqref{stressbalance}. These are the Mason number, Ma, and the electric capillary number, $\mathrm{Ca_{E}}$,
\begin{equation}
    \mathrm{Ma}=\frac{\mu^{+}}{\epsilon^{+}E_{0}^{2}\tau_{MW}},\quad \mathrm{Ca_{E}}=\frac{\epsilon^{+}E_{0}^{2}a}{\gamma},
\end{equation}
representing the ratio of viscous to electric stresses and the ratio of electric to capillary stresses, respectively.

\section{Solution method}\label{problemsolution}

In this section, we outline the solution to the problem formulated in \S\ref{problemformulation}. Calculating the drop shapes and flow fields simultaneously theoretically is very challenging. The solution strategy presented here is based on the well-established small deformation theory valid for nearly-spherical drops \citep{hetsroni1970flow,rallison1984}. First, we calculate the flow fields assuming that the drops are perfectly spherical. These flow fields are then used to determine the drop shapes to first order in $\delta$, the small parameter quantifying the deviation from a spherical drop shape. The deformed drop shape can be then used to obtain the flow fields at the next order in $\delta,$ and so on.

In \S\ref{drop shape}, we describe the parameterization of the drop shape and the calculation of its curvature and unit normal vector. In \S\ref{electrostatics}, we solve Laplace's equation (\ref{laplaceeqn}) for the electric potential. \S\ref{hydrodynamics} addresses the solution of the Stokes equations (\ref{stokes}) for the fluid velocity and pressure. In \S\ref{droptranslation}, we obtain the translational velocities of the drops induced by electrohydrodynamic and DEP interactions. \S\ref{kinematics} applies the kinematic boundary condition to establish relationships between the flow fields and the drop shapes. \S\ref{stress balance} focuses on balancing the electric and hydrodynamic stresses at the drop interface. Finally, in \S\ref{charge conservation}, we derive evolution equations for the surface charge, which are then integrated numerically. In the analysis that follows, we focus on solving the governing equations for the flow fields and shape of drop 1. The corresponding analysis for drop 2 is identical, with appropriate interchanges of the subscripts.

\subsection{Drop shapes}\label{drop shape}
We parameterise the interface of drop 1 as
\begin{equation}\label{interfacedefinition}
    \xi_{1}=r-[1+\delta f_{1}(\theta,\phi,t)]=0, 
\end{equation}
where $\delta\ll 1$ (to be defined in \S\ref{stress balance}) is a small parameter. Deviations of the interface from a spherical shape are represented by $f_{1} (\theta,\phi,t),$ which, in general, can be expanded as an infinite series of spherical harmonics. In anticipation of the form of the electric stress \citep{das2021three}, we retain only the second-order harmonics:
\begin{equation}
f_{1}=\tensor{Q}_{1}^{f}\bcdoubledot\bm{\hat{r}\hat{r}},
\end{equation}
where $\bm{\hat{r}}=\bm{r}/r$ denotes the unit vector in the radial direction, and double dots denote the scalar product of two tensors, i.e., $\tensor{A \bcdoubledot B}=A_{ij}B_{ji}$. The shape coefficients are contained within $\tensor{Q}^{f}_{1}$, a second-order symmetric traceless tensor. The interface is advected with the flow, so the kinematic boundary condition can be expressed as
\begin{equation}\label{kinematiccondition0}
    \frac{\partial \xi_{1}}{\partial t}+\bm{v}_{1}\bcdot\bnabla \xi_{1}=0.
\end{equation}
Using~\eqref{interfacedefinition}, the surface normal $\bm{n}_{1}$ and the curvature $\kappa_{1}$ of the interface of drop 1 are given by
\begin{equation}
    \bm{n}_{1}=\frac{\bnabla\xi_{1}}{|\bnabla\xi_{1}|}=\bm{\hat{r}}-2\delta(\tensor{Q}_{1}^{f}\bcdot\bm{\hat{r}})\bcdot(\tensor{I}-\bm{\hat{r}\hat{r}}),\quad
    \kappa_{1}=\frac{1}{2}\bnabla_{s}\bcdot\bm{n}_{1}=1+2\delta \tensor{Q}_{1}^{f}\bcdoubledot\bm{\hat{r}\hat{r}},
\end{equation}
respectively.

\subsection{Electrostatics}\label{electrostatics}

To satisfy Laplace's equation~\eqref{laplaceeqn}, the electric potential inside and outside drop 1 can be expanded into multipoles. For an isolated drop with no charge convection, the system is axisymmetric, and a dipole moment is sufficient to represent the electric potential. However, higher-order multipoles can arise for two reasons. The first of these is charge convection along the surface of the drop by the interfacial fluid flow, arising both from the flow field of drop 1 itself as well as from the flow field of drop 2. This effect is of order \(\mathcal{O}(\mathrm{Ma}^{-2}\lambda^{-2}+\lambda^{-1}\mathrm{Ma}^{-2}R^{-3})\).
Secondly, because the external electric field experienced by drop 1 is not uniform -- it includes perturbations generated by the electric field of drop 2 -- electrostatic interactions can induce higher-order multipoles. These nonuniformities appear at \(\mathcal{O}(R^{-4})\). However, in this work we disregard both these effects and represent the charge distribution on each drop interface using only a dipole moment. Therefore, the present theory is valid for pairs of drops that are highly viscous relative to the surrounding fluid (\(\lambda \gg 1\)) and widely separated (\(R \gg 1\)). From here on, we retain all physical effects up to and including terms of order \(\mathcal{O}(\lambda^{-1})\) and \(\mathcal{O}(R^{-3})\).  With these assumptions, the potentials inside and outside drop 1 can be written as

\begin{align}\label{potentialinside}
    \phi^{-} = \phi_1^\infty + \bm{P}_{1}\bcdot\bm{r}, \qquad  \phi^{+} = \phi_1^\infty + {r^{-3}}\bm{P}_{1}\bcdot\bm{r},
\end{align}
respectively, where 
\begin{align}\label{potentialoutside}
    \phi_1^\infty = -\bm{\hat{E}}_{0}\bcdot\bm{r}+\frac{\bm{P}_{2}\bcdot(\bm{r-R})}{|\bm{r-R}|^{-3}}
\end{align}
is the effective external electric potential experienced by drop 1. Here, $\bm{P}_{1}(t)$ and $\bm{P}_{2}(t)$ denote the induced dipole moments of drop 1 and drop 2, respectively. The potentials resulting from the drop's own dipole $\bm{P}_{1}$ inside and outside the drop are given by growing and decaying harmonics, respectively. The disturbance in the potential near drop 1 due to interactions with drop 2 is captured by the second term on the right hand side of~\eqref{potentialoutside}. The associated electric fields, $\bm{E} = -\bnabla\phi$, inside and outside drop 1 are given by
\begin{equation}
    \bm{E}_{1}^{-}=\bm{E}^{\infty}_{1}-\bm{P}_{1}, \qquad  \bm{E}^{+}_{1}=\bm{E}^{\infty}_{1}-r^{-3}\bm{P}_{1}\bcdot(\tensor{I}-3\bm{\hat{r}\hat{r}}),
\end{equation}
respectively, where 
\begin{equation}\label{ambientfield}
\bm{E}^{\infty}_{1}=\bm{\hat{E}}_{0}-R^{-3}\bm{P}_{2}\bcdot(\tensor{I}-3\bm{\hat{R}\hat{R}})
\end{equation} 
is the effective external electric field experienced by drop 1. Note that by retaining only the leading order (i.e., the $\mathcal{O}(R^{-3})$) effect of $\bm{P}_{2}$ on the electric field near drop 1, $\bm{E}_{1}^{\infty}$ remains uniform, and so does not induce any higher-order multipoles, as discussed previously. The electric stress tensors inside and outside the drop are then given by
\begin{equation}
    \tensor{\tau}^{E,\pm}_{1}=\epsilon^{\pm}\left(\bm{E}_{1}^{\pm}\bm{E}_{1}^{\pm}-\frac{1}{2}(E_{1}^{\pm})^{2}\tensor{I}\right).
\end{equation}
The discontinuity in the electric traction across the interface of drop 1 is given by
\begin{equation}\label{elecstress}
    \llbracket\bm{f}^{E}\rrbracket_{1}=\llbracket\tensor{\tau}^{E}\rrbracket_{1}\bcdot\bm{\hat{r}}=p_{1}^{E}\bm{\hat{r}}+\tensor{Q}_{1,\bm{\hat{r}}}^{E}\bcdoubledot\bm{\hat{r}\hat{r}\hat{r}}+\left(\tensor{Q}_{1,\bm{\hat{t}}}^{E}\bcdot\bm{\hat{r}}\right)\bcdot(\tensor{I}-\bm{\hat{r}\hat{r}})+\bm{T}^{E}_{1}\times\bm{\hat{r}},
\end{equation}
where
\begin{subequations}\label{elecstresscoefficients}
\begin{align}
        p^{E}_{1} &= -\frac{1-Q}{6}(E_{1}^{\infty})^{2}+\frac{4-Q}{3}\left(\bm{E}_{1}^{\infty}\bcdot\bm{P}_{1}\right)+\frac{Q+2}{6}P_{1}^{2}, \\ 
        \begin{split}
            \tensor{Q}^{E}_{1,\bm{\hat{r}}} &= (1-Q)\bm{E}_{1}^{\infty}\bm{E}_{1}^{\infty}+\frac{1+2Q}{2}\left(\bm{E}_{1}^{\infty}\bm{P}_{1}+\bm{P}_{1}\bm{E}_{1}^{\infty}\right)+\frac{5-2Q}{2}\bm{P}_{1}\bm{P}_{1} \\
         & +\left[ -\frac{1-Q}{3}(E_{1}^{\infty})^{2}-\frac{1+2Q}{3}\left(\bm{E}_{1}^{\infty}\bcdot\bm{P}_{1}\right)-\frac{5-2Q}{6}P_{1}^{2}\right]\tensor{I},
        \end{split} \\
        \begin{split}
        \tensor{Q}^{E}_{1,\bm{\hat{t}}} &= (1-Q)\bm{E}_{1}^{\infty}\bm{E}_{1}^{\infty}+\frac{1+2Q}{2}\left(\bm{E}_{1}^{\infty}\bm{P}_{1}+\bm{P}_{1}\bm{E}_{1}^{\infty}\right)-(2+Q)\bm{P}_{1}\bm{P}_{1} \\
        &+\left[-\frac{1-Q}{3}(E_{1}^{\infty})^{2}-\frac{1+2Q}{3}\left(\bm{E}_{1}^{\infty}\bcdot\bm{P}_{1}\right)+\frac{2+Q}{3}P_{1}^{2}\right]\tensor{I},
        \end{split} \\
        \bm{T}_{1}^{E}&=\frac{3}{2}\bm{P}_{1}\times\bm{E}_{1}^{\infty}. \label{elecstresscoefficients_d}
    \end{align}
    \end{subequations}
In~\eqref{elecstress}, $p_{1}^{E}(t)$ is a scalar representing an isotropic electric pressure, $\tensor{Q}^{E}_{1,\bm{\hat{r}}}(t)$ and $\tensor{Q}^{E}_{1,\bm{\hat{t}}}(t)$ are symmetric, traceless second-order tensors representing the radial and tangential electric stresses driving drop deformation and straining flows, and $\bm{T}_{1}^{E}(t)$ is a vector representing the net electric torque on the drop. 

\subsection{Hydrodynamics}\label{hydrodynamics}

The general solution of the Stokes equations~\eqref{stokes} in spherical coordinates was derived by Lamb \citep{lamb1924hydrodynamics}. Inside drop 1, the pressure and velocity fields are given by growing harmonics,
\begin{subequations} \label{lambin}
\begin{align} 
     p_{1}^{-} &= \sum_{n=0}^{\infty}p_{1,n}, \label{lambinp}
    \\ \bm{v}_{1}^{-} &= \sum_{n=1}^{\infty}\left[\bnabla\times(\bm{r}\chi_{1,n})+\bnabla\phi_{1,n}+\frac{n+3}{2(n+1)(2n+3)}r^{2}\bnabla p_{1,n}-\frac{n}{(n+1)(2n+3)}\bm{r}p_{1,n}\right]\label{lambinv}.
\end{align}
\end{subequations}
Outside drop 1, the pressure and velocity fields are given by decaying harmonics,

\begin{subequations}\label{lambout}
\begin{align}
    p_{1}^{+} &= p_{1}^{\infty} + \sum_{n=-1}^{\infty}p_{1,-n-1},\label{lamboutp} \\ 
    \bm{v}_{1}^{+} &= \bm{v}^{\infty}_{1} + \sum_{n=1} ^{\infty}\left[\bnabla\times(\bm{r}\chi_{1,-n-1})+\bnabla\phi_{1,-n-1}+\frac{2-n}{2n(2n-1)\lambda}r^{2}\bnabla p_{1,-n-1}+\frac{n+1}{n(2n-1)\lambda}\bm{r}p_{1,-n-1}\right]\label{lamboutv}.
\end{align}\end{subequations}

Here, we have included an external pressure $p^{\infty}_{1}(\bm{r},t)$ and velocity $\bm{v}^{\infty}_{1}(\bm{r},t),$ which represent the hydrodynamic effect of drop 2 on drop 1, which we will derive shortly. Each of $p_{1,n}(\bm{r},t),$ $\chi_{1,n}(\bm{r},t)$ and $\phi_{1,n}(\bm{r},t)$ is a sum of spherical harmonics of order $n$. They represent the various elements of the flow field produced by drop 1, corresponding to pressure, toroidal flow and velocity potential, respectively, \citep{kim2013microhydrodynamics,happelbrenner}. The first few modes, corresponding to $n=1,2,-2,-3$, are
\begin{subequations}\label{lambharmonics}
\begin{align}
& p_{1,1}=r\bm{d}^{p}_{1}\bcdot\bm{\hat{r}}, \quad p_{1,2}=r^{2}\tensor{q}^{p}_{1}\bcdoubledot\bm{\hat{r}\hat{r}}, \quad p_{1,-2}=r^{-2}\bm{D}^{p}_{1}\bcdot\bm{\hat{r}},\quad p_{1,-3}=r^{-3}\tensor{Q}_{1}^{p}\bcdoubledot\bm{\hat{r}\hat{r}},\label{pharmonics} \\
& \phi_{1,1}=r\bm{d}^{\phi}_{1}\bcdot\bm{\hat{r}}, \quad \phi_{1,2}=r^{2}\tensor{q}^{\phi}_{1}\bcdoubledot\bm{\hat{r}\hat{r}}, \quad \phi_{1,-2}=r^{-2}\bm{D}_{1}^{\phi}\bcdot\bm{\hat{r}},\quad \phi_{1,-3}=r^{-3}\tensor{Q}_{1}^{\phi}\bcdoubledot\bm{\hat{r}\hat{r}}, \\
& \chi_{1,1}=r\bm{d}^{\chi}_{1}\bcdot\bm{\hat{r}}, \quad \chi_{1,2}=r^{2}\tensor{q}^{\chi}_{1}\bcdoubledot\bm{\hat{r}\hat{r}}, \quad \chi_{1,-2}=r^{-2}\bm{D}^{\chi}_{1}\bcdot\bm{\hat{r}},\quad \chi_{1,-3}=r^{-3}\tensor{Q}_{1}^{\chi}\bcdoubledot\bm{\hat{r}\hat{r}}.
\end{align}
\end{subequations}
We refer to the vectors $\bm{D}^{p}_{1}(t)$, $\bm{D}_{1}^{\phi}(t)$, $\bm{D}^{\chi}_{1}(t)$, $\bm{d}_{1}^{p}(t)$, $\bm{d}_{1}^{\phi}(t)$, $\bm{d}_{1}^{\chi}(t)$ and tensors $\tensor{Q}_{1}^{p}(t)$, $\tensor{Q}_{1}^{\phi}(t)$, $\tensor{Q}^{\chi}_{1}(t)$, $\tensor{q}^{p}_{1}(t)$, $\tensor{q}_{1}^{\phi}(t)$, $\tensor{q}^{\chi}_{1}(t)$ appearing in~\eqref{lambharmonics} as flow multipoles, and the $p,\phi,\chi$ superscripts denote the element of the fluid flow described by the vector or tensor. For example, $\tensor{q}_{1}^{p}$ represents the $n=2$ pressure distribution inside the drop, and $\bm{D}^{\chi}_{1}$ represents the $n=1$ rotational flow (called the rotlet) outside the drop. As with the electric potential, growing and decaying harmonics are retained inside and outside the drops, respectively. In addition to the harmonics given in~\eqref{pharmonics}, we also include a monopole for the hydrostatic pressure in each domain. Inside drop 1, the monopole is given by the $n=0$ term in~\eqref{lambinp} and is denoted by $\tilde{p}_{1}^{-}$. Outside drop 1, the monopole is given by the $n=-1$ term in~\eqref{lamboutp} and is denoted by $\tilde{p}_{1}^{+}$. Using the pressures and velocities given by~\eqref{lambin}--\eqref{lambout}, the associated hydrodynamic stress tensors can be calculated in each domain:
\begin{equation}
    \tensor{\tau}^{H,\pm}_{1}=-p^{\pm}_{1}\tensor{I}+\mu^{\pm}\left(\bnabla\bm{v}^{\pm}_{1}+\bnabla\bm{v}_{1}^{\pm T}\right).
\end{equation}
After calculating the traction discontinuity given by $\llbracket\bm{f}^{H}\rrbracket_{1}=\llbracket\tensor{\tau}^{H}\rrbracket_{1}\bcdot\bm{\hat{r}},$ we retain only those modes which are excited by the electric tractions in~\eqref{elecstress} (specifically, $n=1,2,-2,-3$). The velocities in~\eqref{lambin} and~\eqref{lambout} are then written
\begin{equation}
\begin{aligned}\label{interiorflow0}
        \bm{v}_{1}^{-}=\bm{d}_{1}^{\chi}\times\bm{r}+\bm{d}_{1}^{\phi}+2\tensor{q}^{\phi}_{1}\bcdot\bm{r}+\frac{r^{2}\tensor{q}^{p}_{1}\bcdot\bm{r}}{21\lambda}\bcdot(5\tensor{I}-2\bm{\hat{r}\hat{r}}),
    \end{aligned}
    \end{equation}
\begin{equation}\begin{aligned}\label{exteriorflow0}
\bm{v}^{+}_{1}=
r^{-3}\bm{D}^{\chi}_{1}\times\bm{r}+r^{-5}\frac{\tensor{Q}^{p}_{1}\bcdoubledot\bm{rrr}}{2}+r^{-5}\left(\tensor{Q}^{\phi}_{1}\bcdot\bm{r}\right)\bcdot(2\tensor{I}-5\bm{\hat{r}\hat{r}})+\bm{v}^{\infty}_{1}.
\end{aligned}\end{equation}
To calculate $p^{\infty}_{1}$ and $\bm{v}^{\infty}_{1}$, we apply the procedure outlined in this section to calculate the leading-order flows generated by drop 2 while neglecting hydrodynamic interactions with drop 1. The leading-order exterior pressure and velocity fields produced by drop 2 are denoted by $p_{2,0}^{+}$ and $\bm{v}_{2,0}^{+}.$ To approximate their values near drop 1, $p_{2,0}^{+}$ and $\bm{v}_{2,0}^{+}$ are evaluated at $\bm{r-R}$ and expanded for small $r/R.$ Keeping terms up to $\mathcal{O}(R^{-3}),$ the pressure and velocity fields arising from drop 2 are
\begin{equation}
    p^{\infty}_{1}=p_{2,0}^{+}|_{\bm{r-R}}=\tilde{p}_{2,0}^{+}+R^{-3}\tensor{Q}^{p}_{2,0}\bcdoubledot\bm{\hat{R}\hat{R}}+\mathcal{O}(R^{-4})
\end{equation}
and
\begin{equation}
    \bm{v}^{\infty}_{1}=\bm{v}^{+}_{2,0}|_{\bm{r-R}}=R^{-2}\bm{N}^{\infty}_{1}+R^{-3}\tensor{S}^{\infty}_{1}\bcdot\bm{r}+R^{-3}\boldsymbol{\omega}^{\infty}_{1}\times\bm{r}+\mathcal{O}(R^{-4}),
\end{equation}
where
\begin{subequations}
    \begin{align}
    \bm{N}^{\infty}_{1}&=-\bm{D}^{\chi}_{2,0}\times\bm{\hat{R}}-\frac{1}{2}\tensor{Q}^{p}_{2,0}\bcdoubledot\bm{\hat{R}\hat{R}\hat{R}}, \label{ambientflowN} \\
    \begin{split}
    \tensor{S}^{\infty}_{1}&=-\frac{3}{2}\left((\bm{D}^{\chi}_{2,0}\times\bm{\hat{R}})\bm{\hat{R}}+\bm{\hat{R}}(\bm{D}^{\chi}_{2,0}\times\bm{\hat{R}})\right)-\frac{5}{2}\tensor{Q}^{p}_{2,0}\bcdoubledot\bm{\hat{R}\hat{R}\hat{R}\hat{R}} \\ 
        &+\frac{1}{2}\left(\bm{\hat{R}}(\tensor{Q}^{p}_{2,0}\bcdot\bm{\hat{R}})+(\tensor{Q}^{p}_{2,0}\bcdot\bm{\hat{R}})\bm{\hat{R}}\right)+\frac{1}{2}\left(\tensor{Q}^{p}_{2,0}\bcdoubledot\bm{\hat{R}\hat{R}}\right)\tensor{I}, \label{ambientflowS}
    \end{split} \\
    \boldsymbol{\omega}^{\infty}_{1} &= -\frac{1}{2}\bm{D}^{\chi}_{2,0}\bcdot(\tensor{I}-3\bm{\hat{R}\hat{R}})+\frac{1}{2}(\tensor{Q}^{p}_{2,0}\bcdot\bm{\hat{R}})\times\bm{\hat{R}}. \label{ambientflowomega}
    \end{align}
\end{subequations}
Here, $\tilde{p}^{+}_{2,0}$, $\bm{D}^{\chi}_{2,0}$, and $\tensor{Q}^{p}_{2,0}$ represent the exterior pressure monopole of drop 2, the rotlet, and the exterior pressure distribution ($n=2$ mode), respectively, without hydrodynamic interactions. $\bm{N}_{1}^{\infty}$ is a uniform flow field containing contributions from the rotational and straining flows produced by drop 2, and is the leading-order translational velocity of drop 1 in the limit of large $R.$ $\tensor{S}_{1}^{\infty}$ is a second-order tensor containing the straining component of the ambient flow field. $\boldsymbol{\omega}_{1}^{\infty}$ is the angular velocity of drop 1 resulting from the ambient flow field $\bm{v}^{\infty}_{1}$, and equals half the vorticity of $\bm{v}^{\infty}_{1}$. The hydrodynamic traction associated with the ambient flow is
\begin{equation}
\llbracket\tensor{\tau}^{\infty}\rrbracket_{1}\bcdot\bm{\hat{r}}=\left(-p_{1}^{\infty}\tensor{I}+\nabla\bm{v}_{1}^{\infty}+\nabla\bm{v}_{1}^{\infty T}\right)\bcdot\bm{\hat{r}}=-\left(\tilde{p}_{2,0}^{+}+R^{-3}\tensor{Q}^{p}_{2,0}\bcdoubledot\bm{\hat{R}\hat{R}}\right)\bm{\hat{r}}+2R^{-3}\tensor{S}_{1}^{\infty}\bcdot\bm{\hat{r}},
\end{equation}
which is added to the hydrodynamic traction of the flow field of drop 1
to obtain the total hydrodynamic traction on drop 1 as
\begin{equation}\begin{split}\label{hdtractioninitial}
\llbracket\bm{f}^{H}\rrbracket_{1}=&\left(\tilde{p}^{-}_{1}-\tilde{p}_{1}^{+}-p_{1}^{\infty}\right)\bm{\hat{r}}+\left(24\tensor{Q}_{1}^{\phi}-3\tensor{Q}^{p}_{1}-4\lambda\tensor{q}_{1}^{\phi}+\frac{\tensor{q}^{p}_{1}}{7}+2R^{-3}\tensor{S}_{1}^{\infty}\right)\bcdoubledot\bm{\hat{r}\hat{r}\hat{r}}
\\&+\left[\left(-16\tensor{Q}_{1}^{\phi}+\tensor{Q}_{1}^{p}-\frac{16\tensor{q}^{p}_{1}}{21}-4\lambda\tensor{q}_{1}^{\phi}+2R^{-3}\tensor{S}_{1}^{\infty}\right)\bcdot\bm{\hat{r}}\right]\bcdot(\tensor{I}-\bm{\hat{r}\hat{r}})-3\bm{D}_{1}^{\chi}\times\bm{\hat{r}}.
\end{split}\end{equation}

The hydrodynamic interactions between the drops are captured by $p^{\infty}_{1}$ and $\tensor{S}_{1}^{\infty}.$ While hydrodynamic interactions do not generate a hydrodynamic torque on drop 1~\citep{hetsroni1970flow}, they still alter the drop angular velocity through external flow vorticity ($\boldsymbol{\omega}^{\infty}_{1}$) and by modifying the interfacial velocity in charge convection.

\subsection{Translational velocity}\label{droptranslation}
The translational velocity of drop 1, denoted by $\bm{U}_{1}(t),$ consists of two parts: one originating from hydrodynamics, found using Fax\'en's law for drops \citep{hetsroni1970flow} and another originating from electrostatics via the DEP force $\bm{f}_{1}^{\mathrm{DEP}}$:
\begin{equation}\label{faxen}
    \bm{U}_{1}=\left(1+\frac{\lambda}{2(2+3\lambda)}\nabla^{2}\right)\bm{v}_{1}^{\infty}|_{\bm{r=0}}-\frac{1}{2\pi \mathrm{Ma}}\frac{1+\lambda}{2+3\lambda}\bm{f}^{\mathrm{DEP}}_{1}.
\end{equation}
The DEP force exerted on the dipole of drop 1 \citep{jones1979} is $\bm{f}_{1}^{DEP}=4\pi\bm{P}_{1}\bcdot\nabla\bm{E}^{\infty}_{1}|_{\bm{r=0}},$ which can be expanded as
\begin{equation}
    \bm{f}_{1}^{DEP}=-12\pi R^{-4}\left[(\bm{P}_{1}\bcdot\bm{\hat{R}})\bm{P}_{2}+(\bm{P}_{2}\bcdot\bm{\hat{R}})\bm{P}_{1}+(\bm{P}_{1}\bcdot\bm{P}_{2})\bm{\hat{R}}-5(\bm{P}_{1}\bcdot\bm{\hat{R}})(\bm{P}_{2}\bcdot\bm{\hat{R}})\bm{\hat{R}}\right].
\end{equation}
We retain the DEP force although it is of order $\mathcal{O}(R^{-4})$ since it causes the important effect of chaining in real systems. The translational velocity of drop 2, denoted by $\bm{U}_{2}(t)$, is found by applying the same procedure as that outlined above to drop 2. The relative velocity of the pair of drops, denoted by $\bm{U}_{21}=\bm{U}_{2}-\bm{U}_{1},$ and hence the evolution of the separation vector $\bm{R},$ can then be easily found to be 
\begin{equation}\label{dRdt}
\bm{U}_{21}=\frac{\mbox{d}\bm{R}}{\mbox{d}t}=\bm{U}_{2}-\bm{U}_{1}.
\end{equation}

\subsection{Kinematic boundary condition}\label{kinematics}

Applying the kinematic boundary condition~\eqref{kinematiccondition0} yields relations between the flow coefficients inside and outside drop 1 given in~\eqref{interiorflow0} and~\eqref{exteriorflow0}. To do so, we recall that the velocity is continuous across the drop interface. In particular, imposing continuity of velocity tangential to the interface and invoking the orthogonality of spherical harmonics gives the relations
\begin{equation}\begin{aligned}\label{tangentialkcrelations}
R^{-3}\bm{\omega}^{\infty}_{1}+\bm{D}^{\chi}_{1}=\bm{d}^{\chi}_{1},\qquad R^{-2}\bm{N}^{\infty}_{1}=\bm{d}^{\phi}_{1},\qquad R^{-3}\tensor{S}^{\infty}_{1}+2\tensor{Q}^{\phi}_{1}=2\tensor{q}_{1}^{\phi}+\frac{5\tensor{q}^{p}_{1}}{21\lambda}.
\end{aligned}\end{equation}
 Substituting the expression for $\xi_{1}$ given in~\eqref{interfacedefinition} into~\eqref{kinematiccondition0} gives the radial component of the kinematic condition, evaluated at the interface of drop 1:
\begin{equation}\begin{aligned}\label{radialkc}
\bm{v}^{-}_{1}\bcdot\bm{\hat{r}}-\delta\bm{v}_{1}^{-}\bcdot\bnabla f_{1}=\bm{v}_{1}^{+}\bcdot\bm{\hat{r}}-\delta\bm{v}_{1}^{+}\bcdot\bnabla f_{1}
=\delta\frac{\partial f_{1}}{\partial t}.
\end{aligned}\end{equation}
Equation~\eqref{radialkc} has two $\mathcal{O}(\delta)$ terms, which means $\bm{v}^{\pm}_{1}\bcdot\bm{\hat{r}}$ must also be $\mathcal{O}(\delta)$ for the equation to hold. Since the radial velocities $\bm{v}^{\pm}_{1}\bcdot\bm{\hat{r}}$ scale as $\lambda^{-1}$, we also require that $\lambda^{-1}$ be $\mathcal{O}(\delta).$ Then, in the velocities $\bm{v}^{\pm}_{1}$ appearing in the nonlinear convective products $\bm{v}^{\pm}_{1}\bcdot\bnabla f_{1},$ we include only the leading-order flow components, i.e., $\bm{D}^{\chi}_{1}$ and $\bm{d}^{\chi}_{1}$, which are independent of $\lambda$.  According to~\eqref{tangentialkcrelations}, $\bm{d}^{\chi}_{1}=\bm{D}_{1}^{\chi}$, and hence $\delta\bm{v}^{+}_{1}\bcdot\bnabla f=\delta\bm{v}^{-}_{1}\bcdot\bnabla f$. Therefore, at leading order the kinematic condition~\eqref{radialkc} can be written
\begin{equation}\begin{aligned}\label{radialkcfin}
\bm{v}_{1}^{-}\bcdot\bm{\hat{r}}=\bm{v}_{1}^{+}\bcdot\bm{\hat{r}}
=\delta\left(\frac{\partial f_{1}}{\partial t}+\bm{v}_{1}\bcdot\bnabla f_{1}\right),
\end{aligned}\end{equation}
where, using the Einstein summation convention, the second term in the right hand side can be written as
\begin{equation}\begin{aligned}\label{vnablafsymmetric}
\bm{v}_{1}\bcdot\bnabla f_{1}=\left(\varepsilon_{lkj}Q^{f}_{1,li}D^{\chi}_{1,k}+\varepsilon_{lki}Q^{f}_{1,lj}D^{\chi}_{1,k}\right)\hat{r}_{i}\hat{r}_{j}=\tensor{Q}_{1}^{\chi f}\bcdoubledot\bm{\hat{r}\hat{r}}.
\end{aligned}
\end{equation} 
Here, $\varepsilon_{lkj}$ and $\varepsilon_{lki}$ are the Levi--Civita tensor and $Q_{1,ij}^{\chi f} = \varepsilon_{lkj}Q^{f}_{1,li}D^{\chi}_{1,k}+\varepsilon_{lki}Q^{f}_{1,lj}D^{\chi}_{1,k}$. Substituting the fluid velocities given in~\eqref{interiorflow0} and~\eqref{exteriorflow0}, the drop shape given in~\eqref{interfacedefinition} and $\bm{v}_{1}\bcdot\bnabla f_{1}$ given in~\eqref{vnablafsymmetric} into the kinematic condition~\eqref{radialkcfin} and combining the result with the relations given in~\eqref{tangentialkcrelations} allows us to express the flow coefficients $\tensor{Q}_{1}^{\phi},$ $\tensor{q}_{1}^{\phi}$ and $\tensor{q}_{1}^{p}$ in terms of $\tensor{Q}^{p}_{1},$ $\partial \tensor{Q}^{f}_{1}/\partial t+\tensor{Q}_{1}^{\chi f}$ and $\tensor{S}_{1}^{\infty}$:

\begin{subequations}
\begin{align}
\tensor{Q}^{\phi}_{1} &= \frac{1}{6}\tensor{Q}^{p}_{1}-\frac{\delta}{3}\left(\frac{\partial \tensor{Q}^{f}_{1}}{\partial t}+\tensor{Q}^{\chi f}_{1}\right)+\frac{1}{3}R^{-3}\tensor{S}_{1}^{\infty}, \label{eq:Qphi1} \\
\tensor{q}^{p}_{1} &= \frac{7\lambda}{2}\tensor{Q}^{p}_{1}-\frac{35\lambda\delta}{2}\left(\frac{\partial \tensor{Q}^{f}_{1}}{\partial t}+\tensor{Q}^{\chi f}_{1}\right)+\frac{35\lambda}{2}R^{-3}\tensor{S}_{1}^{\infty}, \label{eq:qp1} \\
\tensor{q}^{\phi}_{1} &= -\frac{1}{4}\tensor{Q}^{p}_{1}+\frac{7\delta}{4}\left(\frac{\partial \tensor{Q}^{f}_{1}}{\partial t}+\tensor{Q}^{\chi f}_{1}\right)-\frac{5}{4}R^{-3}\tensor{S}_{1}^{\infty}. \label{eq:qphi1}
\end{align}
\end{subequations}

With these relations, we can rewrite the hydrodynamic traction~\eqref{hdtractioninitial} as
\begin{equation}\label{finalhdstress}
    \llbracket\bm{f}^{H}\rrbracket_{1}=\llbracket\tensor{\tau}_{1}^{H}\rrbracket\bcdot\bm{\hat{r}}=p_{1}^{H}\bm{\hat{r}}+\tensor{Q}_{1,\bm{\hat{r}}}^{H}\bcdoubledot\bm{\hat{r}\hat{r}\hat{r}}+\left(\tensor{Q}_{1,\bm{\hat{t}}}^{H}\bcdot\bm{\hat{r}}\right)\bcdot(\tensor{I}-\bm{\hat{r}\hat{r}})+\bm{T}_{1}^{H}\times\bm{\hat{r}},
\end{equation}
where

\begin{subequations}
\begin{align}
p^{H}_{1} &= \tilde{p}_{1}^{-}-\tilde{p}_{1}^{+}-\tilde{p}_{2,0}^{+}-R^{-3}\tensor{Q}^{p}_{2,0}\bcdoubledot\bm{\hat{R}\hat{R}}, \label{hdpressure} \\
\tensor{Q}^{H}_{1,\bm{\hat{r}}} &= \frac{2+3\lambda}{2}\tensor{Q}_{1}^{p}-\frac{16+19\lambda}{2}\delta\left(\frac{\partial \tensor{Q}_{1}^{f}}{\partial t}+\tensor{Q}_{1}^{\chi f}\right)+\frac{20+15\lambda}{2}R^{-3}\tensor{S}_{1}^{\infty}, \label{hdstresscoeffQr} \\
\tensor{Q}^{H}_{\bm{\hat{t}},1} &= -\frac{5(1+\lambda)}{3}\tensor{Q}^{p}_{1}+\frac{16+19\lambda}{3}\delta\left(\frac{\partial \tensor{Q}^{f}_{1}}{\partial t}+\tensor{Q}_{1}^{\chi f}\right)-\frac{10+25\lambda}{3}R^{-3}\tensor{S}_{1}^{\infty}, \label{hdstresscoeffQt} \\
\bm{T}_{1}^{H} &= -3\bm{D}_{1}^{\chi}. \label{hdtorque}
\end{align}
\end{subequations}

\subsection{Stress balance}\label{stress balance}

By enforcing a balance between electric, hydrodynamic, and surface tension stresses at the interface of drop 1, it is possible to express the flow coefficients $\tensor{Q}^{p}_{1}$ and $\bm{D}^{\chi}_{1}$ in terms of the dipole moments and derive a system of coupled ODEs for the shape coefficients $\tensor{Q}_{1}^{f}$. In nondimensional form, the stress balance~\eqref{stressbalance} reads
\begin{equation}\label{dynamic}
\llbracket\bm{f}^{E}\rrbracket_{1}+\mathrm{Ma}\llbracket\bm{f}^{H}\rrbracket_{1}=\frac{2}{\mathrm{Ca_{E}}}\left[\bm{\hat{r}}-2\delta({\tensor{Q}_{1}^{f}}\bcdot\bm{\hat{r}})\bcdot(\tensor{I}-2\bm{\hat{r}\hat{r}})\right].
\end{equation}
In the radial direction, we retain $\mathcal{O}(\delta)$ terms to capture the perturbation of the drop shape associated with the electric and hydrodynamic stresses calculated in the preceding subsections. After substituting~\eqref{elecstress} and~\eqref{finalhdstress} into~\eqref{dynamic}, we invoke the orthogonality of spherical harmonics to the radial component of the equation to obtain the two equations
\begin{equation}
    \begin{aligned}\label{pressurebal}
       p_{1}^{E}+\mathrm{Ma}p^{H}_{1}=\frac{2}{\mathrm{Ca_{E}}},
       \end{aligned}\end{equation}
\begin{equation}\begin{aligned}\label{stressbal1}
\tensor{Q}_{1,\bm{\hat{r}}}^{E}+\mathrm{Ma}\tensor{Q}^{H}_{1,\bm{\hat{r}}}=\frac{4\delta}{\mathrm{Ca_{E}}}\tensor{Q}^{f}_{1}.
    \end{aligned}
\end{equation}
Equation~\eqref{pressurebal} is a scalar equation describing the balance of isotropic pressure across the interface. Equation~\eqref{stressbal1} relates the second-order tensors making up the radial components of the quadrupolar electric and hydrodynamic stresses and requires $\delta=\mathcal{O}(\mathrm{Ca_{E}}),$ and for consistency with previous works on the subject, including that of Taylor \citep{taylor1966studies}, we define
\begin{equation}
    \delta=\frac{3\mathrm{Ca_{E}}}{4(1+2S)^{2}}.
\end{equation}
Next, examining the tangential components of~\eqref{dynamic}, we find
\begin{equation}
\begin{aligned}\label{torquebal}
\bm{T}_{1}^{E}+\mathrm{Ma}\bm{T}_{1}^{H}=\bm{0},
    \end{aligned}
\end{equation}
\begin{equation}
\begin{aligned}\label{tangorthog}
\tensor{Q}^{E}_{1,\bm{\hat{t}}}+\mathrm{Ma}\tensor{Q}^{H}_{1,\bm{\hat{t}}}=\bm{0}.
\end{aligned}
\end{equation}
Equation~\eqref{torquebal} is the torque balance on drop 1, while~\eqref{tangorthog} relates the second-order tensors making up the tangential components of the quadrupolar stresses. Substituting the values of 
$\bm{T}^{E}_{1}$ and $\bm{T}^{H}_{1}$ given in~\eqref{elecstresscoefficients_d} and~\eqref{hdtorque} into~\eqref{torquebal} yields
\begin{equation}\label{dchi}
    \bm{D}^{\chi}_{1}=\frac{1}{2\mathrm{Ma}}\bm{P}_{1}\times\bm{E}_{1}^{\infty}.
\end{equation}
Combining~\eqref{stressbal1} and~\eqref{tangorthog} gives an expression for the flow coefficients $\tensor{Q}^{p}_{1}$ as well as a system of ODEs for the shape coefficients $\tensor{Q}_{1}^{f}, $ which are coupled to one another through $\tensor{Q}^{\chi f}_{1}$: 

\begin{equation}\label{straining}
    \tensor{Q}^{p}_{1}=\frac{2}{\mathrm{Ma}(3+2\lambda)}\left[-\frac{3}{(1+2S)^{2}}\tensor{Q}_{1}^{f}+\tensor{Q}_{1,\bm{\hat{r}}}^{E}+\frac{3}{2}\tensor{Q}^{E}_{1,\bm{\hat{t}}}\right]+\frac{10(1-\lambda)}{3+2\lambda}R^{-3}\tensor{S}_{1}^{\infty},
\end{equation}

\begin{equation}\label{shapetrans}
   \frac{\partial \tensor{Q}_{1}^{f}}{\partial t}=\frac{10(1+\lambda)}{\delta \mathrm{Ma}(16+19\lambda)(3+2\lambda)}\left[-\frac{3}{(1+2S)^{2}}\tensor{Q}^{f}_{1}+\tensor{Q}^{E}_{1,\bm{\hat{r}}}+\frac{6+9\lambda}{10(1+\lambda)}\tensor{Q}^{E}_{1,\bm{\hat{t}}}\right]+\frac{5}{\delta(3+2\lambda)}R^{-3}\tensor{S}_{1}^{\infty}-\tensor{Q}^{\chi f}_{1}.
\end{equation}
The flow coefficients $\tensor{Q}^{p}_{1}$ and the ODEs for the shape coefficients $\tensor{Q}^{f}_{1}$ are defined in terms of the dipole moments $\bm{P}_{1}$ and $\bm{P}_{2}$, which are functions of time. The flow and shape coefficients defining $\bm{v}^{\infty}_{1}$ and $p^{\infty}_{1}$ are given by the corresponding equations for drop 2 calculated without any hydrodynamic interactions (i.e., equations~\eqref{dchi},~\eqref{straining} and~\eqref{shapetrans} without $\tensor{S}^{\infty}$),
\begin{subequations}
\begin{align}
\bm{D}^{\chi}_{2,0} &= \frac{1}{2\mathrm{Ma}}\bm{P}_{2}\times\bm{E}_{2}^{\infty}, \label{eq:Dchi} \\
\tensor{Q}^{p}_{2,0} &= \frac{2}{\mathrm{Ma}(3+2\lambda)}\left[-\frac{3}{(1+2S)^{2}}\tensor{Q}_{2}^{f}+\tensor{Q}_{2,\bm{\hat{r}}}^{E}+\frac{3}{2}\tensor{Q}^{E}_{2,\bm{\hat{t}}}\right], \label{eq:ambientstraining} \\
\frac{\partial \tensor{Q}_{2,0}^{f}}{\partial t} &= \frac{10(1+\lambda)}{\delta \mathrm{Ma}(16+19\lambda)(3+2\lambda)}\left[-\frac{3}{(1+2S)^{2}}\tensor{Q}^{f}_{2}+\tensor{Q}^{E}_{2,\bm{\hat{r}}}+\frac{6+9\lambda}{10(1+\lambda)}\tensor{Q}^{E}_{2,\bm{\hat{t}}}\right]-\tensor{Q}^{\chi f}_{2}. \label{eq:ambientshapetrans}
\end{align}
\end{subequations}

\subsection{Charge conservation}\label{charge conservation}
The charge on the interface of drop 1 evolves according to the transport equation~\eqref{chargecons}, which in nondimensional form is given by

\begin{equation}\label{chargeconsnondim}
    \frac{\partial q_{1}}{\partial t}+\frac{2+Q}{1+2S}\llbracket\bm{j}\bcdot\bm{\hat{r}}\rrbracket_{1}+\bnabla_{s,1}\bcdot(q_{1}\bm{v}_{1})=0,
\end{equation}
where 
\begin{equation}\label{currentjump}
\llbracket\bm{j}\bcdot\bm{\hat{r}}\rrbracket_{1}=S\bm{E}_{1}^{+}\bcdot\bm{\hat{r}}-\bm{E}_{1}^{-}\bcdot\bm{\hat{r}}\end{equation}
is the discontinuity in the current across the interface of drop 1 in the radial direction.
The dimensionless charge distribution $q_{1}$ introduced in~\eqref{gauss} is
\begin{equation}\label{chargedensity}
    q_{1}=(1-Q)\bm{E}_{\infty}\bcdot\bm{\hat{r}}+(2+Q)\bm{P}_{1}\bcdot\bm{\hat{r}}.\end{equation}
The interfacial fluid velocity $\bm{v}_{1}$ appearing in~\eqref{chargeconsnondim} is $\bm{v}_{1}=\bm{v}^{+}_{1}|_{r=1}=\bm{v}^{-}_{1}|_{r=1}.$ Substituting the charge distribution~\eqref{chargedensity}, the current jump~\eqref{currentjump} and fluid velocity~\eqref{exteriorflow0} into~\eqref{chargeconsnondim} allows us to derive a system of ODEs for the dipole moment $\bm{P}_{1}$:
\begin{equation}\begin{split}\label{dipoleevolution}
    \frac{\mbox{d}\bm{P}_{1}}{\mbox{d} t}=&-\bm{P}_{1}-\frac {S-1}{1+2S}\bm{E}_{1}^{\infty}+\bm{\Omega}_{1}\times\left(\frac{1-Q}{2+Q}\bm{E}_{1}^{\infty}+\bm{P}_{1}\right)
        \\&+\left(\frac{1-Q}{2+Q}\bm{E}_{1}^{\infty}+\bm{P}_{1}\right)\bcdot\left(\frac{1}{5}\tensor{Q}_{1}^{p}-\frac{6}{5}\delta\left(\frac{\partial \tensor{Q}_{1}^{f}}{\partial t}+\tensor{Q}^{\chi f}_{1}\right)+R^{-3}\tensor{S}^{\infty}_{1}\right)+R^{-3}\frac{1-Q}{2+Q}\frac{\mbox{d}\bm{P}_{2}}{\mbox{d} t}\bcdot(\tensor{I}-3\bm{\hat{R}\hat{R}}),\end{split}
\end{equation}
where $\boldsymbol{\Omega}_{1}=\bm{D}_{1}^{\chi}+R^{-3}\boldsymbol{\omega}_{1}^{\infty}$ is the total rotational velocity of drop 1. If the electric field is sufficiently weak, drop 1 rotates only as a result of electrohydrodynamic interactions with drop 2. We refer to this type of behaviour as the Taylor regime, in which the impact of charge convection scales like the straining flow as $\mathcal{O}(\mathrm{Ma}^{-1}\lambda^{-1}),$ as seen in~\eqref{straining}. In sufficiently strong fields, self-induced rotation is possible in both drops due to the Quincke instability. We refer to this type of behaviour as the Quincke regime, in which drop rotation is the dominant contribution to the fluid velocity, and convection scales like the solid-body rotation as $\mathcal{O}(\mathrm{Ma}^{-1}).$ 
An equation similar to~\eqref{dipoleevolution} can be obtained for the dipole moment of drop 2. Then, the system of ODEs describing the evolution of the dipole moments and shape coefficients of drops 1 and 2 can be written succinctly as

\begin{equation}\label{eqnssummary}
    \frac{\mbox{d}}{\mbox{d}t}\begin{bmatrix}
        \bm{P}_{1}
        \\\bm{P}_{2}
        \\\tensor{Q}^{f}_{1}
        \\\tensor{Q}^{f}_{2}
        \\\bm{R}
    \end{bmatrix}=\mathcal{F}\left(\bm{P}_{1},\bm{P}_{2},\tensor{Q}^{f}_{1},\tensor{Q}^{f}_{2},\bm{R};\mathrm{Ca_{E}},\mathrm{Ma},Q,S,\lambda\right).
\end{equation}
This system of ODEs is integrated numerically with initial conditions corresponding to initially spherical drops, $\tensor{Q}^{f}_{1}(0)=\tensor{Q}^{f}_{2}(0)=\bm{0}$,
and zero surface charge density, $q_{1}(\bm{r},0)=q_{2}(\bm{r},0)=0$. When substituted into~\eqref{chargedensity}, the charge density conditions yield initial conditions for both dipole moments,
\begin{equation}\label{dipoleICs}
\bm{P}_{1}(0)=\bm{P}_{2}(0)=\frac{R^{3}}{\frac{Q+2}{Q-1}R^{3}+1}\bm{\hat{E}}_{0}\bcdot\left(\tensor{I}+\frac{3}{\frac{Q+2}{Q-1}R^{3}-2}\bm{\hat{R}\hat{R}}\right). 
\end{equation}

\section{Results and discussion}\label{results}

The nonlinearities associated with charge convection and the convective product $\bm{v}\bcdot\nabla f$ make it unfeasible to find closed-form solutions to the system of equations~\eqref{eqnssummary}. Hence, in this section, we solve the equations numerically using a fourth-order explicit Runge-Kutta method. Unless otherwise stated, we use the parameter values from the experiments of Salipante and Vlahovska \citep{salipante2010electrohydrodynamics} (also used by Das and Saintillan \citep{das2021three}), as given in Table~\ref{parametervalues}. First, in \S\ref{hysteresissection}, we discuss the occurrence of hysteresis during the onset of Quincke rotation of an isolated drop reported in the experiments of~\citet{salipante2010electrohydrodynamics}. To analyze an isolated drop, we take the limit of widely separated drops, i.e., $R\rightarrow\infty$, recovering the theory for an isolated drop developed by \citet{das2021three}. Hysteresis occurs in both isolated drops and interacting pairs of drops, but here, for simplicity, we consider only the former case. Second, in \S\ref{freelysuspended}, we consider the effect of electrohydrodynamic interactions between a pair of drops if the drops are free to translate. Finally, in \S\ref{fixeddrops}, we consider interactions between a pair of drops that are fixed in space. The MATLAB codes that solve \eqref{eqnssummary} numerically to produce these results are provided in the Supplementary Material \cite{SM}.

\renewcommand{\arraystretch}{2}
\begin{table}[]
    \centering
    \begin{tabular}{|M{4em}|M{4em}|M{8em}|M{8em}|M{6em}|M{8em}|M{7em}|}
         \hline
         $\epsilon^{+}/\epsilon_{0}$ & $\epsilon^{-}/\epsilon_{0}$ & $\sigma^{+}$ & $\sigma^{-}$ & $\mu^{+}$ & $\gamma$ &$a$   \\ \hline
         $5.3$  & $3$ & \SI{4.5e-11}{\siemens/\meter} & \SI{1.23e-12}{\siemens/\meter} & \SI{0.69}{\Pa/\second} & \SI{4.5e-3}{\N/\meter} &\SI{0.9e-3}{\meter}\\ \hline
    \end{tabular}
    \caption{Material properties of Salipante and Vlahovska \citep{salipante2010electrohydrodynamics}. $\epsilon_{0}=$ \SI{8.8542e-12}{\farad/\meter} is the permittivity of free space.}
    \label{parametervalues}
\end{table}

\subsection{Hysteresis in isolated drops}\label{hysteresissection}

\begin{figure}
    \centering
    \includegraphics[width=\textwidth]{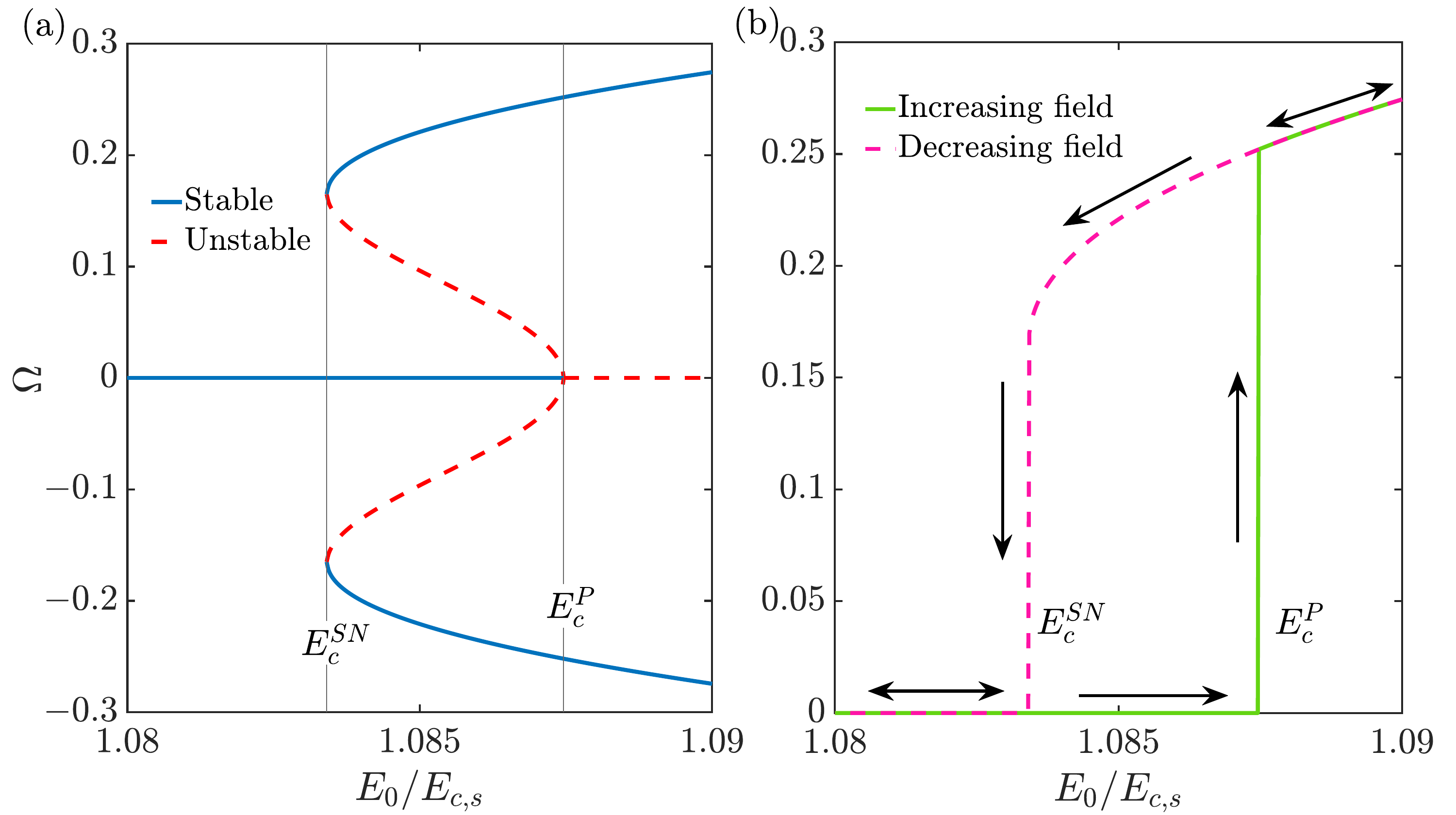}
    \caption{(a) Bifurcation diagram and (b) corresponding hysteresis loop for the onset of rotation in an isolated drop when $\lambda=14.1.$ In subfigure (a), the blue solid lines denote stable solutions, while the red dashed lines denote unstable ones. In subfigure (b), the green solid line marks the angular velocity of the drop when the field is increased, and the pink dashed line marks the angular velocity of the drop when the field is decreased.}
    \label{hysteresis_and_bifurcation}
\end{figure}

It is well known that the onset of Quincke rotation of an isolated solid sphere occurs via a supercritical pitchfork bifurcation \citep{turcu1987,jones1984quincke} at a critical electric field strength $E_{c,s}$, as defined in~\eqref{critfield}. For an isolated drop, the behavior is slightly more complicated. Experiments by~\citet{salipante2010electrohydrodynamics} have revealed the occurrence of hysteresis in the onset of Quincke rotation for a drop which becomes more significant as the viscosity ratio is decreased or the drop size is increased.
This corresponds to the supercritical pitchfork bifurcation that occurs for solid spheres being replaced by a subcritical pitchfork bifurcation and two saddle-node bifurcations in the case of drops. As we now show, our model is able to successfully capture this hysteresis effect.

Figure~\ref{hysteresis_and_bifurcation}(a) displays the associated bifurcation diagram, where $\Omega$ denotes the rotational velocity of an isolated drop. Solid blue lines denote stable solutions, while red dashed lines denote unstable solutions. Figure~\ref{hysteresis_and_bifurcation}(b) displays the corresponding hysteresis loop, in which the green solid and pink dashed lines mark the angular velocity of the drop in increasing and decreasing electric fields, respectively. In particular, Figure~\ref{hysteresis_and_bifurcation} shows that increasing the electric field strength gradually from zero eventually causes the angular velocity of the drop to jump discontinuously from the stable zero rotation solution branch to a stable nonzero rotation solution branch at the critical electric field strength of the pitchfork bifurcation (denoted by $E_{c}^{P},$ and indicated with a vertical line). However, if the electric field strength is then decreased, the rotation persists until the critical electric field strength of the saddle node bifurcations is reached (denoted by $E_{c}^{SN},$ and indicated with another vertical line), causing the angular velocity to discontinuously jump back to the stable zero rotation solution branch \citep{strogatz}. Since the time-stepping algorithm cannot determine the nonzero unstable solution branches, Figure~\ref{hysteresis_and_bifurcation}(a) was obtained by seeking steady solutions of~\eqref{eqnssummary} for an isolated drop (i.e., $\bm{R} \rightarrow \infty$) using the MATLAB nonlinear least-squares solver {\it lsqnonlin}~\cite{lsqnonlin}.  

\begin{figure}
    \centering
\includegraphics[width=0.8\textwidth]{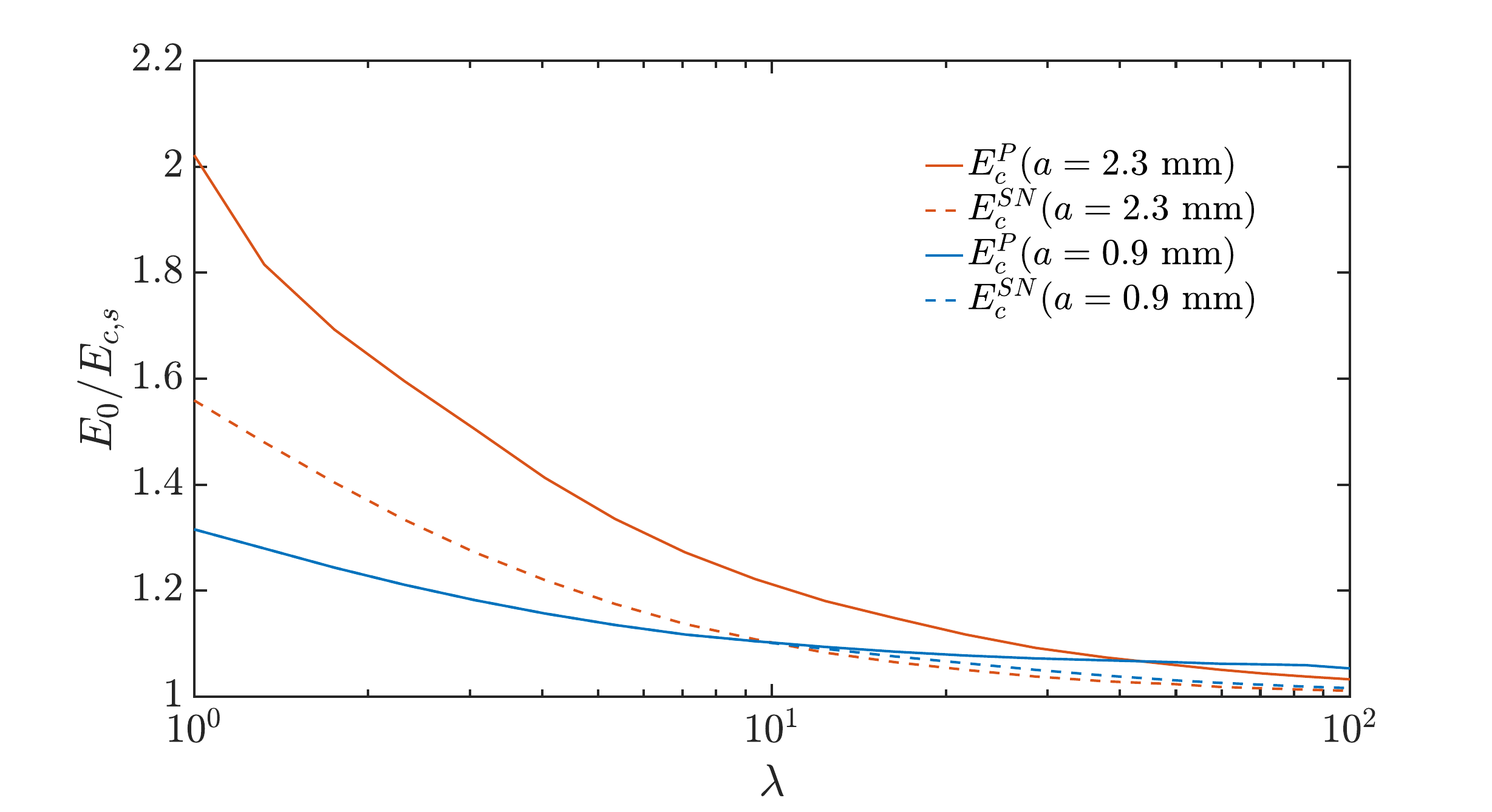}
    \caption{Critical electric field strengths (scaled by $E_{c,s}$) for saddle node and subcritical pitchfork bifurcations of an isolated drop for two different drop sizes as functions of the viscosity ratio. Dotted lines denote the critical electric field strength of the saddle node bifurcations $E_{c}^{SN}$ and solid lines denote the critical electric field strength of the subcritical pitchfork bifurcation $E_{c}^{P}$. }
    \label{hysteresis}
\end{figure}

Figure~\ref{hysteresis} displays the critical electric field strengths (scaled by $E_{c,s}$) for the pitchfork bifurcation $E^{P}_{c}$ and saddle node bifurcations $E^{SN}_{c}$ as functions of the viscosity ratio for two different drop sizes. Dotted and solid lines of the same color correspond to $E^{SN}_{c}$ and $E^{P}_{c},$ respectively, for the same drop-fluid system. The present model, which reduces to that of~\citet{das2021three} for an isolated drop, correctly predicts the increasing significance of hysteresis as the drop size is increased. In particular, Figure~\ref{hysteresis} shows that a drop with $a=$ \SI{2.3}{\milli\meter} (red lines) exhibits hysteresis across the entire range of viscosity ratios studied. Furthermore, the hysteresis becomes slightly more significant as the viscosity ratio is decreased, in agreement with the experimental results of~\citet{salipante2010electrohydrodynamics}. 
However, for a drop with $a=$ \SI{0.9}{\milli\meter} the present model predicts that the hysteresis appears only for drops when $\lambda$ is greater than approximately $10,$ as shown by the blue lines in Figure~\ref{hysteresis}, in contrast to the experimental results~\citep{salipante2010electrohydrodynamics}.
This disagreement is likely to be a consequence of the limitations of the small deformation theory, which is accurate for large $\lambda$ and small $Ca_E$ (using the parameters given in Table \ref{parametervalues}, $\mathrm{Ca}_{E}$ is around 1.1 for the drop of radius $a=$ \SI{0.9}{\milli\meter} in an electric field of strength $E_{0}=1.5E_{c,s}$). It is remarkable then, but likely coincidental, that the theory captures the correct trend for the drop of radius $a=$ \SI{2.3}{\milli\meter} (again using the parameters in Table~\ref{parametervalues}, $\mathrm{Ca}_{E}$ is close to 7 for the drop of radius $a=$ \SI{2.3}{\milli\meter} in an electric field of strength $E_{0}=2E_{c,s}$). We note that hysteresis is found to occur only for deforming drops, and more specifically as a result of the presence of the convective product $\tensor{Q}^{\chi f}$, as defined in~\eqref{vnablafsymmetric}, in the system of ODEs~\eqref{eqnssummary}. 

\begin{figure}
    \centering
\includegraphics[width=0.9\textwidth]{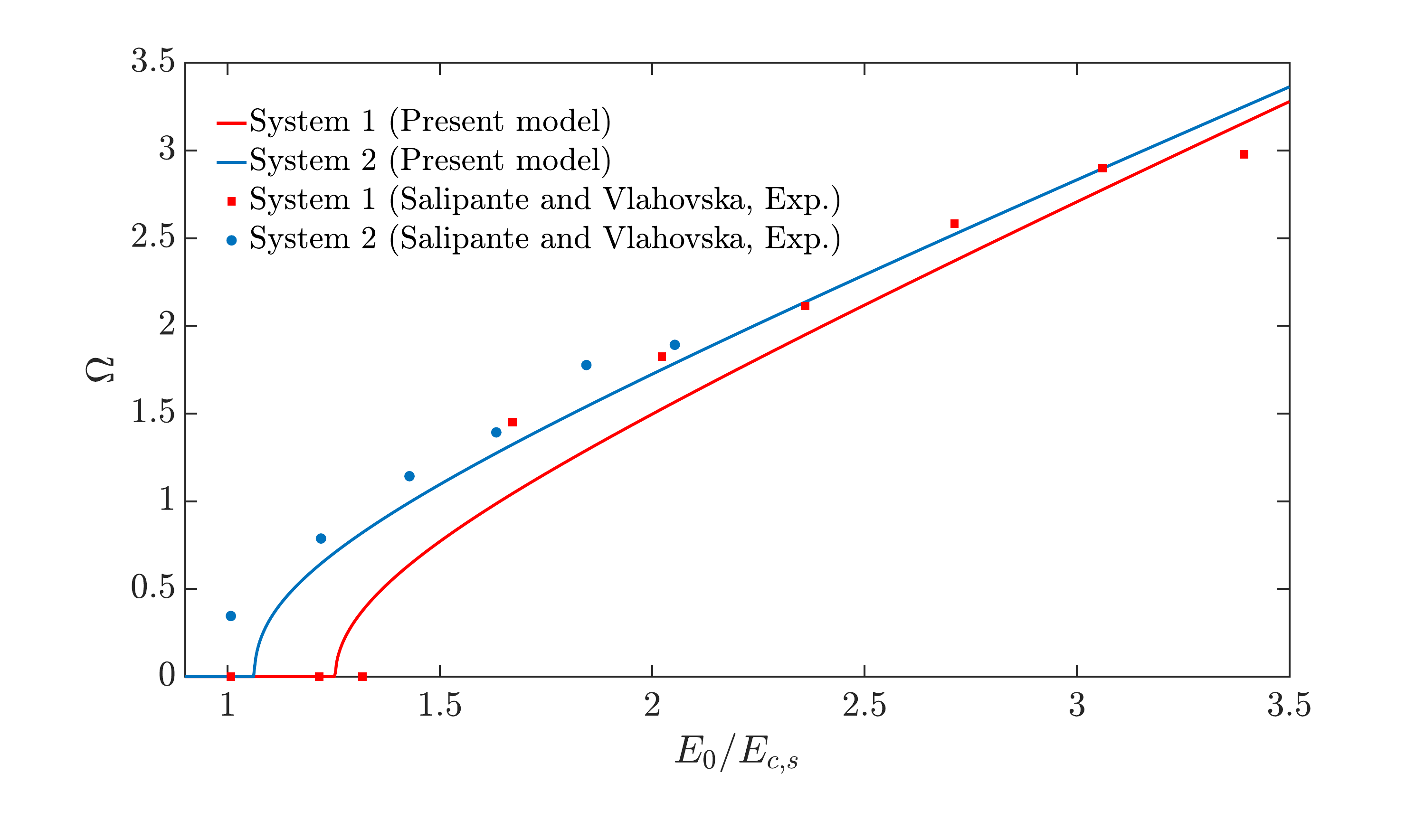}
    \caption{Angular velocity of an isolated drop as functions of the scaled applied electric field strength for two sets of material parameters: System 1, for which $\lambda=1.4$ and $a=$ \SI{0.76}{\milli\meter} (red line and squares) and System 2, for which $\lambda=14$ and $a=$ \SI{0.565}{\milli\meter} (blue line and dots).}
    \label{singledropomega}
\end{figure}
Figure~\ref{singledropomega} displays the dependence of the angular velocity of an isolated drop on the applied electric field strength for two different sets of material properties, which correspond to those studied by \citet{salipante2010electrohydrodynamics}. For System 1 ($\lambda=1.4$ and $a=$ \SI{0.76}{\milli\meter}), indicated by the red line and squares, the model underestimates the experimentally observed angular velocities in most instances. The sharp jump between the third and fourth experimental data points suggests possible hysteresis, but the present model does not predict such behaviour at these parameter values. The agreement with the experimental results is better for System 2 ($\lambda=14$ and $a=$ \SI{0.565}{\milli\meter}), indicated by the blue line and dots, which has a higher viscosity ratio and lower electric capillary number. The experimental results suggest that a high-viscosity drop has a critical electric field strength which is less than that of a solid sphere, which is corroborated by the numerical simulations of~\citet{das2017sims} and~\citet{firouznia2023spectral}, but is not captured by the present small deformation theory.

\subsection{Freely suspended pairs of drops}\label{freelysuspended}

Next, we examine the behaviour of pairs of drops that are free to translate as a result of hydrodynamic and DEP interactions. Specifically, the rotlet flow of one drop provides the dominant contribution to the translational velocity of the other, while straining flow and DEP forces provide higher-order corrections. In practical settings, the drops may come into contact and coalesce, but the present model does not account for coalescence and will predict unphysical behaviour when the separation distance between the drops is smaller than two drop radii, i.e., when the drops overlap. Hence, to prevent the drops from coming into contact or overlapping, we impose a short-range repulsive force \citep{bossis1984dynamic,takamura1981microrheology} along the line joining the centres of the two drops, denoted by $\bm{F}_{R}$ and given by
\begin{equation}\label{repulsiveforce}
    \bm{F}_{R} = -A\frac{e^{-B\left(R-2.01\right)}}{1-e^{-B\left(R-2.01\right)}}\bm{\hat{R}},
\end{equation}
where $A$ and $B$ are adjustable non-dimensional parameters. In our simulations, we found the values $A = B = 200$ to be sufficient to avoid overlaps. In most cases when a pair of drops attract one another, this repulsive force will balance the attractive electrohydrodynamic or DEP interactions and the separation distance $R$ will settle to a value slightly above 2. 
We separate our results into two categories based on the electric field strength: weak fields (i.e., the Taylor regime), wherein the drop trajectories are driven solely by induced straining flows and DEP forces, and strong fields (i.e., the Quincke regime), wherein drop rotation is also present and the drop trajectories are characterised by complex spiraling motions due to the rotlet flows. First, we validate the present model against the results reported in previous publications concerning drop trajectories in the Taylor regime. Next, we report new results for drop trajectories in the Quincke regime. In simulating the trajectories of freely suspended drops in the Quincke regime, we apply small perturbations to $\bm{P}_{1}(0)$ and $\bm{P}_{2}(0)$ in order to induce Quincke rotation in different directions, leading to a range of different behaviours. In practical settings, specific perturbations may be deliberately applied via an imposed fluid flow, electric field asymmetry or mechanical disturbance, or random perturbations may be introduced by weak residual flows or electrostatic disturbances.

\subsubsection{Taylor regime}

\begin{figure}
    \centering
\includegraphics[width=\textwidth]{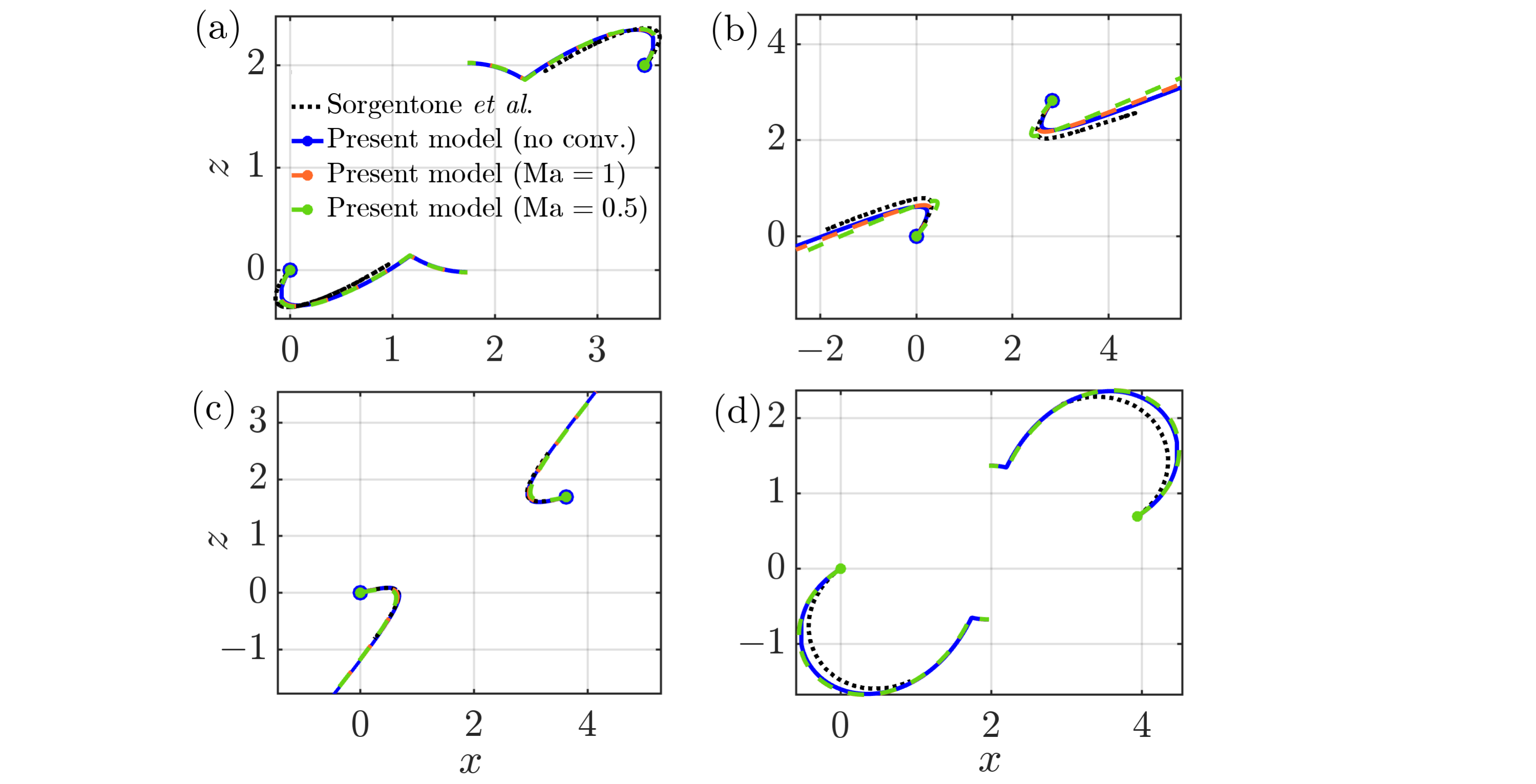}
    \caption{Four examples of two-dimensional drop trajectories when the electric field is applied in the $z$ direction. In each case, $R(t=0)=4,$ $\lambda=1$ and $Ca_{E}=0.1.$ Other parameter values are as follows: (a) $S=10,$ $Q=1,$ $\theta_{R}(t=0)=60\degree$; (b) $S=1,$ $Q=10,$ $\theta_{R}(t=0)=45\degree$; (c) $S=1,$ $Q=0.1,$ $\theta_{R}(t=0)=65\degree$; (d) $S=0.01,$ $Q=1,$ $\theta_{R}(t=0)=80\degree.$ Solid blue lines denote the trajectories predicted by the present small deformation theory, with the dots marking the initial positions of the drops. Black dotted lines denote the corresponding trajectories predicted by the boundary element simulations of \citet{sorgentone_kach_khair_walker_vlahovska_2021} (reproduced from Figure 9 of Ref.~\citep{sorgentone_kach_khair_walker_vlahovska_2021}). Orange and green dashed lines denote the corresponding trajectories when charge convection is retained, with $\mathrm{Ma}=1$ and $\mathrm{Ma}=0.5,$ respectively.}
    \label{sorgentonevalidation}
\end{figure}

\begin{figure}
    \centering
    \includegraphics[width=\textwidth]{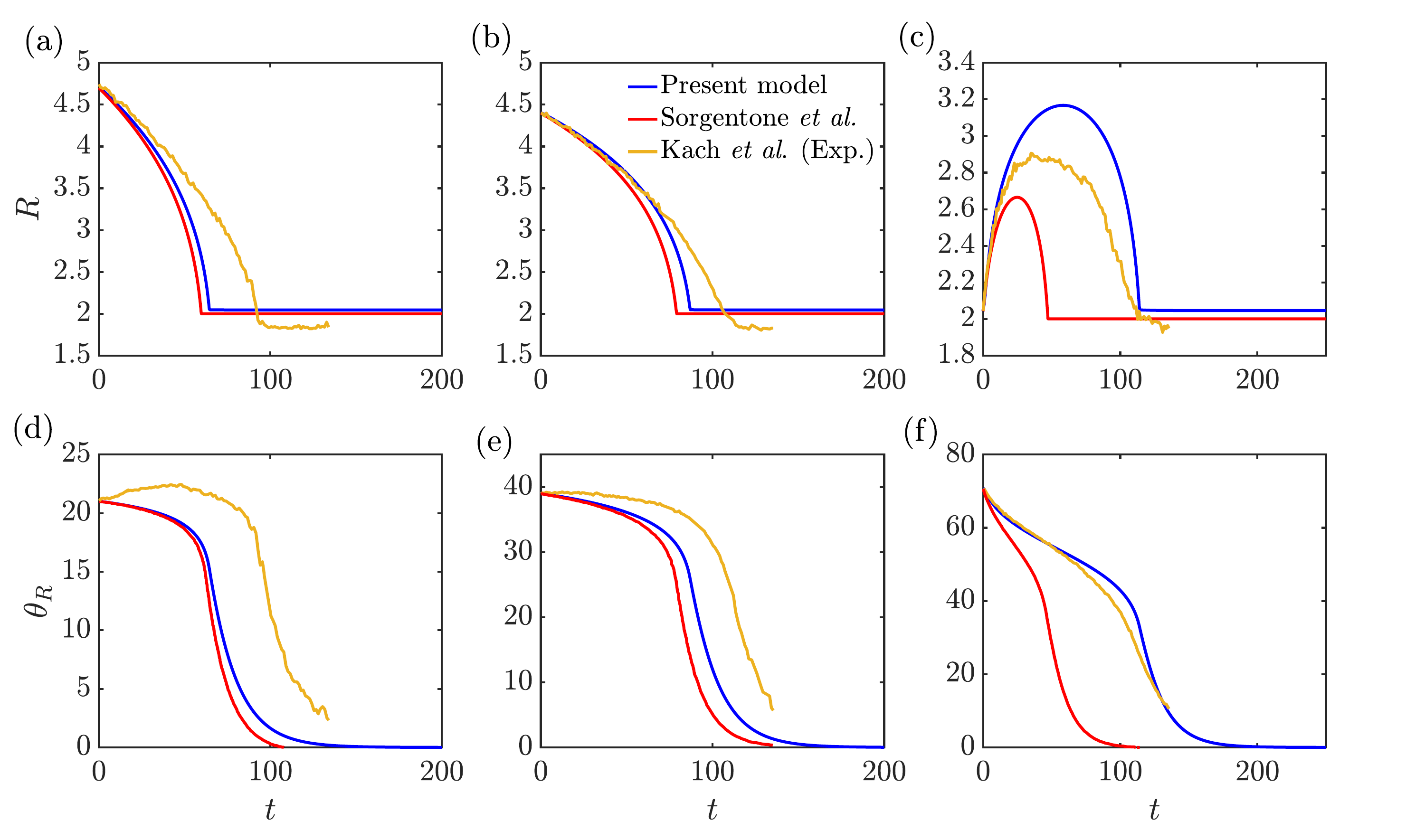}
    \caption{Evolution of the separation distance $R$ and angle $\theta_{R}$ of the separation vector for a pair of drops in the Taylor regime. In each case, $\lambda=0.39,$ $Q=0.57,$ $S=29$, $\mathrm{
Ma}=0.75$ and $\mathrm{Ca}_{E}=0.19.$ The initial spatial configurations of the drops are as follows: (a) $\theta_{R}(t=0)=21\degree$, $R(t=0)=4.7$; (b) $\theta_{R}(t=0)=39\degree$, $R(t=0)=4.4$; (c) $\theta_{R}(t=0)=71\degree$, $R(t=0)=2.04$. Blue lines denote the predictions of the present small deformation theory, red lines denote the predictions of the small deformation theory of \citet{sorgentone_kach_khair_walker_vlahovska_2021}, and yellow lines denote the experimental findings of \citet{kach2022prediction}.}
    \label{kach_comparison_plot}
\end{figure}
We first compare our results with previous work concerning drop trajectories in the Taylor regime, in which Quincke rotation is absent, and drop translation is driven by induced straining flows and DEP. As a result of the axisymmetry of the flow fields and shapes of the drops, the trajectories of pairs of drops in the Taylor regime are two-dimensional. Specifically, they remain in the plane containing the initial drop separation vector and the applied field. Figure~\ref{sorgentonevalidation} compares the drop trajectories predicted by the present theory with those predicted by the boundary element simulations of \citet{sorgentone_kach_khair_walker_vlahovska_2021} (taken from their Figure 9). In the present simulations, we forgo the parameter values of Table~\ref{parametervalues} and instead directly prescribe values of the nondimensional parameters $Q,$ $S,$ $\lambda$ and $\mathrm{Ca}_{E}$ in order to match the values used in Ref.~\citep{sorgentone_kach_khair_walker_vlahovska_2021}. The permittivity and conductivity ratios $Q$ and $S$ change between the four different cases shown in Figure~\ref{sorgentonevalidation}, while $\lambda=1$ and $\mathrm{Ca}_{E}=0.1$ in all cases.
Furthermore, since the simulations of \citet{sorgentone_kach_khair_walker_vlahovska_2021} neglect charge relaxation and convection, we initially remove these effects from the present model for the purposes of the comparison. Instead, we impose the boundary condition, $\llbracket\bm{j}\bcdot\bm{\hat{r}}\rrbracket=0$, which simply prescribes continuity of current across the interface. In this case, the dipoles of the two drops are identical and are given by
\begin{equation}\label{steadydipoles}
\bm{P}_{1} = \bm{P}_{2} = \frac{1}{\frac{1+2S}{1-S}+R^{-3}}\bm{\hat{E}}_{0}\bcdot\left(\tensor{I} + \frac{3R^{-3}}{\frac{1+2S}{1-S}-2R^{-3}}\bm{\hat{R}\hat{R}}\right). 
\end{equation}

As Figure~\ref{sorgentonevalidation} shows, there is good agreement between the two-dimensional trajectories predicted by the simulations of \citet{sorgentone_kach_khair_walker_vlahovska_2021}, shown in black dotted lines, and those predicted by the present small deformation theory, shown in blue solid lines, when charge convection and relaxation are neglected. This is perhaps surprising given the small initial separation distances ($R=4$) and drop viscosities ($\lambda=1$) involved, since our theory is valid for widely separated and highly viscous drops. In the cases shown in subfigures (a) and (d), for which (a) $S=10,$ $Q=1,$ $\theta_{R}(t=0)=60\degree$ and (d) $S=0.01,$ $Q=1,$ $\theta_{R}(t=0)=80\degree$, the drops initially repel one another and the line joining their centres rotates towards the direction of the applied field (i.e., the $z$ direction), before the drops change direction and attract one another. Ultimately, the drops reach a steady state in which they are aligned parallel to the applied field. Conversely, in the cases shown in subfigures (b) and (c), for which (b) $S=1,$ $Q=10,$ $\theta_{R}(t=0)=45\degree$ and (c) $S=1,$ $Q=0.1,$ $\theta_{R}(t=0)=65\degree$, the drops exhibit so-called ``kiss and run'' dynamics, initially drifting towards and then away from each other. 

To investigate the role of charge convection on the drop trajectories, we numerically integrate the complete system of ODEs \eqref{eqnssummary} for both $\mathrm{Ma}=1$ and $\mathrm{Ma}=0.5$. The resulting drop trajectories are denoted with orange ($\mathrm{Ma}=1$) and green ($\mathrm{Ma}=0.5$) dashed lines in Figure~\ref{sorgentonevalidation}.The effect of convection is small in all cases, but is most visible in case (b), where including convection causes the drops to come slightly closer together before repelling than they do when convection is neglected. The strength of convection can be increased by decreasing $\mathrm{Ma}$ further, and when $\mathrm{Ma}$ is sufficiently low, the present model predicts that the ``kiss and run'' behaviour no longer occurs in cases (b) and (c) (not shown in plots). Instead, the lines joining the centres of the drops rotate counterclockwise and the drops initially repel before attracting one another, in a manner similar to cases (a) and (d). We note that when $\mathrm{Ma}$ is very low, charge convection is very strong, giving rise to additional charge multipoles which have been neglected in the present model. Hence, full simulations are better suited for predicting drop dynamics in such systems with strong charge convection. In cases (a) and (b), if the electric field is sufficiently strong and charge convection is present, Quincke rotation may occur as we discuss in the next section.
 
In addition to their boundary element simulations, \citet{sorgentone_kach_khair_walker_vlahovska_2021} developed an asymptotic theory neglecting some physical effects to allow for closed-form solutions for the fluid velocities and DEP forces. In particular, the only electrostatic interactions present in their theory are the DEP forces on the drops, and the only hydrodynamic interactions are those driving drop translation through the Fax\'en's law. Charge convection and relaxation are absent, as is drop deformation, although the authors extended their theory to account for transient deformations, as described in their appendix. The theory of~\citet{sorgentone_kach_khair_walker_vlahovska_2021} neglecting drop deformation was subsequently compared with experiments by \citet{kach2022prediction}. Here, we include additional physical effects (namely, charge convection, relaxation, transient deformation and higher-order electrostatic and hydrodynamic interactions), and demonstrate that doing so improves the agreement with the experimental drop trajectories. Figure~\ref{kach_comparison_plot} shows the predictions of the present theory (blue lines) with experimental results of \citet{kach2022prediction} (yellow lines) and the theory of \citet{sorgentone_kach_khair_walker_vlahovska_2021} (red lines). Three cases with different initial spatial configurations of the drops are shown. Subfigures (a), (b) and (c) display the evolution of the separation distance $R$ in each case, while subfigures (d), (e) and (f) display the corresponding evolutions of the angle between the applied field direction and the line of centres of the drops in each case. In these simulations, we use the same parameter values as those used by \citet{kach2022prediction}, namely, $\lambda=0.39,$ $Q=0.57$, $S=29,$ $\mathrm{Ma}=0.75$ and $\mathrm{Ca}_{E}=0.19$. In the first two cases (subfigures (a) and (d), and subfigures (b) and (e)), the initial separation angle $\theta_{R}(t=0)$ is sufficiently small such that the drops monotonically attract while rotating around one another so that the separation angle approaches zero. In these cases, the inclusion of the additional physical effects mentioned above provides a modest improvement over the theory of \citet{sorgentone_kach_khair_walker_vlahovska_2021} when compared to the experimental results, but underpredicts the time it takes for the separation distance to reach the steady value $R\approx 2$. In the third case (subfigures (c) and (f)), the initial separation angle is larger than in the first two cases, and the drops initially repel one another while the separation angle again decreases, but eventually, the drops change direction and begin to attract. Including drop deformation and charge convection leads to a slight overprediction compared to experiments. However, the agreement between theory and experiment is substantially improved by retaining the additional physical effects. 

\subsubsection{Quincke regime}\label{sec:quinckeregime}

\begin{figure}
    \centering
    \includegraphics[width=0.9\textwidth]{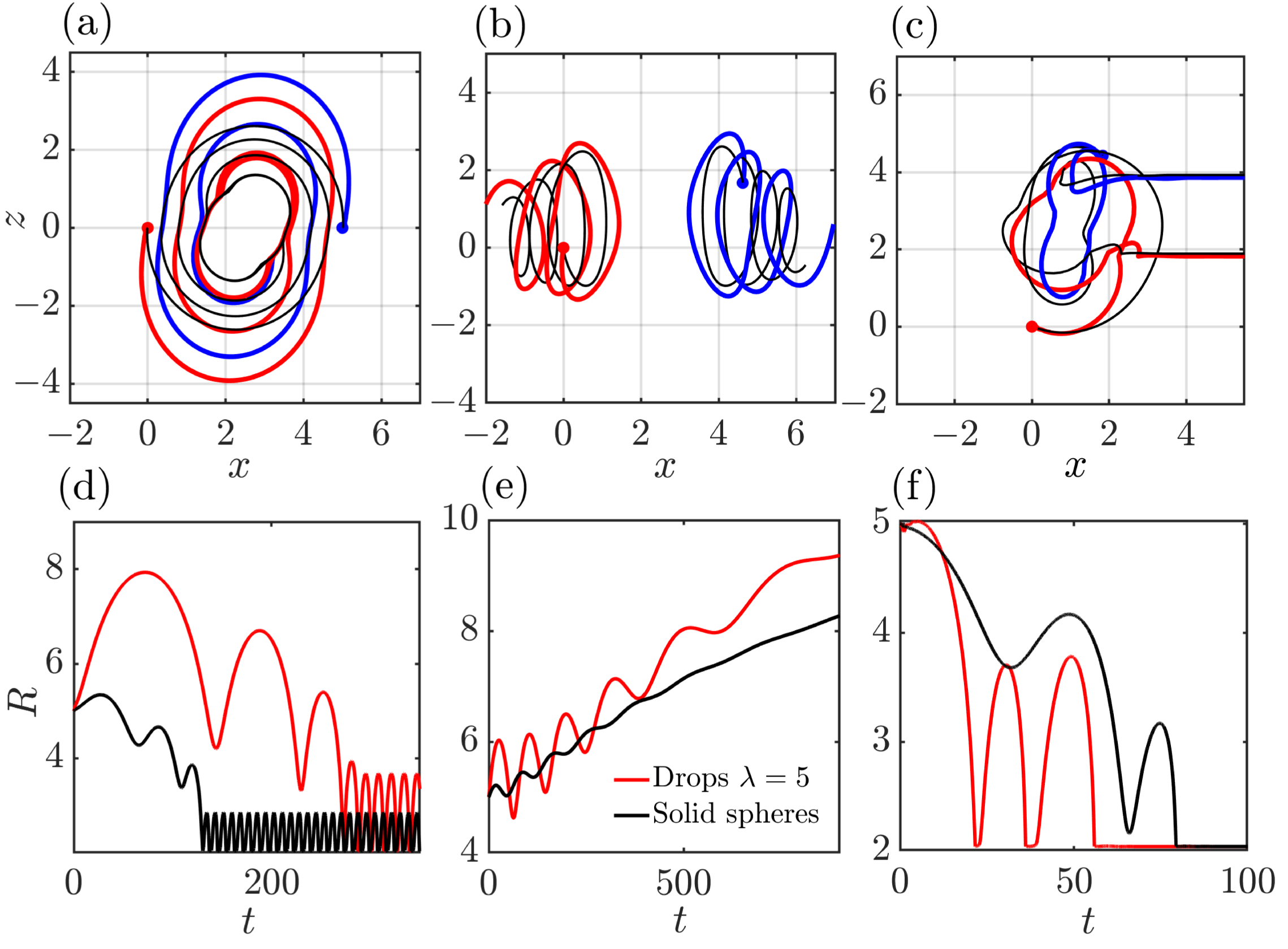}
    \caption{Trajectories of freely suspended drops and spheres with random perturbations to their dipole moments (a-c), and the corresponding evolutions of the separation distance $R$ (d-f) plotted as functions of time $t$. The electric field is applied in the $z$ direction. The thick blue and red lines denote the trajectories of drops with $\lambda=5$, and the thin black lines denote the trajectories of solid spheres with the same initial conditions. Dots denote the initial positions of the drops. In all cases, $E_{0}/E_{c,s}=3.$ 
    }
    \label{trajectoryplot}
\end{figure}

In the Quincke regime, the drop trajectories are primarily influenced by the rotlet flows due to Quincke rotation, almost always leading to complex spiraling motions. A variety of qualitatively different trajectories are obtained for a pair of drops of given material properties depending on the initial positions of the drops and the perturbations to their initial surface charge densities. Figure~\ref{trajectoryplot}(a--c) shows three representative examples of different types of behaviour, with the trajectories of the drops shown with thick blue and red lines, and those of solid spheres with identical initial conditions shown with thin black lines. Figure~\ref{trajectoryplot}(d-f) show the corresponding plots of the separation distance $R$ as functions of time $t$. In the case shown in Figure~\ref{trajectoryplot}(a), the drops and spheres spiral around and attract one another while remaining confined to the $x$-$z$ plane. The solid spheres, which attract one another more quickly than the drops, follow spiraling trajectories around one another. In contrast, the drops exhibit peanut-shaped trajectories due to the straining flows either pushing the drops further apart or pulling them closer together as they spiral around one another. Figure~\ref{trajectoryplot}(d) illustrates the corresponding oscillations in the separation distance $R,$ as a function of time and clearly shows larger oscillations for the drops than for the spheres. This type of behavior can occur when identical perturbations are applied to both drops, leading to corotation, or when no perturbation is applied, in which case hydrodynamic interactions lead to rotation, with the spirals confined to the plane defined by the applied field vector $\bm{\hat{E}}_{0}$ and the separation vector $\bm{\hat{R}}$. The special planar case shown in Figure~\ref{trajectoryplot}(a) will occur less frequently in practical settings than the other two cases shown in Figure~\ref{trajectoryplot}, as any perturbation in the third dimension will cause the drops and spheres to move out of the plane, leading to one of the other two cases discussed subsequently. In the case shown in Figure~\ref{trajectoryplot}(b), the drops and spheres undergo a spiraling motion while moving apart. Figure~\ref{trajectoryplot}(e) shows the separation distance increasing with time. The drops separate more quickly than the spheres, as the additional influence of the straining flow is repulsive when the angle between the separation vector and the applied field is greater than approximately $54.7^{\circ}$, as discussed later in \S\ref{attractionrepulsion}. In the case shown in Figure~\ref{trajectoryplot}(c), the drops and spheres attract one another again but follow complex trajectories around one another before settling into a steady configuration in which they counter-rotate while translating as a pair in a direction perpendicular to the applied field. This behavior is closely related to the self-propulsion of solid spheres under Quincke rotation on surfaces (Quincke rollers), which has garnered significant interest in the active matter community \citep{bricard2013,bricard2015}. The counter-rotating self-propelling case has been explored for two-dimensional drops recently in the simulations of \citet{dong24}. As shown in Figure~\ref{trajectoryplot}(f), the separation distance between the drops or spheres eventually settles at $R \gtrapprox 2$ due to the repulsive force given in~\eqref{repulsiveforce}. 

\subsection{Fixed pairs of drops}\label{fixeddrops}

Although the case of moving drops is the most general and relevant to experiments and other practical settings, steady solutions to the system of equations~\eqref{eqnssummary} do not exist in most cases because the separation vector $\bm{R}$ between the drops changes continuously. 
To gain a better understanding of the electrohydrodynamic interactions in a simplified setting, the relative positions of the two drops can be fixed, so that the separation vector $\bm{R}$ is constant. Doing so allows steady solutions of the resulting simplified system of equations to be obtained.

In the Taylor regime, the flow fields and shape of an isolated drop are axisymmetric, so the problem of a pair of drops can be formulated in two dimensions without any loss of generality. The spatial configuration can be fully described using only the separation distance $R$ and the angle $\theta_R$ between the applied field and the line connecting the centres of the two drops. In the Quincke regime, however, the situation is more complex because the flow fields and shape of an isolated drop are not axisymmetric after the onset of rotation. The dynamics of a pair of drops therefore depend significantly on the relative directions of the rotational axes of the drops, which are determined by the perturbations to the dipole moments. For simplicity, we consider the behaviour of pairs of unperturbed drops, except in the symmetric configurations of $\theta_{R}=0$ and $\theta_{R}=\pi/2$. In these symmetric configurations, unperturbed drops do not rotate, so we introduce perturbations leading to corotation of the drops. In any other configuration, unperturbed drops will naturally tend to corotate (i.e., $\boldsymbol{\Omega}_{1} = \boldsymbol{\Omega}_{2}$) in the plane containing the separation vector $\bm{R}$ due to the straining flows.
Since the drops are identical and all interactions, except for the translational velocities of the drops, are quadratic in $\bm{\hat{R}}$, the dipole moments, shapes, and flow fields of both drops are identical. Thus, we write $\bm{P} = \bm{P}_{1} = \bm{P}_{2}$, $\tensor{Q}^{f} = \tensor{Q}^{f}_{1} = \tensor{Q}^{f}_{2}$, $\tensor{Q}^{p} = \tensor{Q}^{p}_{1} = \tensor{Q}^{p}_{2}$, $\boldsymbol{\Omega} = \boldsymbol{\Omega}_{1} = \boldsymbol{\Omega}_{2}$, and so on for the remaining variables to simplify the following presentation. We can simplify the analysis of the unperturbed system by invoking symmetry arguments. Firstly, there is periodicity in $\pi$; by setting $\theta_{R} = \pi + \theta_{R}^{\prime}$, we have $\bm{\hat{R}} = -\bm{\hat{R}}^{\prime}$, and we need only interchange the names of the drops to recover the original problem. Secondly, moving drop 2 from $\theta_{R} = \theta_{R}^{\prime}$ to $\theta_{R} = \pi - \theta_{R}^{\prime}$, i.e., switching the sign of $\hat{R}_{z}$, results in a change of sign in the initial condition of $P_{x}$, as seen from~\eqref{dipoleICs}. However, $P_{z}$ remains unchanged. As a result, the shape of both drops as well as their dipole moments tilt in the opposite direction, and the sign of the rotation of the drops is reversed. Due to these symmetries, it is sufficient to restrict our analysis to spatial configurations corresponding to $0\leq\theta_{R}\leq\pi/2$. In the following subsections, we report results pertaining to the onset of rotation of interacting pairs of drops, as well as the shapes of the drops and the attractive or repulsive nature of the drop interactions.

\subsubsection{Angular velocity}\label{angular velocity}

\begin{figure}
    \centering
    \includegraphics[width=0.75\linewidth]{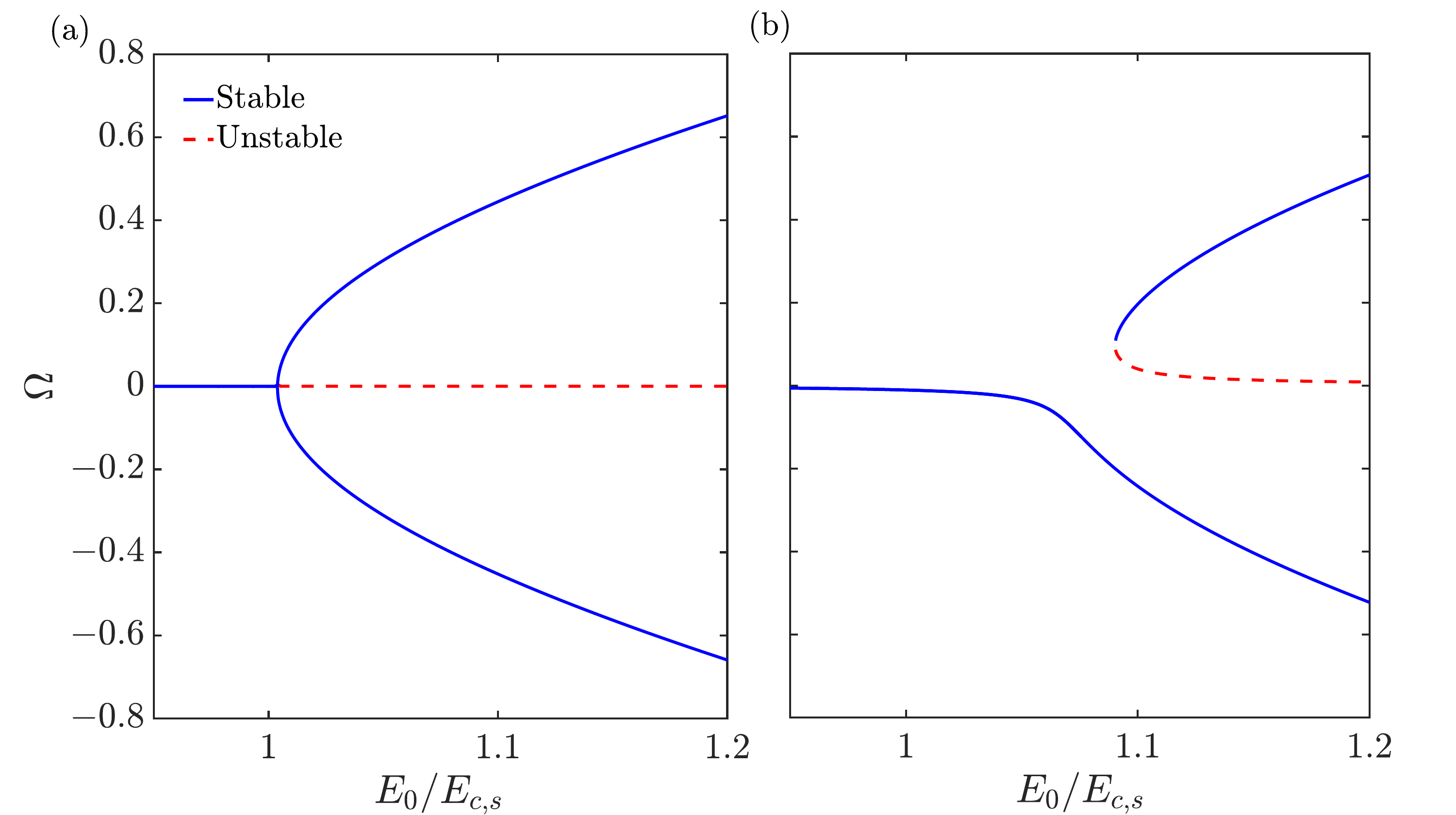}
    \caption{Bifurcation diagrams for pairs of (a) solid spheres and (b) spherical drops with $\lambda=5$, both with $\theta_R=\pi/4$ and $R=5.$ } \label{brokenbifurcation}
\end{figure} 

The onset of Quincke rotation in interacting pairs of solid spheres is largely similar to that of an isolated sphere; the instability corresponds to a supercritical pitchfork bifurcation, but the critical field at which bifurcation occurs is offset from that for an isolated sphere depending on the spatial configuration of the pair of spheres \citep{das2013electrohydrodynamic}. 
However, the onset of rotation in interacting pairs of drops is rather different in that of an isolated drop. When two identical drops interact, the angular velocity of each drop undergoes a pitchfork bifurcation only in certain symmetric configurations, specifically when the separation vector $\bm{R}$ between the drops is either perfectly parallel ($\theta_{R}=0$) or perpendicular ($\theta_{R}=\pi/2$) to the applied field. In more general configurations, the symmetry around drop 1 is broken by the straining flow produced by drop 2, and vice versa, leading to an imperfect bifurcation.

\begin{figure}
    \centering
    \includegraphics[width=0.9\textwidth]{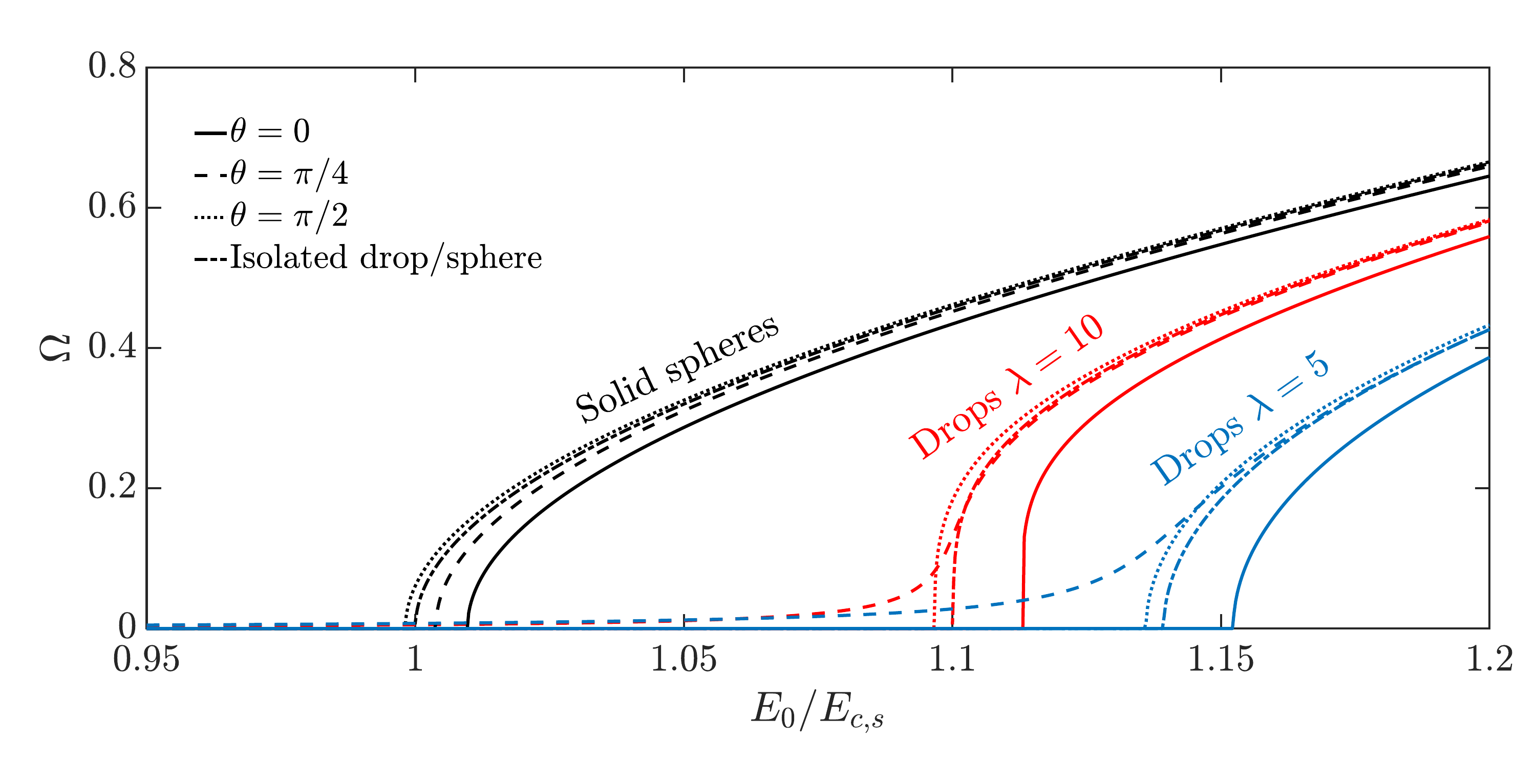}
    \caption{Steady angular velocity magnitudes for corotating solid spheres (black lines) and drops with $\lambda=10$ (red lines) and $\lambda=5$ (blue lines) at different values of $\theta_R$, as well as the values for an isolated sphere and drop. In all cases involving interactions, $R=5.$} \label{omegavse0ecdifftheta}
\end{figure}

The straining flow, which is present even in subcritical fields, produces weak rotation in drop 1 through $\boldsymbol{\omega}^{\infty}$ and $\tensor{S}^{\infty}$, which both scale as $\mathcal{O}(\mathrm{Ma}^{-1}\lambda^{-1}R^{-3})$). This straining flow convects charge around the surface of drop 1 and misaligns its dipole with the local electric field. 

Figure \ref{brokenbifurcation} displays bifurcation diagrams for pairs of (a) solid spheres and (b) spherical drops, with $\theta_R=\pi/4$ in both cases. In the former case, a supercritical pitchfork bifurcation occurs at some critical field which is offset slightly from $E_{c,s}$ because of electrostatic interactions, scaling as $\mathcal{O}(R^{-3})$. In the latter case, no critical field exists and instead the angular velocities of the drops increase steadily as $E_{0}/E_{c,s}$ is increased. At a certain electric field, a saddle node bifurcation occurs, giving rise to new unstable and stable branches. To capture the unsteady solution branches, Figure \ref{brokenbifurcation} is obtained using the nonlinear MATLAB solver {\it fsolve}~\citep{lsqnonlin}.

\begin{figure}
    \centering
    \includegraphics[width=0.8\textwidth]{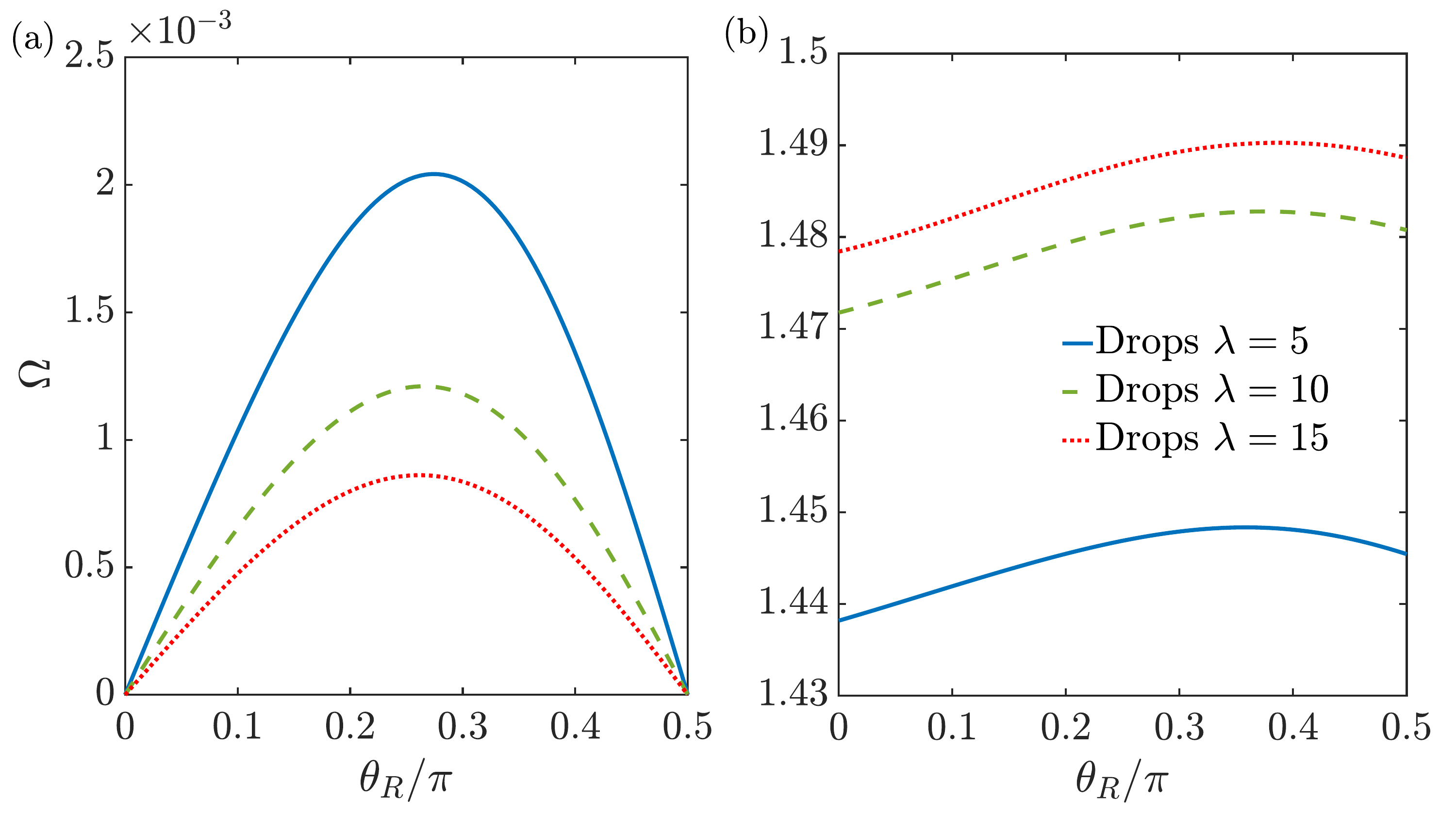} 
    \caption{Steady angular velocity as functions of $\theta_{R}$ with $R=5$ at (a) $E_{0}/E_{c,s}=0.8$ and (b) $E_{0}/E_{c,s}=1.8$ for three values of $\lambda.$ }
    \label{subcriticalomega}
\end{figure}

\begin{figure}
    \centering
    \includegraphics[width=0.7\textwidth]{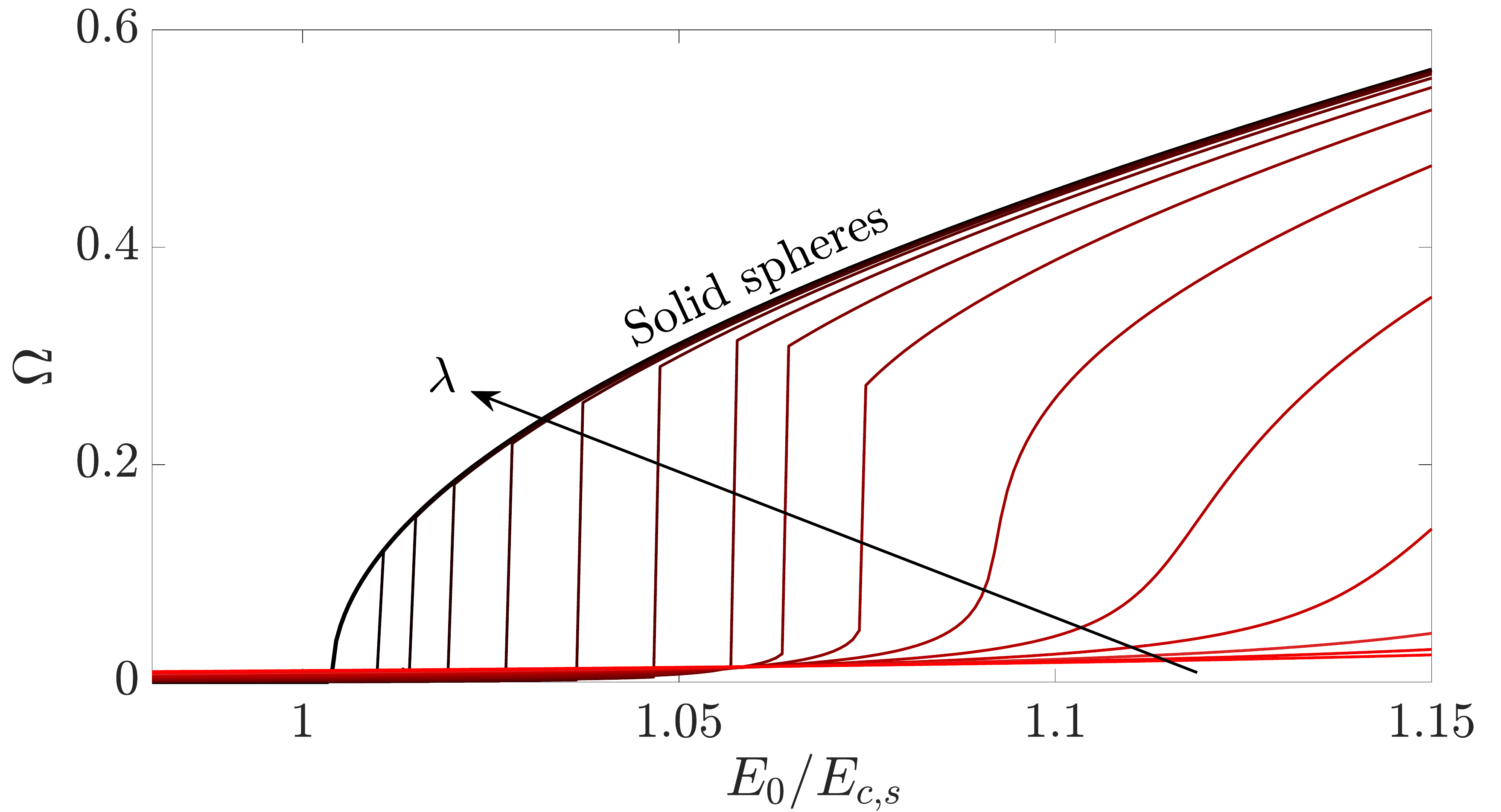}
    \caption{Steady angular velocity of corotating drops at different values of $\lambda,$ corresponding to $\lambda=10^{3n/14}$ for $n=0,1,2,...,13,14.$ In all cases $R=5,$ $\theta_{R}=\pi/4.$}
    \label{higherlambda}
\end{figure}

Figure~\ref{omegavse0ecdifftheta} displays the magnitude of the steady state angular velocity attained by deforming drops for three values of $\theta_{R}$ as well as for an isolated drop and sphere as functions of the electric field strength. Black lines, corresponding to solid spheres, clearly exhibit a pitchfork bifurcation for each value of $\theta_{R}=0$ (solid lines), $\pi/4$ (dashed lines), and $\pi/2$ (dotted lines), as well as for the isolated case. On the other hand, pitchfork bifurcations occur only for the drops in the isolated case and for $\theta_{R}=0$ and $\theta_{R}=\pi/2.$ However, when $\theta_{R}=\pi/4,$ the angular velocity of the drops is nonzero for all electric field strengths regardless of the value of $\lambda$, corresponding to an imperfect bifurcation. For $\theta_{R}=0$ and $\theta_{R}=\pi/2,$ the critical field for rotation of drops (shown with blue and red lines) is greater than that for solid spheres (shown with black lines). The critical field is greater for the less viscous drop (shown with the blue line) when compared to the more viscous one (shown with the red line). The magnitude of the angular velocity at a given electric field strength is always smaller for drops than for solid spheres in all spatial configurations.This is primarily because the straining flows produced by drops convect charge from the poles to the equator of the drop and hence weaken the dipole moment. We note also that the angular velocity of solid spheres or drops is higher for $\theta_{R}=\pi/2$, and lower for $\theta_{R}=0,$ and the critical field for rotation is higher for $\theta_{R}=0$, and lower for $\theta_{R}=\pi/2,$ than for an isolated sphere or drop. Hence, the interactions between drops in the parallel configuration ($\theta_{R}=0$) has a stabilising effect while interactions between drops in the perpendicular configuration ($\theta_{R}=\pi/2$) has a destabilising effect on the pair of drops, when compared to an isolated drop.

Figure~\ref{subcriticalomega} displays the magnitude of the angular velocity of each drop as a function of $\theta_{R}$ for two electric field strengths, namely $E_{0}/E_{c,s}=0.8$ and $E_{0}/E_{c,s}=1.8.$ As shown in Figure~\ref{omegavse0ecdifftheta}, the angular velocity of the drop in the weaker field vanishes at $\theta_{R}=0$ and $\theta_{R}=\pi/2$, and peaks shortly after $\theta_{R}$ exceeds $\pi/4.$ Since rotation of one drop is driven by the straining flow produced by the other drop in this regime, decreasing the viscosity ratio increases the magnitude of the angular velocity. Conversely, in the stronger field, the trend with respect to $\lambda$ is reversed, and drops with greater viscosity attain a greater angular velocity. This is because the dominant contribution to the angular velocity in this case is Quincke rotation, which is stronger for high viscosity drops than low viscosity ones due to the weaker straining flow.

As the drop viscosity becomes large, we expect to recover the supercritical pitchfork bifurcation that occurs for solid spheres. Figure~\ref{higherlambda} displays the magnitude of the angular velocity $\Omega$ as functions of $E_{0}/E_{c,s}$ for logarithmically spaced values of $\lambda$ between $10^{0}$ and $10^{3},$ specifically $\lambda=10^{3n/14}$ for $n=0,1,2,...,13,14.$ At low values of the viscosity ratio, the angular velocity increases smoothly and continuously as the electric field strength is increased. As the viscosity ratio is increased, discontinuous jumps in the angular velocity characteristic of the hysteresis discussed in \S\ref{hysteresissection} occur. Finally, as the viscosity ratio becomes large, the jumps become progressively smaller until the supercritical bifurcation is eventually recovered.

\subsubsection{Drop shape}\label{dropshape}

We now turn our attention to the shapes of the drops, focusing on the effects of the interactions between the drops. First, we validate the results of the present theory against those of \citet{zabarankin}. In that work, the separation vector between the drops is perfectly parallel to the applied field (i.e., $\theta_{R}=0$) and charge convection is absent, so the system is axisymmetric. Since the work of \citet{zabarankin} is valid for dissimilar drops, it also applies to the identical drops considered here. In the axisymmetric configuration, $\tensor{Q}^{f}$ is a diagonal traceless tensor, and the shapes of the drops are characterized by a single shape coefficient, $Q^{f}_{zz}$. Thus,~\eqref{interfacedefinition} reduces to

\begin{equation}
    \xi_{1} = r - \left[1 + \delta Q^{f}_{zz}P_{2}(\cos\theta)\right] = 0,
\end{equation}
where $P_{2}(\cos\theta) = \frac{1}{2}(3\cos^{2}\theta - 1)$ is the Legendre polynomial of degree two. In the absence of charge convection, the dipole moments (given by~\eqref{steadydipoles}) can be approximated by the asymptotic expansion

\begin{equation}
    \bm{P}_{1} = \bm{P}_{2} = \frac{1 - S}{1 + 2S}\left(1 + 2R^{-3}\frac{1 - S}{1 + 2S}\right)\bm{\hat{E}}_{0} + \mathcal{O}(R^{-6}).
\end{equation}
The electric stresses in~\eqref{elecstresscoefficients} can then be calculated using these expressions. Solving for $\tensor{S}^{\infty}_{1}$, $\tensor{Q}^{p}_{1}$, and $\tensor{Q}^{f}_{1}$ via~\eqref{ambientflowS},~\eqref{straining}, and~\eqref{shapetrans} gives the required shape coefficient as

\begin{equation}
Q_{1,zz}^{f} = \mathcal{F}\left(1 + 4R^{-3}\frac{1 - S}{1 + 2S}\right) + R^{-3}\frac{3S(QS - 1)(16 + 19\lambda)}{5(1 + \lambda)^{2}} + \mathcal{O}(R^{-6}),
\end{equation}
precisely as found by \citet{zabarankin}. Since the present model can handle arbitrary drop configurations, we explore the effects of varying $\theta_{R}$ on the drop shape. For arbitrarily positioned pairs of drops, electrostatic and hydrodynamic interactions give rise to non-axisymmetric drop shapes even in the Taylor regime. We measure the degree of deformation and the tilt angle, defined as the angle between the major axes of the drops and the $x$-axis. To quantify the deformation, we define the deformation parameter
\begin{equation}
    D_{Q} = \frac{l - b}{l + b},
\end{equation}
where $l$ and $b$ are the lengths of the semi-major and semi-minor axes in the $xz$ plane. Using the shape coefficients in $\tensor{Q}^{f}$, the tilt angle $\theta_T$ is given by
\begin{equation}
    \theta_T = \frac{1}{2} \arctan\left(\frac{2Q_{xz}^{f}}{Q_{xx}^{f} - Q^{f}_{zz}}\right).
\end{equation}
It is important to emphasise that $\theta_{T}$ is not measured in the same way as other occurrences of $\theta$ in the present work. Measuring $\theta_{T}$ from the positive $x$-axis ensures consistency with previous works on isolated drops~\citep{salipante2010electrohydrodynamics,das2021three}, but it must be borne in mind that the angle denoted by $\theta_{T}$ is given by $\pi/2-\theta$ in our spherical coordinate system, in which $\theta$ is measured from the positive $z$ direction.
We calculate the semi-major and semi-minor axes of the drop as $l = 1 + \delta f({\theta = \pi/2-\theta_T, \phi = 0})$ and $b = 1 + \delta f({\theta = -\theta_T, \phi = 0})$, and hence determine the deformation parameter $D_{Q}$ to be
\begin{equation}\label{dqdef}
    D_{Q} = \frac{1}{2} \delta |Q^{f}_{1,xx} - Q_{1,zz}^{f}| \sqrt{1 + \left(\frac{2Q^{f}_{1,xz}}{Q_{1,xx}^{f} - Q_{1,zz}^{f}}\right)^{2}} + \mathcal{O}(\delta^{2}).
\end{equation}
The tilt angle is zero for an isolated drop in the Taylor regime and for axisymmetric pairs of drops. However, it is nonzero for a rotating isolated drop \citep{das2021three} or for non-axisymmetric pairs of drops. 

\begin{figure}
    \centering
    \includegraphics[width=0.8\textwidth]{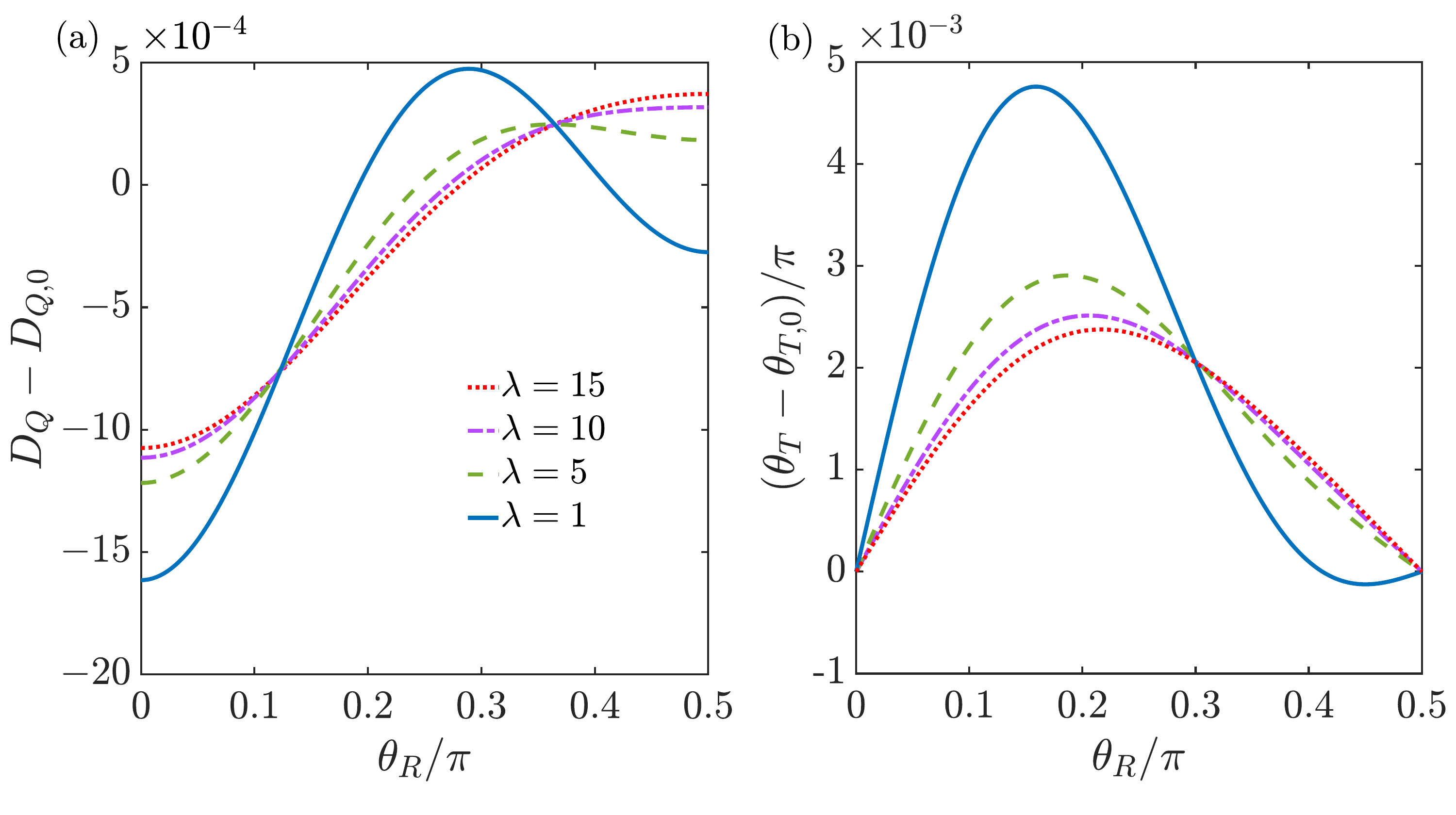}
    \caption{Deviation of (a) deformation parameter $D_{Q}$ and (b) tilt angle $\theta_T$ for interacting drops in the Taylor regime from the corresponding values for an isolated drop, $D_{Q,0}$ and $\theta_{T,0}$, as functions of $\theta_{R}$ for $E_{0}/E_{c,s}=0.8$ for various viscosity ratios. See Table~\ref{isolateddropdeformationandtilt} for values of $D_{Q,0}$ and $\theta_{T,0}$ for each $\lambda$. In all cases, $R=5$.}
    \label{taylordq-dq0}
\end{figure}

Figure~\ref{taylordq-dq0}(a) shows the difference between the deformation parameter for interacting drops, $D_{Q}$, and that of an isolated drop in the Taylor regime ($E_{0}/E_{c,s}=0.8$), denoted $D_{Q,0}$, as functions of $\theta_{R}$ for a range of viscosity ratios. Values of $D_{Q,0}$ are listed in Table \ref{isolateddropdeformationandtilt}. For all viscosity ratios studied, the deformation is smallest when the drops are aligned perfectly with the field. However, the separation angle $\theta_{R}$ corresponding to the maximum deformation depends on the viscosity ratio. More viscous drops (e.g.\,$\lambda=10,15$) deform most when perpendicular to the field, while less viscous drops (e.g.\,$\lambda=1,5$) experience maximum deformation at intermediate angles, which decrease with the viscosity ratio. Note that as a consequence of the symmetries discussed at the beginning of \S\ref{fixeddrops}, $Q_{xx}^{f}$ and $Q_{zz}^{f}$ are symmetric and $Q_{xz}^{f}$ is antisymmetric around $\theta_{R}=\pi/2$, so from~\eqref{dqdef} we see that $D_{Q}$ is also symmetric around $\theta_{R}=\pi/2$. 

Figure~\ref{taylordq-dq0}(b) shows the tilt angle in a subcritical electric field. Note that the tilt angle of an isolated drop, denoted by $\theta_{T,0}$ in the Taylor regime is zero. The value of $\theta_{R}$ that maximizes the tilt angle also depends on the viscosity ratio. High-viscosity ratio drops tilt most around $\theta_{R}=\pi/4$, while for lower viscosity ratios, this angle decreases towards alignment with the field (i.e., towards $\theta_{R}=0$). For all cases except $\lambda=1$, the tilt angle is greater than that for the isolated drop across the full range of $\theta_{R}$; for $\lambda=1$, there is a short range of $\theta_{R}$ near perpendicular alignment within which the tilt angle is smaller than that of an isolated drop.

\begin{figure}
    \centering
    \includegraphics[width=0.8\textwidth]{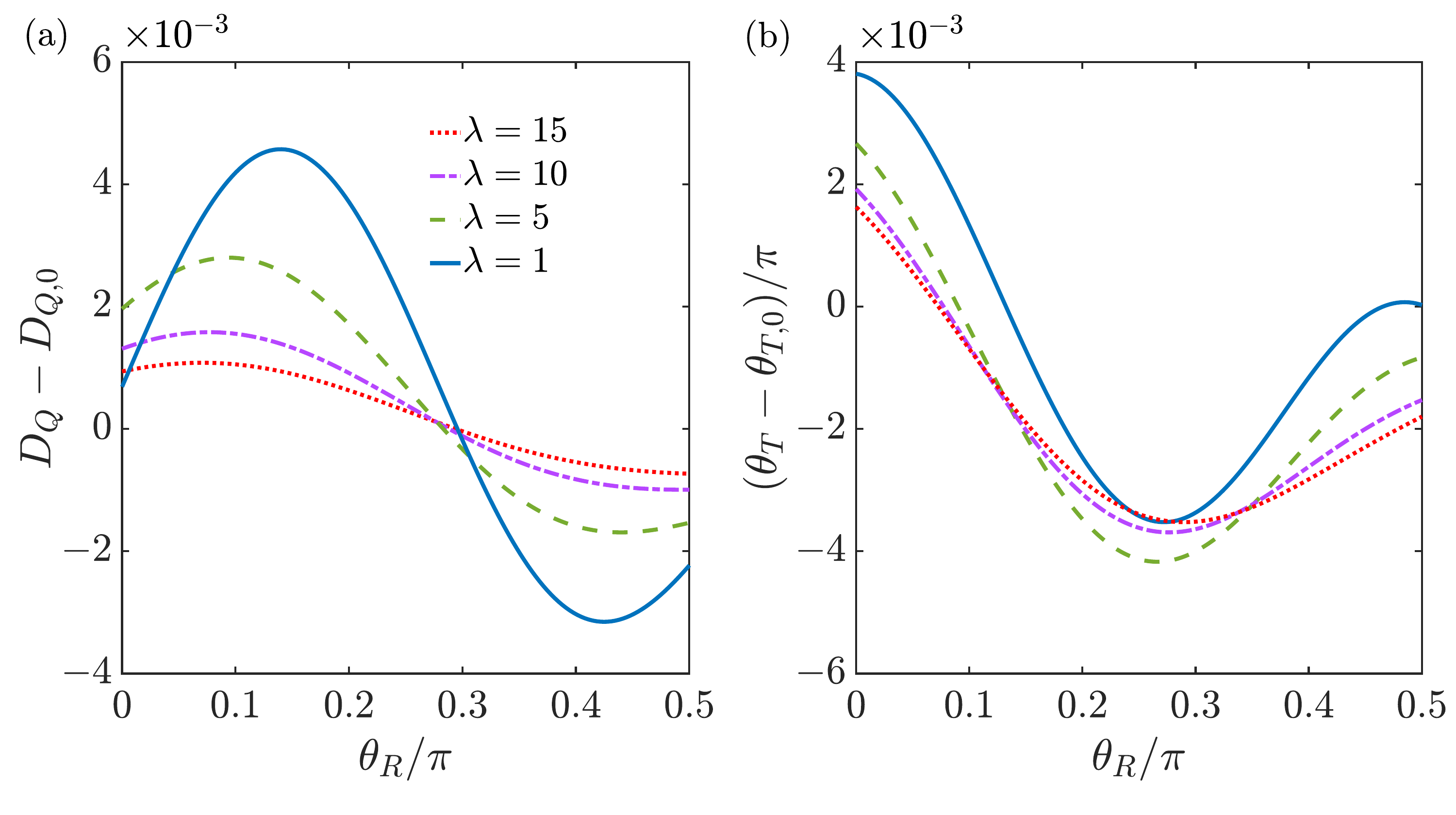}
    \caption{Deviation of (a) deformation parameter $D_{Q}$ and (b) tilt angle $\theta_T$ of corotating drops in Quincke rotation from the corresponding values for an isolated drop, $D_{Q,0}$ and $\theta_{T,0}$, as functions of $\theta_{R}$ for $E_{0}/E_{c,s}=1.8$ for various viscosity ratios. See Table~\ref{isolateddropdeformationandtilt} for values of $D_{Q,0}$ and $\theta_{T,0}$ for each $\lambda$. In all cases, $R=5$.}
    \label{Dqcorotating}
\end{figure}

Figure~\ref{Dqcorotating}(a) displays the difference in the deformation parameter for corotating drops and for an isolated drop in the Quincke regime ($E_{0}/E_{c,s}=1.8$). In general, interactions between the drops increase their deformation compared to that of an isolated drop when the pair of drops is aligned nearly parallel to the applied field (i.e., near $\theta_{R}=0$) and decrease it when the alignment is close to perpendicular (i.e., near $\theta_{R}=\pi/2$). The precise value of $\theta_{R}$ at which this behaviour changes is close to $\pi/4$ and depends on the viscosity ratio. The values of $\theta_{R}$ corresponding to the maximum and minimum deformation also depend on the viscosity ratio.

The effect of interactions on the tilt angle of corotating drops is shown in Figure~\ref{Dqcorotating}(b). For all viscosity ratios studied, the largest tilt angle occurs when $\theta_{R}=0$. The alignment corresponding to the smallest tilt is near to $\theta_{R}=\pi/4$, although this depends nonlinearly on the viscosity ratio. Drops with $\lambda=5$ reach their minimum tilt angle at a smaller value of $\theta_{R}$ than either lower viscosity drops ($\lambda=1$) or higher viscosity drops ($\lambda=10,15$). While Figures~\ref{taylordq-dq0} and~\ref{Dqcorotating}(a) show the largest impact of interactions on deformation and tilt angle for drops with $\lambda=1$, Figure~\ref{Dqcorotating}(b) shows that the tilt angle for drops with $\lambda=5$ deviates more from the value for an isolated drop than it does for drops with $\lambda=1$.

\begin{table}[]
    \centering

\begin{tabular}{|M{2em}|M{5em}|M{5em}|M{5em}|M{5em}|}
\hline
\multirow{2}{*}{$\lambda$}&\multicolumn{2}{c|}{ $E_{0}/E_{c,s}=0.8$}&\multicolumn{2}{c|}{$E_{0}/E_{c,s}=1.8$}
\\
\cline{2-5}
     & $D_{Q,0}$ &$\theta_{T,0}/\pi$& $D_{Q,0}$ &$\theta_{T,0}/\pi$ \\ \hline
    1& 0.047 &  0 & 0.129  &  0.217
 \\
    5 & 0.058
  & 0 & 0.083  &  0.345 \\
    10 &  0.061
  &
   0&  0.051
  &
   0.385 \\
    15 &   0.062  &  0 &   0.036
  &  0.4
 \\ \hline
\end{tabular}
    \caption{Isolated drop deformation $D_{Q,0}$ and tilt angle $\theta_{T,0}/\pi$ in the Taylor regime ($E_{0}/E_{c,s}=0.8$) and corotating in a stronger field ($E_{0}/E_{c,s}=1.8).$}
    \label{isolateddropdeformationandtilt}
\end{table}

\subsubsection{Attraction and repulsion between pairs of drops}\label{attractionrepulsion}

We now explore the effect of the separation angle $\theta_{R}$ on the attractive or repulsive nature of the interactions between drops, as quantified by the component of the relative velocity of the pair of drops in the $\bm{\hat{R}}$ direction, $\bm{U}_{21}\bcdot\bm{\hat{R}}$. For identical, unperturbed drops, $\bm{U}_{21} = 2\bm{U}_{2} = -2\bm{U}_{1}$. By definition, a positive value of $\bm{U}_{21}\bcdot\bm{\hat{R}}$ indicates repulsion, while a negative value indicates attraction. We also determine the contributions to drop attraction or repulsion from straining flows and DEP. Previous work by \citet{sorgentone_kach_khair_walker_vlahovska_2021} on drops in the Taylor regime (in which charge relaxation and convection are neglected and constant dipole moments $\bm{P} = (1-S)/(1+2S)\bm{\hat{E}}_{0}$ are assumed)  has shown that the radial components of both the induced straining flow and the DEP force vary as $P_{2}(\cos\theta)$ around each drop. From this, it follows that drops with $SQ > 1$ will attract for $\theta_{R} < \arccos(1/3) \approx 54.7^\circ$ and will repel otherwise. Although the rotlet flows associated with Quincke rotation dominate the fluid velocities in strong fields, they do not directly appear in the radial component of the relative velocity of the pair of drops, as is evident from the definition of $\bm{N}^{\infty}$ in~\eqref{ambientflowN}.

\begin{figure}
    \centering
    \includegraphics[width=\textwidth]{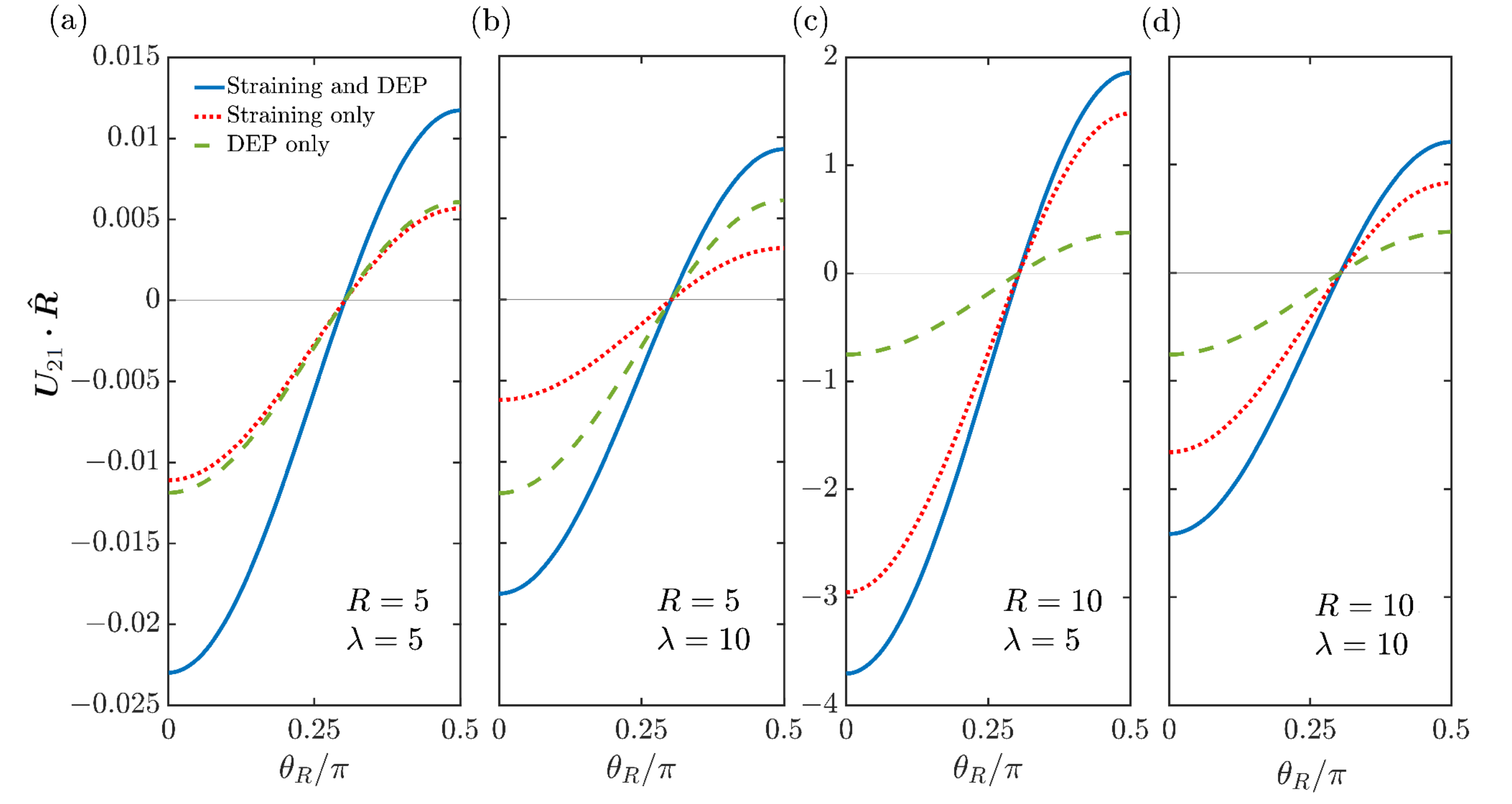}
    \caption{The radial component of the relative velocity of a pair of drops in the Taylor regime ($E_{0}/E_{c,s}=0.8$), plotted as functions of $\theta_{R}$ in four cases: (a) $R=5,$ $\lambda=5$, (b) $R=5,$ $\lambda=10$, (c) $R=10,$ $\lambda=5$, and (d) $R=10,$ $\lambda=10$. Note that the $y$-axis values in subfigure (a) apply to subfigures (a) and (b), and the $y$-axis values in subfigure (c) apply to subfigures (c) and (d). The blue solid line represents the radial component of the total pair velocity $\bm{U}_{21}\cdot\bm{\hat{R}}$, while the red dotted line and green dashed line represent the contributions from straining flow and DEP, respectively.}
    \label{taylorudotr}
\end{figure}

Figure~\ref{taylorudotr} shows the radial component of the relative velocity of a pair of drops in the Taylor regime ($E_{0}/E_{c,s}=0.8$) plotted as functions of $\theta_{R}.$  Figure~\ref{taylorudotr} includes results for four cases: (a) $R=5,$ $\lambda=5$, (b) $R=5,$ $\lambda=10$, (c) $R=10,$ $\lambda=5$, and (d) $R=10,$ $\lambda=10$. The separation angles $\theta_{R}$ corresponding to attraction and repulsion are consistent with the results of earlier work in which charge convection and relaxation were neglected~\citep{sorgentone_kach_khair_walker_vlahovska_2021}. In all cases, the interactions are attractive when the drops are nearly parallel to the field and repulsive when they are nearly perpendicular, with maximum attraction and repulsion occurring at exactly parallel and perpendicular configurations, respectively. More precisely, the interactions switch from attraction to repulsion at approximately $\theta_{R} = 54.4^\circ$ when $R = 5$, and $\theta_{R} = 54.7^\circ$ when $R = 10$, suggesting that charge relaxation and convection have little impact on the attractive or repulsive nature of drop interactions in the Taylor regime. DEP and straining flows act cooperatively for all angles $\theta_{R}$, but their relative strengths depend on $R$ and $\lambda$. Since straining scales as $\mathcal{O}(R^{-2}\mathrm{Ma}^{-1}\lambda^{-1})$ and DEP as $\mathcal{O}(R^{-4}\mathrm{Ma}^{-1})$, DEP is the stronger effect at the smaller separation distance considered (namely, $R=5$, as shown in (a) and (b)), while straining dominates at the larger separation distance (namely, $R=10$, shown in (c) and (d)). For both separation distances, increasing the viscosity ratio reduces the straining flow, weakening the overall attraction or repulsion.

\begin{figure}
    \centering
    \includegraphics[width=\textwidth]{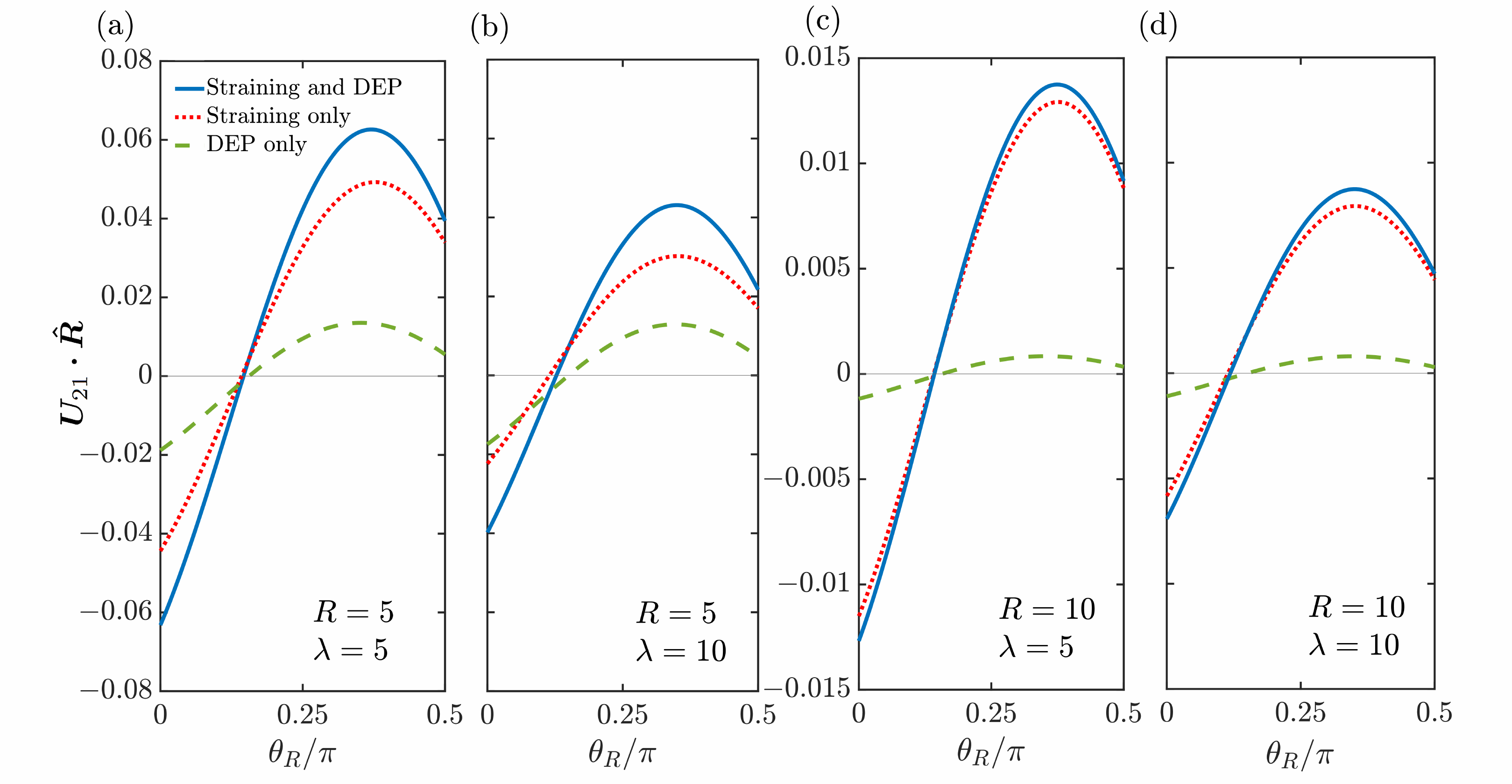}
    \caption{The radial component of the relative velocity of a pair of drops corotating in the Quincke regime ($E_{0}/E_{c,s}=1.8$) plotted as functions of $\theta_{R}$ in four cases: (a) $R=5,$ $\lambda=5$, (b) $R=5,$ $\lambda=10$, (c) $R=10,$ $\lambda=5$, and (d) $R=10,$ $\lambda=10$. Note that the $y$-axis values in subfigure (a) apply to subfigures (a) and (b), and the $y$-axis values in subfigure (c) apply to subfigures (c) and (d).  The blue solid line represents the radial component of the total pair velocity $\bm{U}_{21}\cdot\bm{\hat{R}}$, while the red dotted line and green dashed line represent the contributions from straining flow and DEP, respectively.}
    \label{corotatingudotr}
\end{figure}

Figure~\ref{corotatingudotr} shows $\bm{U}_{21}\bcdot\bm{\hat{R}}$ for corotating drops in the Quincke regime, for the same values of $R$ and $\lambda$ as used in Figure~\ref{taylorudotr}. As in the Taylor regime, the strongest attraction occurs when the drops are exactly parallel to the field (i.e., when $\theta_{R} = 0$), and increasing $\theta_{R}$ reduces the attraction, which eventually switches from attraction to repulsion. However, unlike in the weaker field, the strongest repulsion does not occur when the drops are precisely perpendicular to the field due to the tilted dipole moments. Further increases to the electric field strength enhance the transverse dipole component, shifting the separation angle corresponding to the maximum repulsion away from $\theta_{R} = \pi/2$. Additionally, the separation angle at which interactions switch between attractive and repulsive depends on both $\lambda$ and $R$, but in all cases, the range of separation angles corresponding to repulsion is larger than that corresponding to attraction. For the four cases examined, the switch occurs approximately at (a) $\theta_{R}=26.3^\circ$, (b) $\theta_{R}=22.9^\circ$, (c) $\theta_{R}=27.3^\circ$, and (d) $\theta_{R}=21.3^\circ$. Since DEP and straining contributions do not change sign at the same value of $\theta_{R}$, there are configurations where they oppose each other. This is most apparent in Figure~\ref{corotatingudotr}(b), where there is a small range of angles (around $6^\circ$) within which straining is repulsive and DEP is attractive. Straining is the stronger of the two effects at most separation angles, except near the points where it changes sign.

\section{Summary and Conclusions}\label{conclusions}

In this work, we have presented a three-dimensional nonlinear theory to capture electrohydrodynamic interactions between pairs of identical leaky dielectric drops exposed to a spatially uniform DC electric field. The present model, based on the Taylor--Melcher leaky dielectric framework, is valid for widely separated viscous drops experiencing small deformations. The novelty of this model lies in incorporating charge relaxation and convection in the charge transport equation for arbitrarily positioned deforming drops in three dimensions, enabling us to capture the Quincke rotation instability. We derived a system of nonlinear coupled ODEs for the drop dipole moments, shapes, and positions, which were integrated numerically. The present model demonstrates good agreement with previous theoretical \citep{zabarankin,sorgentone_kach_khair_walker_vlahovska_2021}, numerical \citep{sorgentone_kach_khair_walker_vlahovska_2021}, and experimental \citep{kach2022prediction} studies on drop dynamics in the Taylor regime. The present model successfully reproduces hysteresis in the angular velocity of isolated drops, as observed in experiments \citep{salipante2010electrohydrodynamics}. In the Quincke regime, freely suspended drops exhibit a range of qualitatively different trajectories based on their initial positions and perturbations to their dipole moments. Drop trajectories are primarily influenced by the rotlet flows due to Quincke rotation, leading to attraction, repulsion, and sustained pair translation. 

To gain deeper insights into the electrohydrodynamic interactions, we analysed drops fixed in a plane but still free to rotate and deform. By varying the viscosity ratio and spatial configurations of the drops, we investigate the effects of interactions on angular velocity, drop deformation, and tilt angle. We find that the onset of Quincke rotation, which corresponds to a pitchfork bifurcation for isolated drops or pairs of fixed solid spheres, corresponds instead to an imperfect bifurcation for pairs of fixed drops, which occurs due to symmetry-breaking hydrodynamic interactions arising from electrically induced straining flows. These interactions cause unperturbed drops to corotate in the plane defined by the applied field and the line joining the centres of the drops, with the rotation direction determined by the angle between the line of centres and the applied field. Electrohydrodynamic interactions can either increase or decrease the critical field for Quincke rotation, depending on the relative drop positions. We also find that drop deformation and tilt angle vary non-monotonically with the angle between the applied field and the line connecting the centres of the drops, as well as with the viscosity ratio. Finally, we examine the effects of straining flows and DEP forces on drop attraction and repulsion across different spatial configurations for two different viscosity ratios. We find that the maximum repulsion between the drops no longer occurs when the alignment of the pair of drops is precisely perpendicular to the applied field - as is the case in the Taylor regime - due to the tilted dipole moments characteristic of Quincke rotation.

The present model advances the understanding of electrohydrodynamic drop interactions, particularly in the context of Quincke rotation, and opens multiple avenues for future research. While the present model shows excellent agreement with weak-field trajectories computed numerically, results for pairs of drops in close proximity are unlikely to be accurate because the model relies on the assumption of widely-separated drops. Modelling interactions in close proximity could be achieved using boundary element \citep{das2017sims,firouznia2023spectral}, finite element \citep{wagoner2021}, immersed interface \citep{hu2015hybrid} or volume of fluid \citep{lopez2011} methods. The model developed in this paper can be adapted to systems of three or more drops with varying physical parameters or sizes. Theoretical investigations of drops with dissimilar physical properties but identical sizes could provide valuable insights into the sustained translation of a pair of drops observed in boundary element simulations of such systems~\citep{Sorgentone_Vlahovska_2022}. Extending the model to multiple identical drops would be useful for studying emulsions in strong electric fields undergoing Quincke rotation, where drop rotation significantly influences bulk rheology. Experiments by~\citet{lobry} on suspensions of solid particles undergoing Quincke rotation showed significant reductions in the apparent viscosity of the suspension, but a similar study for drops has not been conducted, either experimentally or theoretically. The theoretical framework presented here represents a first step toward studying such systems.

\begin{acknowledgments}
The authors gratefully acknowledge the financial support of the University of Strathclyde in the form of a Research Excellence Award.
\end{acknowledgments}

\section*{Data availability}

MATLAB codes to reproduce the figures in the article and movies corresponding to Figures \ref{sorgentonevalidation} and \ref{trajectoryplot} are available in the Supplementary Material \cite{SM}.


\bibliography{apssamp}

\providecommand{\noopsort}[1]{}\providecommand{\singleletter}[1]{#1}%
\begin{thebibliography}{59}%
\makeatletter
\providecommand \@ifxundefined [1]{%
 \@ifx{#1\undefined}
}%
\providecommand \@ifnum [1]{%
 \ifnum #1\expandafter \@firstoftwo
 \else \expandafter \@secondoftwo
 \fi
}%
\providecommand \@ifx [1]{%
 \ifx #1\expandafter \@firstoftwo
 \else \expandafter \@secondoftwo
 \fi
}%
\providecommand \natexlab [1]{#1}%
\providecommand \enquote  [1]{``#1''}%
\providecommand \bibnamefont  [1]{#1}%
\providecommand \bibfnamefont [1]{#1}%
\providecommand \citenamefont [1]{#1}%
\providecommand \href@noop [0]{\@secondoftwo}%
\providecommand \href [0]{\begingroup \@sanitize@url \@href}%
\providecommand \@href[1]{\@@startlink{#1}\@@href}%
\providecommand \@@href[1]{\endgroup#1\@@endlink}%
\providecommand \@sanitize@url [0]{\catcode `\\12\catcode `\$12\catcode
  `\&12\catcode `\#12\catcode `\^12\catcode `\_12\catcode `\%12\relax}%
\providecommand \@@startlink[1]{}%
\providecommand \@@endlink[0]{}%
\providecommand \url  [0]{\begingroup\@sanitize@url \@url }%
\providecommand \@url [1]{\endgroup\@href {#1}{\urlprefix }}%
\providecommand \urlprefix  [0]{URL }%
\providecommand \Eprint [0]{\href }%
\providecommand \doibase [0]{https://doi.org/}%
\providecommand \selectlanguage [0]{\@gobble}%
\providecommand \bibinfo  [0]{\@secondoftwo}%
\providecommand \bibfield  [0]{\@secondoftwo}%
\providecommand \translation [1]{[#1]}%
\providecommand \BibitemOpen [0]{}%
\providecommand \bibitemStop [0]{}%
\providecommand \bibitemNoStop [0]{.\EOS\space}%
\providecommand \EOS [0]{\spacefactor3000\relax}%
\providecommand \BibitemShut  [1]{\csname bibitem#1\endcsname}%
\let\auto@bib@innerbib\@empty
\bibitem [{\citenamefont {Melcher}\ and\ \citenamefont
  {Taylor}(1969)}]{melcher1969electrohydrodynamics}%
  \BibitemOpen
  \bibfield  {author} {\bibinfo {author} {\bibfnamefont {J.~R.}\ \bibnamefont
  {Melcher}}\ and\ \bibinfo {author} {\bibfnamefont {G.~I.}\ \bibnamefont
  {Taylor}},\ }\bibfield  {title} {\bibinfo {title} {Electrohydrodynamics: a
  review of the role of interfacial shear stresses},\ }\href@noop {} {\bibfield
   {journal} {\bibinfo  {journal} {Annual Review of Fluid Mechanics}\ }\textbf
  {\bibinfo {volume} {1}},\ \bibinfo {pages} {111} (\bibinfo {year}
  {1969})}\BibitemShut {NoStop}%
\bibitem [{\citenamefont {Collins}\ \emph {et~al.}(2008)\citenamefont
  {Collins}, \citenamefont {Jones}, \citenamefont {Harris},\ and\ \citenamefont
  {Basaran}}]{collins2008electrohydrodynamic}%
  \BibitemOpen
  \bibfield  {author} {\bibinfo {author} {\bibfnamefont {R.~T.}\ \bibnamefont
  {Collins}}, \bibinfo {author} {\bibfnamefont {J.~J.}\ \bibnamefont {Jones}},
  \bibinfo {author} {\bibfnamefont {M.~T.}\ \bibnamefont {Harris}},\ and\
  \bibinfo {author} {\bibfnamefont {O.~A.}\ \bibnamefont {Basaran}},\
  }\bibfield  {title} {\bibinfo {title} {Electrohydrodynamic tip streaming and
  emission of charged drops from liquid cones},\ }\href@noop {} {\bibfield
  {journal} {\bibinfo  {journal} {Nature Physics}\ }\textbf {\bibinfo {volume}
  {4}},\ \bibinfo {pages} {149} (\bibinfo {year} {2008})}\BibitemShut {NoStop}%
\bibitem [{\citenamefont {Basaran}\ \emph {et~al.}(2013)\citenamefont
  {Basaran}, \citenamefont {Gao},\ and\ \citenamefont
  {Bhat}}]{basaran2013nonstandard}%
  \BibitemOpen
  \bibfield  {author} {\bibinfo {author} {\bibfnamefont {O.~A.}\ \bibnamefont
  {Basaran}}, \bibinfo {author} {\bibfnamefont {H.}~\bibnamefont {Gao}},\ and\
  \bibinfo {author} {\bibfnamefont {P.~P.}\ \bibnamefont {Bhat}},\ }\bibfield
  {title} {\bibinfo {title} {Nonstandard inkjets},\ }\href@noop {} {\bibfield
  {journal} {\bibinfo  {journal} {Annual Review of Fluid Mechanics}\ }\textbf
  {\bibinfo {volume} {45}},\ \bibinfo {pages} {85} (\bibinfo {year}
  {2013})}\BibitemShut {NoStop}%
\bibitem [{\citenamefont {Seyed-Yagoobi}(2005)}]{seyed2005electrohydrodynamic}%
  \BibitemOpen
  \bibfield  {author} {\bibinfo {author} {\bibfnamefont {J.}~\bibnamefont
  {Seyed-Yagoobi}},\ }\bibfield  {title} {\bibinfo {title} {Electrohydrodynamic
  pumping of dielectric liquids},\ }\href@noop {} {\bibfield  {journal}
  {\bibinfo  {journal} {Journal of Electrostatics}\ }\textbf {\bibinfo {volume}
  {63}},\ \bibinfo {pages} {861} (\bibinfo {year} {2005})}\BibitemShut
  {NoStop}%
\bibitem [{\citenamefont {Eow}\ and\ \citenamefont
  {Ghadiri}(2002)}]{eow2002electrostatic}%
  \BibitemOpen
  \bibfield  {author} {\bibinfo {author} {\bibfnamefont {J.~S.}\ \bibnamefont
  {Eow}}\ and\ \bibinfo {author} {\bibfnamefont {M.}~\bibnamefont {Ghadiri}},\
  }\bibfield  {title} {\bibinfo {title} {Electrostatic enhancement of
  coalescence of water droplets in oil: a review of the technology},\
  }\href@noop {} {\bibfield  {journal} {\bibinfo  {journal} {Chemical
  Engineering Journal}\ }\textbf {\bibinfo {volume} {85}},\ \bibinfo {pages}
  {357} (\bibinfo {year} {2002})}\BibitemShut {NoStop}%
\bibitem [{\citenamefont {Ptasinski}\ and\ \citenamefont
  {Kerkhof}(1992)}]{ptasinski1992electric}%
  \BibitemOpen
  \bibfield  {author} {\bibinfo {author} {\bibfnamefont {K.~J.}\ \bibnamefont
  {Ptasinski}}\ and\ \bibinfo {author} {\bibfnamefont {P.~J. A.~M.}\
  \bibnamefont {Kerkhof}},\ }\bibfield  {title} {\bibinfo {title} {Electric
  field driven separations: phenomena and applications},\ }\href@noop {}
  {\bibfield  {journal} {\bibinfo  {journal} {Separation Science and
  Technology}\ }\textbf {\bibinfo {volume} {27}},\ \bibinfo {pages} {995}
  (\bibinfo {year} {1992})}\BibitemShut {NoStop}%
\bibitem [{\citenamefont {O'Konski}\ and\ \citenamefont
  {Thacher}(1953{\natexlab{a}})}]{okonski_thacher}%
  \BibitemOpen
  \bibfield  {author} {\bibinfo {author} {\bibfnamefont {C.~T.}\ \bibnamefont
  {O'Konski}}\ and\ \bibinfo {author} {\bibfnamefont {H.~C.}\ \bibnamefont
  {Thacher}},\ }\bibfield  {title} {\bibinfo {title} {The distortion of aerosol
  droplets by an electric field},\ }\href@noop {} {\bibfield  {journal}
  {\bibinfo  {journal} {The Journal of Physical Chemistry}\ }\textbf {\bibinfo
  {volume} {57}},\ \bibinfo {pages} {955} (\bibinfo {year}
  {1953}{\natexlab{a}})}\BibitemShut {NoStop}%
\bibitem [{\citenamefont {Allan}\ and\ \citenamefont
  {Mason}(1962)}]{allan1962particle}%
  \BibitemOpen
  \bibfield  {author} {\bibinfo {author} {\bibfnamefont {R.~S.}\ \bibnamefont
  {Allan}}\ and\ \bibinfo {author} {\bibfnamefont {S.~G.}\ \bibnamefont
  {Mason}},\ }\bibfield  {title} {\bibinfo {title} {Particle behaviour in shear
  and electric fields {I}. {D}eformation and burst of fluid drops},\
  }\href@noop {} {\bibfield  {journal} {\bibinfo  {journal} {Proceedings of the
  Royal Society of London. Series A. Mathematical and Physical Sciences}\
  }\textbf {\bibinfo {volume} {267}},\ \bibinfo {pages} {45} (\bibinfo {year}
  {1962})}\BibitemShut {NoStop}%
\bibitem [{\citenamefont {O'Konski}\ and\ \citenamefont
  {Thacher}(1953{\natexlab{b}})}]{o1953distortion}%
  \BibitemOpen
  \bibfield  {author} {\bibinfo {author} {\bibfnamefont {C.~T.}\ \bibnamefont
  {O'Konski}}\ and\ \bibinfo {author} {\bibfnamefont {H.~C.}\ \bibnamefont
  {Thacher}},\ }\bibfield  {title} {\bibinfo {title} {The distortion of aerosol
  droplets by an electric field},\ }\href@noop {} {\bibfield  {journal}
  {\bibinfo  {journal} {The Journal of Physical Chemistry}\ }\textbf {\bibinfo
  {volume} {57}},\ \bibinfo {pages} {955} (\bibinfo {year}
  {1953}{\natexlab{b}})}\BibitemShut {NoStop}%
\bibitem [{\citenamefont {Garton}\ and\ \citenamefont
  {Krasucki}(1964)}]{garton1964bubbles}%
  \BibitemOpen
  \bibfield  {author} {\bibinfo {author} {\bibfnamefont {C.~G.}\ \bibnamefont
  {Garton}}\ and\ \bibinfo {author} {\bibfnamefont {Z.}~\bibnamefont
  {Krasucki}},\ }\bibfield  {title} {\bibinfo {title} {Bubbles in insulating
  liquids: stability in an electric field},\ }\href@noop {} {\bibfield
  {journal} {\bibinfo  {journal} {Proceedings of the Royal Society of London.
  Series A. Mathematical and Physical Sciences}\ }\textbf {\bibinfo {volume}
  {280}},\ \bibinfo {pages} {211} (\bibinfo {year} {1964})}\BibitemShut
  {NoStop}%
\bibitem [{\citenamefont {Taylor}(1966)}]{taylor1966studies}%
  \BibitemOpen
  \bibfield  {author} {\bibinfo {author} {\bibfnamefont {G.~I.}\ \bibnamefont
  {Taylor}},\ }\bibfield  {title} {\bibinfo {title} {Studies in
  electrohydrodynamics. {I}. {T}he circulation produced in a drop by an
  electric field},\ }\href@noop {} {\bibfield  {journal} {\bibinfo  {journal}
  {Proceedings of the Royal Society of London. Series A. Mathematical and
  Physical Sciences}\ }\textbf {\bibinfo {volume} {291}},\ \bibinfo {pages}
  {159} (\bibinfo {year} {1966})}\BibitemShut {NoStop}%
\bibitem [{\citenamefont {Saville}(1997)}]{saville1997electrohydrodynamics}%
  \BibitemOpen
  \bibfield  {author} {\bibinfo {author} {\bibfnamefont {D.~A.}\ \bibnamefont
  {Saville}},\ }\bibfield  {title} {\bibinfo {title} {Electrohydrodynamics: the
  {T}aylor--{M}elcher leaky dielectric model},\ }\href@noop {} {\bibfield
  {journal} {\bibinfo  {journal} {Annual Review of Fluid Mechanics}\ }\textbf
  {\bibinfo {volume} {29}},\ \bibinfo {pages} {27} (\bibinfo {year}
  {1997})}\BibitemShut {NoStop}%
\bibitem [{\citenamefont {Papageorgiou}(2019)}]{papageorgiou2019}%
  \BibitemOpen
  \bibfield  {author} {\bibinfo {author} {\bibfnamefont {D.~T.}\ \bibnamefont
  {Papageorgiou}},\ }\bibfield  {title} {\bibinfo {title} {Film flows in the
  presence of electric fields},\ }\href@noop {} {\bibfield  {journal} {\bibinfo
   {journal} {Annual Review of Fluid Mechanics}\ }\textbf {\bibinfo {volume}
  {51}},\ \bibinfo {pages} {155} (\bibinfo {year} {2019})}\BibitemShut
  {NoStop}%
\bibitem [{\citenamefont {Vlahovska}(2019)}]{vlahovska2019}%
  \BibitemOpen
  \bibfield  {author} {\bibinfo {author} {\bibfnamefont {P.~M.}\ \bibnamefont
  {Vlahovska}},\ }\bibfield  {title} {\bibinfo {title} {Electrohydrodynamics of
  drops and vesicles},\ }\href@noop {} {\bibfield  {journal} {\bibinfo
  {journal} {Annual Review of Fluid Mechanics}\ }\textbf {\bibinfo {volume}
  {51}},\ \bibinfo {pages} {305} (\bibinfo {year} {2019})}\BibitemShut
  {NoStop}%
\bibitem [{\citenamefont {Torza}\ \emph {et~al.}(1971)\citenamefont {Torza},
  \citenamefont {Cox},\ and\ \citenamefont
  {Mason}}]{torza1971electrohydrodynamic}%
  \BibitemOpen
  \bibfield  {author} {\bibinfo {author} {\bibfnamefont {S.}~\bibnamefont
  {Torza}}, \bibinfo {author} {\bibfnamefont {R.~G.}\ \bibnamefont {Cox}},\
  and\ \bibinfo {author} {\bibfnamefont {S.~G.}\ \bibnamefont {Mason}},\
  }\bibfield  {title} {\bibinfo {title} {Electrohydrodynamic deformation and
  bursts of liquid drops},\ }\href@noop {} {\bibfield  {journal} {\bibinfo
  {journal} {Philosophical Transactions of the Royal Society of London. Series
  A, Mathematical and Physical Sciences}\ }\textbf {\bibinfo {volume} {269}},\
  \bibinfo {pages} {295} (\bibinfo {year} {1971})}\BibitemShut {NoStop}%
\bibitem [{\citenamefont {Ajayi}(1978)}]{ajayi1978note}%
  \BibitemOpen
  \bibfield  {author} {\bibinfo {author} {\bibfnamefont {O.~O.}\ \bibnamefont
  {Ajayi}},\ }\bibfield  {title} {\bibinfo {title} {A note on {T}aylor’s
  electrohydrodynamic theory},\ }\href@noop {} {\bibfield  {journal} {\bibinfo
  {journal} {Proceedings of the Royal Society of London. A. Mathematical and
  Physical Sciences}\ }\textbf {\bibinfo {volume} {364}},\ \bibinfo {pages}
  {499} (\bibinfo {year} {1978})}\BibitemShut {NoStop}%
\bibitem [{\citenamefont {Lanauze}\ \emph {et~al.}(2015)\citenamefont
  {Lanauze}, \citenamefont {Walker},\ and\ \citenamefont
  {Khair}}]{lanauze2015}%
  \BibitemOpen
  \bibfield  {author} {\bibinfo {author} {\bibfnamefont {J.~A.}\ \bibnamefont
  {Lanauze}}, \bibinfo {author} {\bibfnamefont {L.~M.}\ \bibnamefont
  {Walker}},\ and\ \bibinfo {author} {\bibfnamefont {A.~S.}\ \bibnamefont
  {Khair}},\ }\bibfield  {title} {\bibinfo {title} {Nonlinear
  electrohydrodynamics of slightly deformed oblate drops},\ }\href@noop {}
  {\bibfield  {journal} {\bibinfo  {journal} {Journal of Fluid Mechanics}\
  }\textbf {\bibinfo {volume} {774}},\ \bibinfo {pages} {245} (\bibinfo {year}
  {2015})}\BibitemShut {NoStop}%
\bibitem [{\citenamefont {Das}\ and\ \citenamefont
  {Saintillan}(2017)}]{das2017sims}%
  \BibitemOpen
  \bibfield  {author} {\bibinfo {author} {\bibfnamefont {D.}~\bibnamefont
  {Das}}\ and\ \bibinfo {author} {\bibfnamefont {D.}~\bibnamefont
  {Saintillan}},\ }\bibfield  {title} {\bibinfo {title} {Electrohydrodynamics
  of viscous drops in strong electric fields: numerical simulations},\
  }\href@noop {} {\bibfield  {journal} {\bibinfo  {journal} {Journal of Fluid
  Mechanics}\ }\textbf {\bibinfo {volume} {829}},\ \bibinfo {pages} {127}
  (\bibinfo {year} {2017})}\BibitemShut {NoStop}%
\bibitem [{\citenamefont {Peng}\ \emph {et~al.}(2024)\citenamefont {Peng},
  \citenamefont {Brand{\~a}o}, \citenamefont {Yariv},\ and\ \citenamefont
  {Schnitzer}}]{peng2024}%
  \BibitemOpen
  \bibfield  {author} {\bibinfo {author} {\bibfnamefont {G.~G.}\ \bibnamefont
  {Peng}}, \bibinfo {author} {\bibfnamefont {R.}~\bibnamefont {Brand{\~a}o}},
  \bibinfo {author} {\bibfnamefont {E.}~\bibnamefont {Yariv}},\ and\ \bibinfo
  {author} {\bibfnamefont {O.}~\bibnamefont {Schnitzer}},\ }\bibfield  {title}
  {\bibinfo {title} {Equatorial blowup and polar caps in drop
  electrohydrodynamics},\ }\href@noop {} {\bibfield  {journal} {\bibinfo
  {journal} {Physical Review Fluids}\ }\textbf {\bibinfo {volume} {9}},\
  \bibinfo {pages} {083701} (\bibinfo {year} {2024})}\BibitemShut {NoStop}%
\bibitem [{\citenamefont {Brosseau}\ and\ \citenamefont
  {Vlahovska}(2017)}]{brosseau2017}%
  \BibitemOpen
  \bibfield  {author} {\bibinfo {author} {\bibfnamefont {Q.}~\bibnamefont
  {Brosseau}}\ and\ \bibinfo {author} {\bibfnamefont {P.~M.}\ \bibnamefont
  {Vlahovska}},\ }\bibfield  {title} {\bibinfo {title} {Streaming from the
  equator of a drop in an external electric field},\ }\href@noop {} {\bibfield
  {journal} {\bibinfo  {journal} {Physical Review Letters}\ }\textbf {\bibinfo
  {volume} {119}},\ \bibinfo {pages} {034501} (\bibinfo {year}
  {2017})}\BibitemShut {NoStop}%
\bibitem [{\citenamefont {Wagoner}\ \emph {et~al.}(2021)\citenamefont
  {Wagoner}, \citenamefont {Vlahovska}, \citenamefont {Harris},\ and\
  \citenamefont {Basaran}}]{wagoner2021}%
  \BibitemOpen
  \bibfield  {author} {\bibinfo {author} {\bibfnamefont {B.~W.}\ \bibnamefont
  {Wagoner}}, \bibinfo {author} {\bibfnamefont {P.~M.}\ \bibnamefont
  {Vlahovska}}, \bibinfo {author} {\bibfnamefont {M.~T.}\ \bibnamefont
  {Harris}},\ and\ \bibinfo {author} {\bibfnamefont {O.~A.}\ \bibnamefont
  {Basaran}},\ }\bibfield  {title} {\bibinfo {title} {Electrohydrodynamics of
  lenticular drops and equatorial streaming},\ }\href@noop {} {\bibfield
  {journal} {\bibinfo  {journal} {Journal of Fluid Mechanics}\ }\textbf
  {\bibinfo {volume} {925}},\ \bibinfo {pages} {A36} (\bibinfo {year}
  {2021})}\BibitemShut {NoStop}%
\bibitem [{\citenamefont {Weiler}(1893)}]{weiler}%
  \BibitemOpen
  \bibfield  {author} {\bibinfo {author} {\bibfnamefont {W.}~\bibnamefont
  {Weiler}},\ }\bibfield  {title} {\bibinfo {title} {Zur darstellung
  elektrischer kraftlinien},\ }\href@noop {} {\bibfield  {journal} {\bibinfo
  {journal} {Zeitschrift für den Physikalischen und Chemischen Unterricht}\
  }\textbf {\bibinfo {volume} {6}},\ \bibinfo {pages} {194} (\bibinfo {year}
  {1893})}\BibitemShut {NoStop}%
\bibitem [{\citenamefont {Quincke}(1896)}]{quincke1896ueber}%
  \BibitemOpen
  \bibfield  {author} {\bibinfo {author} {\bibfnamefont {G.}~\bibnamefont
  {Quincke}},\ }\bibfield  {title} {\bibinfo {title} {Ueber rotationen im
  constanten electrischen felde},\ }\href@noop {} {\bibfield  {journal}
  {\bibinfo  {journal} {Annalen der Physik}\ }\textbf {\bibinfo {volume}
  {295}},\ \bibinfo {pages} {417} (\bibinfo {year} {1896})}\BibitemShut
  {NoStop}%
\bibitem [{\citenamefont {Turcu}(1987)}]{turcu1987}%
  \BibitemOpen
  \bibfield  {author} {\bibinfo {author} {\bibfnamefont {I.}~\bibnamefont
  {Turcu}},\ }\bibfield  {title} {\bibinfo {title} {Electric field induced
  rotation of spheres},\ }\href@noop {} {\bibfield  {journal} {\bibinfo
  {journal} {Journal of Physics A: Mathematical and General}\ }\textbf
  {\bibinfo {volume} {20}},\ \bibinfo {pages} {3301} (\bibinfo {year}
  {1987})}\BibitemShut {NoStop}%
\bibitem [{\citenamefont {Jones}(1984)}]{jones1984quincke}%
  \BibitemOpen
  \bibfield  {author} {\bibinfo {author} {\bibfnamefont {T.~B.}\ \bibnamefont
  {Jones}},\ }\bibfield  {title} {\bibinfo {title} {Quincke rotation of
  spheres},\ }\href@noop {} {\bibfield  {journal} {\bibinfo  {journal} {IEEE
  Transactions on Industry Applications}\ }\textbf {\bibinfo {volume}
  {IA-20}},\ \bibinfo {pages} {845} (\bibinfo {year} {1984})}\BibitemShut
  {NoStop}%
\bibitem [{\citenamefont {Das}\ and\ \citenamefont
  {Saintillan}(2013)}]{das2013electrohydrodynamic}%
  \BibitemOpen
  \bibfield  {author} {\bibinfo {author} {\bibfnamefont {D.}~\bibnamefont
  {Das}}\ and\ \bibinfo {author} {\bibfnamefont {D.}~\bibnamefont
  {Saintillan}},\ }\bibfield  {title} {\bibinfo {title} {Electrohydrodynamic
  interaction of spherical particles under {Q}uincke rotation},\ }\href@noop {}
  {\bibfield  {journal} {\bibinfo  {journal} {Physical Review E}\ }\textbf
  {\bibinfo {volume} {87}},\ \bibinfo {pages} {043014} (\bibinfo {year}
  {2013})}\BibitemShut {NoStop}%
\bibitem [{\citenamefont {Dolinsky}\ and\ \citenamefont
  {Elperin}(2012)}]{Dolinsky}%
  \BibitemOpen
  \bibfield  {author} {\bibinfo {author} {\bibfnamefont {Y.}~\bibnamefont
  {Dolinsky}}\ and\ \bibinfo {author} {\bibfnamefont {T.}~\bibnamefont
  {Elperin}},\ }\bibfield  {title} {\bibinfo {title} {Dipole interaction of the
  {Q}uincke rotating particles},\ }\href@noop {} {\bibfield  {journal}
  {\bibinfo  {journal} {Physical Review E}\ }\textbf {\bibinfo {volume} {85}},\
  \bibinfo {pages} {026608} (\bibinfo {year} {2012})}\BibitemShut {NoStop}%
\bibitem [{\citenamefont {He}\ \emph {et~al.}(2013)\citenamefont {He},
  \citenamefont {Salipante},\ and\ \citenamefont {Vlahovska}}]{he2013}%
  \BibitemOpen
  \bibfield  {author} {\bibinfo {author} {\bibfnamefont {H.}~\bibnamefont
  {He}}, \bibinfo {author} {\bibfnamefont {P.~F.}\ \bibnamefont {Salipante}},\
  and\ \bibinfo {author} {\bibfnamefont {P.~M.}\ \bibnamefont {Vlahovska}},\
  }\bibfield  {title} {\bibinfo {title} {Electrorotation of a viscous droplet
  in a uniform direct current electric field},\ }\href@noop {} {\bibfield
  {journal} {\bibinfo  {journal} {Physics of Fluids}\ }\textbf {\bibinfo
  {volume} {25}},\ \bibinfo {pages} {032106} (\bibinfo {year}
  {2013})}\BibitemShut {NoStop}%
\bibitem [{\citenamefont {Das}\ and\ \citenamefont
  {Saintillan}(2021)}]{das2021three}%
  \BibitemOpen
  \bibfield  {author} {\bibinfo {author} {\bibfnamefont {D.}~\bibnamefont
  {Das}}\ and\ \bibinfo {author} {\bibfnamefont {D.}~\bibnamefont
  {Saintillan}},\ }\bibfield  {title} {\bibinfo {title} {A three-dimensional
  small-deformation theory for electrohydrodynamics of dielectric drops},\
  }\href@noop {} {\bibfield  {journal} {\bibinfo  {journal} {Journal of Fluid
  Mechanics}\ }\textbf {\bibinfo {volume} {914}},\ \bibinfo {pages} {A22}
  (\bibinfo {year} {2021})}\BibitemShut {NoStop}%
\bibitem [{\citenamefont {Salipante}\ and\ \citenamefont
  {Vlahovska}(2013)}]{salipante2013electrohydrodynamic}%
  \BibitemOpen
  \bibfield  {author} {\bibinfo {author} {\bibfnamefont {P.~F.}\ \bibnamefont
  {Salipante}}\ and\ \bibinfo {author} {\bibfnamefont {P.}~\bibnamefont
  {Vlahovska}},\ }\bibfield  {title} {\bibinfo {title} {Electrohydrodynamic
  rotations of a viscous droplet},\ }\href@noop {} {\bibfield  {journal}
  {\bibinfo  {journal} {Physical Review E}\ }\textbf {\bibinfo {volume} {88}},\
  \bibinfo {pages} {043003} (\bibinfo {year} {2013})}\BibitemShut {NoStop}%
\bibitem [{\citenamefont {Salipante}\ and\ \citenamefont
  {Vlahovska}(2010)}]{salipante2010electrohydrodynamics}%
  \BibitemOpen
  \bibfield  {author} {\bibinfo {author} {\bibfnamefont {P.~F.}\ \bibnamefont
  {Salipante}}\ and\ \bibinfo {author} {\bibfnamefont {P.~M.}\ \bibnamefont
  {Vlahovska}},\ }\bibfield  {title} {\bibinfo {title} {Electrohydrodynamics of
  drops in strong uniform dc electric fields},\ }\href@noop {} {\bibfield
  {journal} {\bibinfo  {journal} {Physics of Fluids}\ }\textbf {\bibinfo
  {volume} {22}},\ \bibinfo {pages} {112110} (\bibinfo {year}
  {2010})}\BibitemShut {NoStop}%
\bibitem [{\citenamefont {Firouznia}\ \emph {et~al.}(2023)\citenamefont
  {Firouznia}, \citenamefont {Bryngelson},\ and\ \citenamefont
  {Saintillan}}]{firouznia2023spectral}%
  \BibitemOpen
  \bibfield  {author} {\bibinfo {author} {\bibfnamefont {M.}~\bibnamefont
  {Firouznia}}, \bibinfo {author} {\bibfnamefont {S.~H.}\ \bibnamefont
  {Bryngelson}},\ and\ \bibinfo {author} {\bibfnamefont {D.}~\bibnamefont
  {Saintillan}},\ }\bibfield  {title} {\bibinfo {title} {A spectral boundary
  integral method for simulating electrohydrodynamic flows in viscous drops},\
  }\href@noop {} {\bibfield  {journal} {\bibinfo  {journal} {Journal of
  Computational Physics}\ }\textbf {\bibinfo {volume} {489}},\ \bibinfo {pages}
  {112248} (\bibinfo {year} {2023})}\BibitemShut {NoStop}%
\bibitem [{\citenamefont {Varshney}\ \emph {et~al.}(2012)\citenamefont
  {Varshney}, \citenamefont {Ghosh}, \citenamefont {Bhattacharya},\ and\
  \citenamefont {Yethiraj}}]{varshney2012self}%
  \BibitemOpen
  \bibfield  {author} {\bibinfo {author} {\bibfnamefont {A.}~\bibnamefont
  {Varshney}}, \bibinfo {author} {\bibfnamefont {S.}~\bibnamefont {Ghosh}},
  \bibinfo {author} {\bibfnamefont {S.}~\bibnamefont {Bhattacharya}},\ and\
  \bibinfo {author} {\bibfnamefont {A.}~\bibnamefont {Yethiraj}},\ }\bibfield
  {title} {\bibinfo {title} {Self organization of exotic oil-in-oil phases
  driven by tunable electrohydrodynamics},\ }\href@noop {} {\bibfield
  {journal} {\bibinfo  {journal} {Sci. Rep.}\ }\textbf {\bibinfo {volume}
  {2}},\ \bibinfo {pages} {738} (\bibinfo {year} {2012})}\BibitemShut {NoStop}%
\bibitem [{\citenamefont {Varshney}\ \emph {et~al.}(2016)\citenamefont
  {Varshney}, \citenamefont {Gohil}, \citenamefont {Sathe}, \citenamefont {RV},
  \citenamefont {Joshi}, \citenamefont {Bhattacharya}, \citenamefont
  {Yethiraj},\ and\ \citenamefont {Ghosh}}]{varshney2016multiscale}%
  \BibitemOpen
  \bibfield  {author} {\bibinfo {author} {\bibfnamefont {A.}~\bibnamefont
  {Varshney}}, \bibinfo {author} {\bibfnamefont {S.}~\bibnamefont {Gohil}},
  \bibinfo {author} {\bibfnamefont {M.}~\bibnamefont {Sathe}}, \bibinfo
  {author} {\bibfnamefont {S.~R.}\ \bibnamefont {RV}}, \bibinfo {author}
  {\bibfnamefont {J.}~\bibnamefont {Joshi}}, \bibinfo {author} {\bibfnamefont
  {S.}~\bibnamefont {Bhattacharya}}, \bibinfo {author} {\bibfnamefont
  {A.}~\bibnamefont {Yethiraj}},\ and\ \bibinfo {author} {\bibfnamefont
  {S.}~\bibnamefont {Ghosh}},\ }\bibfield  {title} {\bibinfo {title}
  {Multiscale flow in an electro-hydrodynamically driven oil-in-oil emulsion},\
  }\href@noop {} {\bibfield  {journal} {\bibinfo  {journal} {Soft Matter}\
  }\textbf {\bibinfo {volume} {12}},\ \bibinfo {pages} {1759} (\bibinfo {year}
  {2016})}\BibitemShut {NoStop}%
\bibitem [{\citenamefont {Tadavani}\ \emph {et~al.}(2016)\citenamefont
  {Tadavani}, \citenamefont {Munroe},\ and\ \citenamefont
  {Yethiraj}}]{tadavani2016effect}%
  \BibitemOpen
  \bibfield  {author} {\bibinfo {author} {\bibfnamefont {S.~K.}\ \bibnamefont
  {Tadavani}}, \bibinfo {author} {\bibfnamefont {J.~R.}\ \bibnamefont
  {Munroe}},\ and\ \bibinfo {author} {\bibfnamefont {A.}~\bibnamefont
  {Yethiraj}},\ }\bibfield  {title} {\bibinfo {title} {The effect of
  confinement on the electrohydrodynamic behavior of droplets in a microfluidic
  oil-in-oil emulsion},\ }\href@noop {} {\bibfield  {journal} {\bibinfo
  {journal} {Soft Matter}\ }\textbf {\bibinfo {volume} {12}},\ \bibinfo {pages}
  {9246} (\bibinfo {year} {2016})}\BibitemShut {NoStop}%
\bibitem [{\citenamefont {Sozou}(1975)}]{sozou1975electrohydrodynamics}%
  \BibitemOpen
  \bibfield  {author} {\bibinfo {author} {\bibfnamefont {C.}~\bibnamefont
  {Sozou}},\ }\bibfield  {title} {\bibinfo {title} {Electrohydrodynamics of a
  pair of liquid drops},\ }\href@noop {} {\bibfield  {journal} {\bibinfo
  {journal} {Journal of Fluid Mechanics}\ }\textbf {\bibinfo {volume} {67}},\
  \bibinfo {pages} {339} (\bibinfo {year} {1975})}\BibitemShut {NoStop}%
\bibitem [{\citenamefont {Baygents}\ \emph {et~al.}(1998)\citenamefont
  {Baygents}, \citenamefont {Rivette},\ and\ \citenamefont
  {Stone}}]{baygents1998electrohydrodynamic}%
  \BibitemOpen
  \bibfield  {author} {\bibinfo {author} {\bibfnamefont {J.~C.}\ \bibnamefont
  {Baygents}}, \bibinfo {author} {\bibfnamefont {N.~J.}\ \bibnamefont
  {Rivette}},\ and\ \bibinfo {author} {\bibfnamefont {H.~A.}\ \bibnamefont
  {Stone}},\ }\bibfield  {title} {\bibinfo {title} {Electrohydrodynamic
  deformation and interaction of drop pairs},\ }\href@noop {} {\bibfield
  {journal} {\bibinfo  {journal} {Journal of Fluid Mechanics}\ }\textbf
  {\bibinfo {volume} {368}},\ \bibinfo {pages} {359} (\bibinfo {year}
  {1998})}\BibitemShut {NoStop}%
\bibitem [{\citenamefont {Zabarankin}(2020)}]{zabarankin}%
  \BibitemOpen
  \bibfield  {author} {\bibinfo {author} {\bibfnamefont {M.}~\bibnamefont
  {Zabarankin}},\ }\bibfield  {title} {\bibinfo {title} {Small deformation
  theory for two leaky dielectric drops in a uniform electric field},\
  }\href@noop {} {\bibfield  {journal} {\bibinfo  {journal} {Proceedings of the
  Royal Society A}\ }\textbf {\bibinfo {volume} {476}},\ \bibinfo {pages}
  {20190517} (\bibinfo {year} {2020})}\BibitemShut {NoStop}%
\bibitem [{\citenamefont {Sorgentone}\ \emph {et~al.}(2021)\citenamefont
  {Sorgentone}, \citenamefont {Kach}, \citenamefont {Khair}, \citenamefont
  {Walker},\ and\ \citenamefont
  {Vlahovska}}]{sorgentone_kach_khair_walker_vlahovska_2021}%
  \BibitemOpen
  \bibfield  {author} {\bibinfo {author} {\bibfnamefont {C.}~\bibnamefont
  {Sorgentone}}, \bibinfo {author} {\bibfnamefont {J.~I.}\ \bibnamefont
  {Kach}}, \bibinfo {author} {\bibfnamefont {A.~S.}\ \bibnamefont {Khair}},
  \bibinfo {author} {\bibfnamefont {L.~M.}\ \bibnamefont {Walker}},\ and\
  \bibinfo {author} {\bibfnamefont {P.~M.}\ \bibnamefont {Vlahovska}},\
  }\bibfield  {title} {\bibinfo {title} {Numerical and asymptotic analysis of
  the three-dimensional electrohydrodynamic interactions of drop pairs},\
  }\href {https://doi.org/10.1017/jfm.2020.1007} {\bibfield  {journal}
  {\bibinfo  {journal} {Journal of Fluid Mechanics}\ }\textbf {\bibinfo
  {volume} {914}},\ \bibinfo {pages} {A24} (\bibinfo {year}
  {2021})}\BibitemShut {NoStop}%
\bibitem [{\citenamefont {Kach}\ \emph {et~al.}(2022)\citenamefont {Kach},
  \citenamefont {Walker},\ and\ \citenamefont {Khair}}]{kach2022prediction}%
  \BibitemOpen
  \bibfield  {author} {\bibinfo {author} {\bibfnamefont {J.~I.}\ \bibnamefont
  {Kach}}, \bibinfo {author} {\bibfnamefont {L.~M.}\ \bibnamefont {Walker}},\
  and\ \bibinfo {author} {\bibfnamefont {A.~S.}\ \bibnamefont {Khair}},\
  }\bibfield  {title} {\bibinfo {title} {Prediction and measurement of leaky
  dielectric drop interactions},\ }\href@noop {} {\bibfield  {journal}
  {\bibinfo  {journal} {Physical Review Fluids}\ }\textbf {\bibinfo {volume}
  {7}},\ \bibinfo {pages} {013701} (\bibinfo {year} {2022})}\BibitemShut
  {NoStop}%
\bibitem [{\citenamefont {Sorgentone}\ and\ \citenamefont
  {Vlahovska}(2022)}]{Sorgentone_Vlahovska_2022}%
  \BibitemOpen
  \bibfield  {author} {\bibinfo {author} {\bibfnamefont {C.}~\bibnamefont
  {Sorgentone}}\ and\ \bibinfo {author} {\bibfnamefont {P.~M.}\ \bibnamefont
  {Vlahovska}},\ }\bibfield  {title} {\bibinfo {title} {Tandem droplet
  locomotion in a uniform electric field},\ }\href@noop {} {\bibfield
  {journal} {\bibinfo  {journal} {Journal of Fluid Mechanics}\ }\textbf
  {\bibinfo {volume} {951}},\ \bibinfo {pages} {R2} (\bibinfo {year}
  {2022})}\BibitemShut {NoStop}%
\bibitem [{\citenamefont {Kach}\ \emph {et~al.}(2023)\citenamefont {Kach},
  \citenamefont {Walker},\ and\ \citenamefont
  {Khair}}]{kach2023nonequilibrium}%
  \BibitemOpen
  \bibfield  {author} {\bibinfo {author} {\bibfnamefont {J.~I.}\ \bibnamefont
  {Kach}}, \bibinfo {author} {\bibfnamefont {L.~M.}\ \bibnamefont {Walker}},\
  and\ \bibinfo {author} {\bibfnamefont {A.~S.}\ \bibnamefont {Khair}},\
  }\bibfield  {title} {\bibinfo {title} {Nonequilibrium structure formation in
  electrohydrodynamic emulsions},\ }\href@noop {} {\bibfield  {journal}
  {\bibinfo  {journal} {Soft Matter}\ }\textbf {\bibinfo {volume} {19}},\
  \bibinfo {pages} {9179} (\bibinfo {year} {2023})}\BibitemShut {NoStop}%
\bibitem [{\citenamefont {Dong}\ \emph {et~al.}(2024)\citenamefont {Dong},
  \citenamefont {Xie}, \citenamefont {Zhou}, \citenamefont {Lu},\ and\
  \citenamefont {Wang}}]{dong24}%
  \BibitemOpen
  \bibfield  {author} {\bibinfo {author} {\bibfnamefont {Q.}~\bibnamefont
  {Dong}}, \bibinfo {author} {\bibfnamefont {Z.}~\bibnamefont {Xie}}, \bibinfo
  {author} {\bibfnamefont {X.}~\bibnamefont {Zhou}}, \bibinfo {author}
  {\bibfnamefont {J.}~\bibnamefont {Lu}},\ and\ \bibinfo {author}
  {\bibfnamefont {Z.}~\bibnamefont {Wang}},\ }\bibfield  {title} {\bibinfo
  {title} {Collective propulsion of viscous drop pairs based on {Q}uincke
  rotation in a uniform electric field},\ }\href@noop {} {\bibfield  {journal}
  {\bibinfo  {journal} {Physics of Fluids}\ }\textbf {\bibinfo {volume} {36}},\
  \bibinfo {pages} {017134} (\bibinfo {year} {2024})}\BibitemShut {NoStop}%
\bibitem [{\citenamefont {Hetsroni}\ and\ \citenamefont
  {Haber}(1970)}]{hetsroni1970flow}%
  \BibitemOpen
  \bibfield  {author} {\bibinfo {author} {\bibfnamefont {G.}~\bibnamefont
  {Hetsroni}}\ and\ \bibinfo {author} {\bibfnamefont {S.}~\bibnamefont
  {Haber}},\ }\bibfield  {title} {\bibinfo {title} {The flow in and around a
  droplet or bubble submerged in an unbound arbitrary velocity field},\
  }\href@noop {} {\bibfield  {journal} {\bibinfo  {journal} {Rheologica Acta}\
  }\textbf {\bibinfo {volume} {9}},\ \bibinfo {pages} {488} (\bibinfo {year}
  {1970})}\BibitemShut {NoStop}%
\bibitem [{\citenamefont {Rallison}(1984)}]{rallison1984}%
  \BibitemOpen
  \bibfield  {author} {\bibinfo {author} {\bibfnamefont {J.~M.}\ \bibnamefont
  {Rallison}},\ }\bibfield  {title} {\bibinfo {title} {The deformation of small
  viscous drops and bubbles in shear flows},\ }\href@noop {} {\bibfield
  {journal} {\bibinfo  {journal} {Annual Review of Fluid Mechanics}\ }\textbf
  {\bibinfo {volume} {16}},\ \bibinfo {pages} {45} (\bibinfo {year}
  {1984})}\BibitemShut {NoStop}%
\bibitem [{\citenamefont {Lamb}(1924)}]{lamb1924hydrodynamics}%
  \BibitemOpen
  \bibfield  {author} {\bibinfo {author} {\bibfnamefont {H.}~\bibnamefont
  {Lamb}},\ }\href@noop {} {\emph {\bibinfo {title} {Hydrodynamics}}}\
  (\bibinfo  {publisher} {University Press},\ \bibinfo {year}
  {1924})\BibitemShut {NoStop}%
\bibitem [{\citenamefont {Kim}\ and\ \citenamefont
  {Karrila}(2013)}]{kim2013microhydrodynamics}%
  \BibitemOpen
  \bibfield  {author} {\bibinfo {author} {\bibfnamefont {S.}~\bibnamefont
  {Kim}}\ and\ \bibinfo {author} {\bibfnamefont {S.~J.}\ \bibnamefont
  {Karrila}},\ }\href@noop {} {\emph {\bibinfo {title} {Microhydrodynamics:
  principles and selected applications}}}\ (\bibinfo  {publisher} {Courier
  Corporation},\ \bibinfo {year} {2013})\BibitemShut {NoStop}%
\bibitem [{\citenamefont {Happel}\ and\ \citenamefont
  {Brenner}(1965)}]{happelbrenner}%
  \BibitemOpen
  \bibfield  {author} {\bibinfo {author} {\bibfnamefont {J.}~\bibnamefont
  {Happel}}\ and\ \bibinfo {author} {\bibfnamefont {H.}~\bibnamefont
  {Brenner}},\ }\href@noop {} {\emph {\bibinfo {title} {Low Reynolds number
  hydrodynamics: with special applications to particulate media}}}\ (\bibinfo
  {publisher} {Prentice-Hall},\ \bibinfo {year} {1965})\BibitemShut {NoStop}%
\bibitem [{\citenamefont {Jones}(1979)}]{jones1979}%
  \BibitemOpen
  \bibfield  {author} {\bibinfo {author} {\bibfnamefont {T.~B.}\ \bibnamefont
  {Jones}},\ }\bibfield  {title} {\bibinfo {title} {Dielectrophoretic force
  calculation},\ }\href@noop {} {\bibfield  {journal} {\bibinfo  {journal}
  {Journal of Electrostatics}\ }\textbf {\bibinfo {volume} {6}},\ \bibinfo
  {pages} {69} (\bibinfo {year} {1979})}\BibitemShut {NoStop}%
\bibitem [{SM()}]{SM}%
  \BibitemOpen
  \href@noop {} {}\bibinfo {note} {See Supplemental Material at xxx for MATLAB
  codes that solves the system of ODEs equation (58) and reproduces the figures
  in the main text. Movies visualising drop dynamics corresponding to Figures
  5, 7 and isolated drops in Taylor and Quincke Regime are also
  provided.}\BibitemShut {Stop}%
\bibitem [{\citenamefont {Strogatz}(2015)}]{strogatz}%
  \BibitemOpen
  \bibfield  {author} {\bibinfo {author} {\bibfnamefont {S.~H.}\ \bibnamefont
  {Strogatz}},\ }\href@noop {} {\emph {\bibinfo {title} {Nonlinear Dynamics and
  Chaos: With Applications to Physics, Biology, Chemistry, and Engineering (2nd
  ed.)}}}\ (\bibinfo  {publisher} {CRC Press},\ \bibinfo {year}
  {2015})\BibitemShut {NoStop}%
\bibitem [{\citenamefont {{The MathWorks, Inc.}}(2024)}]{lsqnonlin}%
  \BibitemOpen
  \bibfield  {author} {\bibinfo {author} {\bibnamefont {{The MathWorks,
  Inc.}}},\ }\href {https://www.mathworks.com} {\bibinfo {title} {Optimization
  toolbox version: 24.2 (r2024b)}} (\bibinfo {year} {2024}),\ \bibinfo {note}
  {accessed: 2024-12-01}\BibitemShut {NoStop}%
\bibitem [{\citenamefont {Bossis}\ and\ \citenamefont
  {Brady}(1984)}]{bossis1984dynamic}%
  \BibitemOpen
  \bibfield  {author} {\bibinfo {author} {\bibfnamefont {G.}~\bibnamefont
  {Bossis}}\ and\ \bibinfo {author} {\bibfnamefont {J.~F.}\ \bibnamefont
  {Brady}},\ }\bibfield  {title} {\bibinfo {title} {Dynamic simulation of
  sheared suspensions. {I}. {G}eneral method},\ }\href@noop {} {\bibfield
  {journal} {\bibinfo  {journal} {The Journal of chemical physics}\ }\textbf
  {\bibinfo {volume} {80}},\ \bibinfo {pages} {5141} (\bibinfo {year}
  {1984})}\BibitemShut {NoStop}%
\bibitem [{\citenamefont {Takamura}\ \emph {et~al.}(1981)\citenamefont
  {Takamura}, \citenamefont {Goldsmith},\ and\ \citenamefont
  {Mason}}]{takamura1981microrheology}%
  \BibitemOpen
  \bibfield  {author} {\bibinfo {author} {\bibfnamefont {K.}~\bibnamefont
  {Takamura}}, \bibinfo {author} {\bibfnamefont {H.~L.}\ \bibnamefont
  {Goldsmith}},\ and\ \bibinfo {author} {\bibfnamefont {S.~G.}\ \bibnamefont
  {Mason}},\ }\bibfield  {title} {\bibinfo {title} {The microrheology of
  colloidal dispersions: {XII}. {T}rajectories of orthokinetic pair-collisions
  of latex spheres in a simple electrolyte},\ }\href@noop {} {\bibfield
  {journal} {\bibinfo  {journal} {Journal of Colloid and Interface Science}\
  }\textbf {\bibinfo {volume} {82}},\ \bibinfo {pages} {175} (\bibinfo {year}
  {1981})}\BibitemShut {NoStop}%
\bibitem [{\citenamefont {Bricard}\ \emph {et~al.}(2013)\citenamefont
  {Bricard}, \citenamefont {Caussin}, \citenamefont {Desreumaux}, \citenamefont
  {Dauchot},\ and\ \citenamefont {Bartolo}}]{bricard2013}%
  \BibitemOpen
  \bibfield  {author} {\bibinfo {author} {\bibfnamefont {A.}~\bibnamefont
  {Bricard}}, \bibinfo {author} {\bibfnamefont {J.-B.}\ \bibnamefont
  {Caussin}}, \bibinfo {author} {\bibfnamefont {N.}~\bibnamefont {Desreumaux}},
  \bibinfo {author} {\bibfnamefont {O.}~\bibnamefont {Dauchot}},\ and\ \bibinfo
  {author} {\bibfnamefont {D.}~\bibnamefont {Bartolo}},\ }\bibfield  {title}
  {\bibinfo {title} {Emergence of macroscopic directed motion in populations of
  motile colloids},\ }\href@noop {} {\bibfield  {journal} {\bibinfo  {journal}
  {Nature}\ }\textbf {\bibinfo {volume} {503}},\ \bibinfo {pages} {95}
  (\bibinfo {year} {2013})}\BibitemShut {NoStop}%
\bibitem [{\citenamefont {Bricard}\ \emph {et~al.}(2015)\citenamefont
  {Bricard}, \citenamefont {Caussin}, \citenamefont {Das}, \citenamefont
  {Savoie}, \citenamefont {Chikkadi}, \citenamefont {Shitara}, \citenamefont
  {Chepizhko}, \citenamefont {Peruani}, \citenamefont {Saintillan},\ and\
  \citenamefont {Bartolo}}]{bricard2015}%
  \BibitemOpen
  \bibfield  {author} {\bibinfo {author} {\bibfnamefont {A.}~\bibnamefont
  {Bricard}}, \bibinfo {author} {\bibfnamefont {J.-B.}\ \bibnamefont
  {Caussin}}, \bibinfo {author} {\bibfnamefont {D.}~\bibnamefont {Das}},
  \bibinfo {author} {\bibfnamefont {C.}~\bibnamefont {Savoie}}, \bibinfo
  {author} {\bibfnamefont {V.}~\bibnamefont {Chikkadi}}, \bibinfo {author}
  {\bibfnamefont {K.}~\bibnamefont {Shitara}}, \bibinfo {author} {\bibfnamefont
  {O.}~\bibnamefont {Chepizhko}}, \bibinfo {author} {\bibfnamefont
  {F.}~\bibnamefont {Peruani}}, \bibinfo {author} {\bibfnamefont
  {D.}~\bibnamefont {Saintillan}},\ and\ \bibinfo {author} {\bibfnamefont
  {D.}~\bibnamefont {Bartolo}},\ }\bibfield  {title} {\bibinfo {title}
  {Emergent vortices in populations of colloidal rollers},\ }\href@noop {}
  {\bibfield  {journal} {\bibinfo  {journal} {Nature Communications}\ }\textbf
  {\bibinfo {volume} {6}},\ \bibinfo {pages} {7470} (\bibinfo {year}
  {2015})}\BibitemShut {NoStop}%
\bibitem [{\citenamefont {Hu}\ \emph {et~al.}(2015)\citenamefont {Hu},
  \citenamefont {Lai},\ and\ \citenamefont {Young}}]{hu2015hybrid}%
  \BibitemOpen
  \bibfield  {author} {\bibinfo {author} {\bibfnamefont {W.-F.}\ \bibnamefont
  {Hu}}, \bibinfo {author} {\bibfnamefont {M.-C.}\ \bibnamefont {Lai}},\ and\
  \bibinfo {author} {\bibfnamefont {Y.-N.}\ \bibnamefont {Young}},\ }\bibfield
  {title} {\bibinfo {title} {A hybrid immersed boundary and immersed interface
  method for electrohydrodynamic simulations},\ }\href@noop {} {\bibfield
  {journal} {\bibinfo  {journal} {Journal of Computational Physics}\ }\textbf
  {\bibinfo {volume} {282}},\ \bibinfo {pages} {47} (\bibinfo {year}
  {2015})}\BibitemShut {NoStop}%
\bibitem [{\citenamefont {L{\'o}pez-Herrera}\ \emph {et~al.}(2011)\citenamefont
  {L{\'o}pez-Herrera}, \citenamefont {Popinet},\ and\ \citenamefont
  {Herrada}}]{lopez2011}%
  \BibitemOpen
  \bibfield  {author} {\bibinfo {author} {\bibfnamefont {J.~M.}\ \bibnamefont
  {L{\'o}pez-Herrera}}, \bibinfo {author} {\bibfnamefont {S.}~\bibnamefont
  {Popinet}},\ and\ \bibinfo {author} {\bibfnamefont {M.~A.}\ \bibnamefont
  {Herrada}},\ }\bibfield  {title} {\bibinfo {title} {A charge-conservative
  approach for simulating electrohydrodynamic two-phase flows using
  volume-of-fluid},\ }\href@noop {} {\bibfield  {journal} {\bibinfo  {journal}
  {Journal of Computational Physics}\ }\textbf {\bibinfo {volume} {230}},\
  \bibinfo {pages} {1939} (\bibinfo {year} {2011})}\BibitemShut {NoStop}%
\bibitem [{\citenamefont {Lobry}\ and\ \citenamefont {Lemaire}(1999)}]{lobry}%
  \BibitemOpen
  \bibfield  {author} {\bibinfo {author} {\bibfnamefont {L.}~\bibnamefont
  {Lobry}}\ and\ \bibinfo {author} {\bibfnamefont {E.}~\bibnamefont
  {Lemaire}},\ }\bibfield  {title} {\bibinfo {title} {Viscosity decrease
  induced by a {DC} electric field in a suspension},\ }\href@noop {} {\bibfield
   {journal} {\bibinfo  {journal} {Journal of Electrostatics}\ }\textbf
  {\bibinfo {volume} {47}},\ \bibinfo {pages} {61} (\bibinfo {year}
  {1999})}\BibitemShut {NoStop}%
\end{thebibliography}%

\end{document}